%% file: Journal.tex
\documentclass[journal,twocolumn,twoside,final]{IEEEtran}
\usepackage{cite}
\usepackage{graphicx}
\usepackage{psfrag}
\usepackage{url}
\usepackage{amsmath}
\usepackage{array}
\usepackage{amssymb}
\usepackage{amsthm}
\usepackage{mathtools}
\usepackage{amsfonts}
\usepackage{graphicx}
\usepackage{epstopdf}
\usepackage{algorithm}
\usepackage{algorithmic}
\usepackage{setspace}
\usepackage{amscd}
\usepackage{mathrsfs}
\usepackage{epsfig}
\usepackage{color}
\usepackage{textcomp}
\usepackage{tikz}
\usepackage{comment}
\usepackage[T1]{fontenc}
\usepackage{pgfplots}
\usepackage{float}
\usepackage{url}
\usepackage[normalem]{ulem}
\usetikzlibrary{shapes,snakes,plotmarks,positioning,fit,chains,tikzmark,arrows,hobby,arrows.meta}
\usetikzlibrary{decorations.pathreplacing,calligraphy}

\usepackage{caption}
\usepackage{subcaption}
\usepackage{glossaries}
\usepackage{glossaries-extra}

\newcommand{\Herm}[1]{{#1}^{\text{H}}}
\newcommand{\Tran}[1]{{#1}^{\text{T}}}
\newcommand{\Conj}[1]{{#1}^{\text{*}}}

\newcommand{\Brac}[1]{\left(#1\right)}
\newcommand{\Sbrac}[1]{\left[#1\right]}
\newcommand{\Cbrac}[1]{\left\{#1\right\}}

\newcommand{\Exp}[1]{\mathbb{E}\Cbrac{#1}}
\newcommand{\Abs}[1]{\left\vert #1 \right\vert}
\newcommand{\Norm}[1]{\left\Vert #1 \right\Vert}

\DeclareMathOperator{\trace}{Tr}
\DeclareMathOperator*{\argmax}{argmax}
\DeclareMathOperator*{\argmin}{argmin}
\DeclareMathOperator{\Vect}{vec}

\setabbreviationstyle[acronym]{long-short}
\newacronym{fwa}{FWA}{fixed wireless access}
\newacronym{mimo}{MIMO}{multiple-input multiple-output}
\newacronym{siso}{SISO}{single-input single-output}
\newacronym{5g}{5G}{fifth generation}
\newacronym{iid}{i.i.d.}{independent and identically distributed}
\newacronym{bs}{BS}{base station}
\newacronym{ue}{UE}{user equipment}
\newacronym{ap}{AP}{access point}
\newacronym{upa}{UPA}{uniform planar array}
\newacronym{los}{LoS}{line-of-sight}
\newacronym{awgn}{AWGN}{additive white Gaussian noise}
\newacronym{isp}{ISP}{internet service provider}
\newacronym{mmse}{MMSE}{minimum mean square error}
\newacronym{mse}{MSE}{mean square error}
\newacronym{rmse}{RMSE}{root mean square error}
\newacronym{zf}{ZF}{zero-forcing}
\newacronym{mrc}{MRC}{maximum-ratio combiner}
\newacronym{mrt}{MRT}{maximum-ratio transmission}
\newacronym{se}{SE}{spectral efficiency}
\newacronym{snr}{SNR}{signal-to-noise ratio}
\newacronym{sinr}{SINR}{signal-to-interference-plus-noise ratio}
\newacronym{rf}{RF}{radio frequency}
\newacronym{ofdm}{OFDM}{orthogonal frequency division multiplexing}
\newacronym{fcch}{FCCH}{frequency correction channel}
\newacronym{fb}{FB}{frequency correction burst signal}
\newacronym{gsm}{GSM}{global system for mobile}
\newacronym{crb}{CRB}{Cram{\'e}r-Rao lower bound}
\newacronym{fim}{FIM}{Fisher information matrix}
\newacronym{dft}{DFT}{discrete Fourier transform}
\newacronym{svd}{SVD}{singular value decomposition}
\newacronym{cp}{CP}{cyclic prefix}
\newacronym{lo}{LO}{local oscillator}
\newacronym{vco}{VCO}{voltage controlled oscillator}
\newacronym{pll}{PLL}{phase locked loop}
\newacronym{tdd}{TDD}{time division duplexing}
\newacronym{ml}{ML}{maximum likelihood}
\newacronym{csi}{CSI}{channel state information}
\newacronym{pcsi}{PCSI}{perfect channel state information}
\newacronym{cpu}{CPU}{central processing unit}
\newacronym{fgb}{FGB}{fixed grid of beams}
\newacronym{nls}{NLS}{non-linear least squares}
\newacronym{fdma}{FDMA}{frequency division multiple access}

\makeatletter
\newcommand\fs@spaceruled{\def\@fs@cfont{\bfseries}\let\@fs@capt\floatc@ruled
  \def\@fs@pre{\vspace{0.5\baselineskip}\hrule height.7pt depth0pt \kern2pt}%
  \def\@fs@post{\kern2pt\hrule\relax}%
  \def\@fs@mid{\kern2pt\hrule\kern2pt}%
  \let\@fs@iftopcapt\iftrue}
\makeatother

\tikzset{
	bs/.pic = {		
		\draw[line width = 1pt,-round cap] (0,0.4\R) -- (-0.2\R,-0.2\R);
		\draw[line width = 1pt,-round cap] (0,0.4\R) -- (0.2\R,-0.2\R);
		\draw[line width = 1pt] (-0.2\R,-0.2\R) -- (0.133\R,0);
		\draw[line width = 1pt] (0.2\R,-0.2\R) -- (-0.133\R,0);
		\draw[line width = 1pt,-round cap] (-0.133\R,0) -- (0.067\R,0.2\R);
		\draw[line width = 1pt,-round cap] (0.133\R,0) -- (-0.067\R,0.2\R);
		\draw[line width = 1pt,-round cap] (-0.213\R,0.4\R) -- (0.213\R,0.4\R);
		\draw[line width = 1pt,-round cap] \foreach \x in {-0.21, -0.14,...,0.22} {(\x\R,0.4\R) -- (\x\R,0.47\R)};
		\node (dim) at (0,0)  [align=center,minimum width=0.5\R,minimum height=1\R] {};
	},
}

\tikzset{
	user/.pic = {	
		\draw[rounded corners=0.02\R] (-0.09\R,-0.17\R) rectangle (0.09\R,0.17\R) ;	
		\draw[fill=gray] (-0.08\R,-0.12\R) rectangle (0.08\R,0.12\R);	
		\draw[rounded corners=0.005\R] (-0.04\R,0.14\R) rectangle (0.04\R,0.15\R);	
		\draw (0,-0.14\R) circle (0.012\R) ; 
		\node (dim) at (0,0)  [align=center,minimum width=0.2\R,minimum height=0.3\R] {};
	}
}

\allowdisplaybreaks

\begin{document}
\title{BeamSync: Over-The-Air Synchronization for Distributed Massive MIMO Systems
	\thanks{Manuscript received January 25, 2023; revised May 22, 2023 and August 22, 2023; accepted November 18, 2023. 
			This work was supported in part by ELLIIT and in part by the REINDEER project of the European Union`s Horizon 2020 research and innovation programme under grant agreement No. 101013425.
			A part of this article was presented at 25th International ITG Workshop on Smart Antennas (WSA 2021)~\cite{ganesan2021beamsync}.
			The editor coordinating the review of this article and approving it for publication was Dr. Chao-Kai Wen. (\textit{Corresponding author:	Unnikrishnan Kunnath Ganesan}.)}
}
\author{Unnikrishnan Kunnath Ganesan, \IEEEmembership{Graduate Student Member, IEEE,} \\
	Rimalapudi Sarvendranath, \IEEEmembership{Member, IEEE,}  \\
	Erik G. Larsson, \IEEEmembership{Fellow, IEEE}
\thanks{Unnikrishnan Kunnath Ganesan and Erik G. Larsson are with the Department of Electrical Engineering (ISY), Linköping University, 581 83 Linköping, Sweden (e-mail: unnikrishnan.kunnath.ganesan@liu.se; erik.g.larsson@liu.se).}
\thanks{Rimalapudi Sarvendranath is with Department of Electrical Engineering at the Indian Institute of Technology, Tirupati, India (email: sarvendranath@iitp.ac.in).} 
		\thanks{Digital Object Identifier XX}
}

\markboth{IEEE Transactions on Wireless Communications, Vol. XX, No. XX, 202X}{Ganesan \MakeLowercase{\textit{et al.}}: BeamSync: Over-The-Air Synchronization for Distributed Massive MIMO Systems}
\maketitle
\thispagestyle{empty}

%%%%%%%%
\begin{abstract}
In distributed massive \gls{mimo} systems, multiple geographically separated \glspl{ap} communicate simultaneously with a user, leveraging the benefits of multi-antenna coherent \gls{mimo} processing and macro-diversity gains from the distributed setups.
However, time and frequency synchronization of the multiple \glspl{ap} is crucial to achieve good performance and enable joint precoding.
In this paper, we analyze the synchronization requirement among multiple \glspl{ap} from a reciprocity perspective, taking into account the multiplicative impairments caused by mismatches in \gls{rf} hardware. 
We demonstrate that a phase calibration of reciprocity-calibrated \glspl{ap} is sufficient for the joint coherent transmission of data to the user.	
To achieve synchronization, we propose a novel over-the-air synchronization protocol, named BeamSync, to calibrate the geographically separated \glspl{ap} without sending any measurements to the \gls{cpu} through fronthaul.
We show that sending the synchronization signal in the dominant direction of the channel between \glspl{ap} is optimal. Additionally, we derive the optimal phase and frequency offset estimators.
Simulation results indicate that the proposed BeamSync method enhances performance by $3$~dB when the number of antennas at the \glspl{ap} is doubled. 
Moreover, the method performs well compared to traditional beamforming techniques.
\end{abstract}


\glsresetall

\section{Introduction}
\label{sec:Introduction}

\IEEEPARstart{M}{assive} \gls{mimo} introduced in~\cite{marzetta2010noncooperative}, has grown into the leading physical layer technology for \gls{5g} and beyond \gls{5g} wireless systems. 
Massive \gls{mimo} uses many antennas at the \gls{ap} and leverages \gls{tdd} to obtain \gls{csi} by exploiting the channel reciprocity. 
It serves multiple users on the same-time frequency resources and controls the multi-user interference through spatial precoding \cite{marzetta2016fundamentals}. 
To meet the increasing demand of wireless traffic, dense deployment of \glspl{ap} in the same coverage area is considered, creating small cells~\cite{chandrasekhar2008femtocell,ngo2017cell}. 
However, extreme densification causes intercell interference and to mitigate it, distributed networks are used~\cite{foschini2005value,karakayali2006network,caire2010rethinking,gesbert2010multi}. 
Large numbers of coordinated \glspl{ap} are distributed over a large coverage area and connected to a \gls{cpu} through a fronthaul to form a distributed network. Distributed massive \gls{mimo} or cell-free massive \gls{mimo} \cite{ngo2017cell,interdonato2019ubiquitous} is a distributed network topology that blends in the advantages of massive \gls{mimo} and ultra-densification.

Distributed massive \gls{mimo} relies on two important assumptions: channel reciprocity and synchronization of distributed transmitters. 
The \gls{tdd} architecture allows the \gls{csi} requirement only at the \gls{ap} to provide data transmission to users without the requirement of \gls{csi} at the receiver. 
The propagation channel from the antenna to the antenna is reciprocal; however, transmitter and receiver hardware are not and thus introducing a multiplicative gain with an unknown amplitude scaling and phase shift between the uplink and downlink channels. 
This makes the channel non-reciprocal in practice. 
For distributed massive \gls{mimo}, it suffices to calibrate the phases of reciprocity-calibrated \glspl{ap} to achieve coherent downlink transmission. 
A \gls{tdd} reciprocity calibration method was proposed in~\cite{shepard2012argos} for collocated massive \gls{mimo}. 
A channel reciprocity calibration method for industrial massive \gls{mimo} was proposed in \cite{lee2017calibration}, which involves incorporating the \gls{rf} impairment report into the power headroom report. 
In \cite{chen2017distributed}, a novel inter-cluster combining method was proposed for reciprocity calibration in distributed massive \gls{mimo}, which enhances the \gls{snr} of the sounding reference signals. 
In~\cite{kim2022gradual}, a gradual  method for   channel non-reciprocity calibration is proposed, which involves the estimation of uplink channels, after which \glspl{ap} are sequentially calibrated  in a particular order. 
The ordering is computed based on \gls{ap}-user channel strengths.

Synchronizing a receiver and transmitter for coherent data transmission is a well-known problem and efficient solutions exist in the literature~\cite{gsmetsi,moose1994technique,schmidl1997robust,huang2006carrier}. 
However, the synchronization techniques developed for a point-to-point communication system  do not extend directly to a distributed communication system. 
Each distributed \gls{ap} is equipped with a reference crystal oscillator to generate its clocks. 
Due to mismatches in the reference oscillator circuits, different transceivers generate different clocks. 
The drift and the jitter in the clock in different transmitters affect the coherent combining of signals at the users. 
To fully realize the benefits of distributed \gls{mimo}, and especially  perform joint coherent beamforming from multiple \glspl{ap} to users, it is essential that the participating \glspl{ap} are frequency and phase aligned. 
While GPS could be used for frequency alignment  in some deployments, for indoor environments synchronization with GPS may not be practical.
Most theoretical works on distributed massive \gls{mimo} assumes perfect synchronization~\cite{huh2011network,ramprashad2009cellular,boccardi2008network,marsch2008base,ngo2017cell,interdonato2019ubiquitous}. 
However, the assumption of perfect reciprocity calibration, and perfect phase and frequency synchronization of carrier is not very practical. 
The problems of achieving channel reciprocity calibration and synchronization in a distributed massive \gls{mimo} are entangled with each other. 
In this paper, we  consider reciprocity calibration and propose methods to synchronize distributed \glspl{ap} to enable coherent transmission to \gls{ue}.

The \gls{cpu} fronthaul network is unable to provide the timing and phase information to the \glspl{ap} and with unknown multiplicative gains from hardware, it becomes harder to synchronize the distributed \glspl{ap} through a fronthaul network. 
To address this issue, over-the-air synchronization methods were studied in~\cite{tu2002coherent,mudumbai2009distributed,rahman2012fully,quitin2012demonstrating,quitin2012distributed,balan2013airsync,rahman2012distributed,abari2015airshare,rogalin2014scalable}. 
Distributed coherent beamforming from multiple transmitters to a single user is studied in~\cite{tu2002coherent,mudumbai2009distributed} and the implementation of these ideas is studied in~\cite{rahman2012fully,quitin2012demonstrating,quitin2012distributed}. 
These works are based on a master-slave protocol, where a master \gls{ap} transmits a tone to slave \glspl{ap} for frequency synchronization and the \gls{ue} performs feedback signaling for phase synchronization. 
In practice, it is desirable to design a synchronization method without involving \glspl{ue} to enable legacy devices without implementing any new protocols in the new network and to save power. 
In the AirSync technique studied in~\cite{balan2013airsync}, a master \gls{ap} transmits out-of-band pilots continuously to all the slave \glspl{ap}, which receive them through a dedicated receive antenna. 
The slave \glspl{ap} keep track of the incoming phase offset from the master \gls{ap} and compensates it during the data transmission. 
During this phase compensation, a constant phase offset coming from the channel between master and slave \glspl{ap}, which is not compensated, becomes part of channel estimates. 
This assumes that the transmit and receive \gls{rf} chains of each transceiver are perfectly calibrated both in amplitude and phase, which need not necessarily be practical. 
The AirShare technique proposed in~\cite{abari2015airshare} uses a dedicated emitter to transmit two low-frequency tones over the air and the distributed \glspl{ap} use a dedicated circuit to receive these tones and generate their reference signal with the frequency equal to the difference of the two tones. 
This technique is robust to variations in temperature and supply voltage at the emitter. 
However, it does not compensate for phase impairments from the hardware. 
To provide synchronization in large distributed networks, \cite{rogalin2014scalable} considers a few \glspl{ap} as anchor \glspl{ap} that exchange pilot signals with each other, and the frequency and timing estimates are sent to the \gls{cpu} through the fronthaul to estimate the correcting factors. The other \glspl{ap} need to synchronize with the nearest anchor \gls{ap} using the master-slave approach. 
In \cite{vieira2021reciprocity}, the calibration data are collected through a beam-sweep approach between all pairs of the participating \glspl{ap} and are sent to the \gls{cpu}. 
An alternating optimization procedure is proposed to estimate the calibration coefficients and the complexity increases with the number of participating \glspl{ap}.

In this paper, we consider a generic scenario where transmit and receive gains in the same \gls{rf} chain are different due to internal clocking structures and manufacturing differences. 
A relative phase calibration of reciprocity-calibrated \glspl{ap} with one of the \glspl{ap} suffices for the coherent combining of signals at the user, thereby achieving the benefits of distributed systems.
The specific contributions of this paper can be summarized as follows:
\begin{itemize}
\item We propose a novel over-the-air synchronization technique, which we call \textit{BeamSync}, to synchronize the phase and carrier frequency between distributed \glspl{ap} without the involvement of a \gls{ue}. 
Each \gls{ap} is equipped with a single \gls{lo}.
The main practical use case for BeamSync is to calibrate the phases of two different \glspl{ap} relative to one another such that these \glspl{ap} can perform joint coherent beamforming on downlink.         
BeamSync can synchronize the phase and frequency of distributed \glspl{ap} without any estimation of the channel between \glspl{ap} and does not require the transmission of  any measurements to the \gls{cpu} through fronthaul.

\item We analytically derive the \gls{nls} estimator for the phase offset between distributed \glspl{ap} and propose a simple phase offset estimator which performs close to the \gls{nls} estimator at a high \gls{snr}.

\item We analytically derive the estimator for the frequency offset between distributed \glspl{ap} based on BeamSync. (A preliminary version of the same was published in the conference paper \cite{ganesan2021beamsync} considering perfect channel reciprocity.)

\item We analytically show that the synchronization signals should be beamformed in the dominant direction of the effective channel in which the signal is received. 

\item Through simulations, we show that BeamSync can achieve good phase calibration even in the presence of phase noise. 
Also, BeamSync performs well compared to the traditional fixed grid of beams beamforming techniques. 
Moreover, we show that the performance of BeamSync improves by $ 3 $ dB when doubling the number of antennas at the \gls{ap} for a fixed \gls{rmse} requirement.
\end{itemize}
 
The rest of the paper is organized as follows.  Section~\ref{sec:systemModel} discusses the system model considered in this paper. BeamSync for phase synchronization of distributed multi-antenna systems is presented in Section~\ref{sec:BeamSyncProtocol} and the frequency synchronization in Section~\ref{sec:frequencySynchornization}. Numerical results and the performance of the BeamSync protocol are discussed in Section~\ref{sec:Simulations}. Finally, concluding remarks are given in Section~\ref{sec:Conclusion}. 

\textbf{\textit{Notations:}} Bold lowercase letters are used to denote vectors and bold uppercase letters are used to denote matrices. 
$\mathbb{R}$ and $\mathbb{C}$ denote the set of real and complex numbers, respectively. 
For a matrix $\mathbf{A}$, $\Conj{\mathbf{A}}$, $\Tran{\mathbf{A}}$, and $\Herm{\mathbf{A}}$ denotes conjugate, transpose and conjugate transpose, respectively. 
$\mathcal{CN}(0,\sigma^2)$ denotes a circularly symmetric complex Gaussian random variable with zero mean and variance $\sigma^2$. 
The identity matrix of size $ K $ is denoted by $ \mathbf{I}_K $. 
$ (\cdot)_\text{R} $ and $ (\cdot)_\text{I} $ denote the real and imaginary parts, respectively. 
For a complex number $ a $, $ \Abs{a} $ and $ \angle{a} $ denotes the absolute value and the angle of $ a $. 
The operation $ \Vect(\cdot) $ denotes vectorization. 
The notations $ \Norm{\cdot}_\text{F} $ and  $ \Norm{\cdot}_2 $ denote the Frobenius and $ l_2 $ norms, respectively.

\section{System Model}
\label{sec:systemModel}

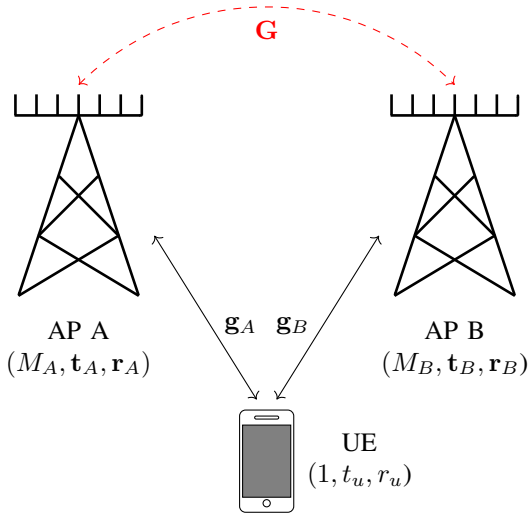
\begin{figure}[!t]
	\centering
	\begin{tikzpicture}
		\newdimen\R
		\R=4cm
		\draw (0,0) pic (bsA) {bs} ;
		\draw (5,0) pic (bsB) {bs} ;
		\draw (2.5,-3) pic (ue) {user} ;
		\draw[<->,shorten >= 0.25cm] (bsAdim.east) to node [align=center,midway,xshift=4mm,yshift=0mm]{$ \mathbf{g}_A $} (uedim.north) ;
		\draw[<->,shorten >= 0.25cm] (bsBdim.west) to node [align=center,midway,xshift=-4mm,yshift=0mm]{$ \mathbf{g}_B $}  (uedim.north) ;

		\draw[red,dashed,<->] (bsAdim.north) to[out=45,in=135] node [align=center,midway,xshift=0mm,yshift=-3mm]{$ \mathbf{G} $} (bsBdim.north) ;
		
		\node  [align=center,above=0.01mm of bsAdim.south] {AP A \\ $(M_A,\mathbf{t}_A,\mathbf{r}_A)$};
		\node  [align=center,above=0.01mm of bsBdim.south] {AP B \\ $(M_B,\mathbf{t}_B,\mathbf{r}_B$)};
		
		\node  [align=center,right=0.05mm of uedim.east] {UE \\ $(1,t_u,r_u$)};
		
	\end{tikzpicture}
	\caption{A distributed massive MIMO system with two \glspl{ap} serving a \gls{ue}. The triple $ (\cdot,\cdot,\cdot) $ refers to (number of antennas, transmit chain gains, receive chain gains) at each unit.}
	\label{fig:distributedMassiveMimo}
\end{figure}

Consider two multi-antenna \glspl{ap} A and B, equipped with $ M_A $ and $ M_B $ antennas, respectively, serving a single antenna \gls{ue} as shown in Fig.~\ref{fig:distributedMassiveMimo}. 
Each antenna at the \gls{ap} has a dedicated \gls{rf} chain and all the \gls{rf} chains in a given \gls{ap} are driven by a single \gls{lo}.
Let $ \mathbf{G}~\in~\mathbb{C}^{M_A \times M_B} $ be the reciprocal channel between A and B. 
Let $ \mathbf{t}_A~=~\Tran{\Sbrac{t^A_1   ~  t^A_2 ~ \cdots ~ t^A_{M_A} }} $, and $ \mathbf{r}_A~=~\Tran{\Sbrac{r^A_1   ~  r^A_2 ~ \cdots ~ r^A_{M_A} }} $ be the transmit and receive gains at all the \gls{rf} chains of \gls{ap}~A. 
Similarly let $ \mathbf{t}_B~=~\Tran{\Sbrac{t^B_1   ~  t^B_2 ~ \cdots ~ t^B_{M_B} }} $ and 
$ \mathbf{r}_B~=~\Tran{\Sbrac{r^B_1   ~  r^B_2 ~ \cdots ~ r^B_{M_B} }} $ be the transmit and receive gains at all the \gls{rf} chains of \gls{ap}~B. 
We assume that both \glspl{ap} are individually reciprocity calibrated according to~\cite{shepard2012argos}.
For clarity reciprocity calibration of a multi-antenna \gls{ap} is presented in the Appendix. 
During the transmission, each \gls{ap} multiplies with the internal calibration coefficient corresponding to each of the antennas. 
For example, for the $ m $th antenna at \gls{ap}~A, during the transmission of the signal, the calibration coefficient $ \frac{t^A_1 r^A_m}{r^A_1 t^A_m}$ will be multiplied with. 
This ensures that both the \glspl{ap} are reciprocity calibrated to their first antenna.  
For the coherent combining of signals from \glspl{ap} A and B, we require a phase calibration between the two \glspl{ap}. 
Mathematically, we need to estimate the phase offset $ (\angle{t^A_1} - \angle{r^A_1}) - (\angle{t^B_1} -\angle{r^B_1}) $ between the \glspl{ap} and correct it during the downlink transmission to the \gls{ue}.

\section{BeamSync: Phase Synchronization}
\label{sec:BeamSyncProtocol}
We propose an over-the-air synchronization procedure, BeamSync, to phase calibrate the reciprocity-calibrated \glspl{ap}.
We sent an omnidirectional pilot signal from \gls{ap} A, from which \gls{ap} B estimates the dominant direction of the reception of the signal. 
Then, \gls{ap} B beamforms the phase synchronization signal to \gls{ap} A in the dominant direction between the two \glspl{ap}. 
\gls{ap} A sends back this signal to \gls{ap} B for the estimation of phase offset. 
%With BeamSync, we beamform the phase synchronization signals from \gls{ap} A to \gls{ap} B and \gls{ap} B estimates and compensates the phase offset. 
BeamSync is a fully digital synchronization technique, and hence, we can beamform the signals in any $ 3 $-dimensional direction. 
Let $\varphi_A = \angle{t^A_1} - \angle{r^A_1}$ and $\varphi_B = \angle{t^B_1} - \angle{r^B_1}$ and $\varphi = \varphi_A - \varphi_B$. 
Our proposed over-the-air phase synchronization protocol between two \glspl{ap} consists of three stages as follows:

\subsubsection*{Stage-I}
Let $\boldsymbol{\Phi} = \Sbrac{\boldsymbol{\phi}_{1}, \boldsymbol{\phi}_{2}, \cdots, \boldsymbol{\phi}_{M_A} } \in \mathbb{C}^{L \times M_A}$ be a pilot matrix at \gls{ap} A such that $\Herm{\mathbf{\Phi}} \mathbf{\Phi}=\mathbf{I}_{M_A}$ and $ L \geq M_A $ is the pilot signal length. 
In the first stage, \gls{ap} A sends the pilot matrix $ \boldsymbol{\Phi} $ and the received signal at \gls{ap} B, $ \mathbf{Y}_{B1} \in \mathbb{C}^{M_B\times L} $ is given by
\begin{equation}
	\begin{aligned}
		\mathbf{Y}_{B1} & = \frac{t^A_1}{r^A_1} \mathbf{D}_{\mathbf{r}_B} \Tran{\mathbf{G}} 	 	\mathbf{D}_{\mathbf{r}_A} 	\Tran{\mathbf{\Phi}} + \mathbf{W}_{B1} \\
						& = \frac{t^A_1}{r^A_1} \mathbf{G}_e^\text{T} \Tran{\mathbf{\Phi}} + \mathbf{W}_{B1}, 
	\end{aligned}
\end{equation}
where $ \mathbf{G}_e = \mathbf{D}_{\mathbf{r}_A} \mathbf{G} \mathbf{D}_{\mathbf{r}_B} $ is the effective channel matrix and  $\mathbf{W}_{B1}$ is \gls{awgn} with \gls{iid} $\mathcal{CN}(0,\sigma^2)$ entries with $\sigma^2$ as the noise power. 
The notation $ \mathbf{D}_\mathbf{a} $ denotes a diagonal matrix with the elements of the vector $ \mathbf{a} $ as   diagonal entries.
%The term $  \frac{t^A_1}{r^A_1} \mathbf{D}_{\mathbf{r}_A} 	\mathbf{D}_{\mathbf{t}_A}^{-1} $ corresponds to the reciprocity calibration of \gls{ap}~A.

\subsubsection*{Stage-II}
\Gls{ap} B determines a unitary beamforming vector $\mathbf{a} \in\mathbb{C}^{M_B\times1}$ from the received signal $\mathbf{Y}_{B1}$. Let $\mathbf{x} \in \mathbb{C}^{N\times1}$ be phase synchronization signal such that $ \Norm{\mathbf{x}}^2 = N$. In the second stage, \gls{ap} B beamforms $\mathbf{x}$ in the direction $\mathbf{a}$. The signal received at \gls{ap} A $, \mathbf{Y}_{A1} \in \mathbb{C}^{M_A\times N} $ can be written as 
\begin{equation}
	\mathbf{Y}_{A1} = \frac{t^B_1}{r^B_1} \mathbf{G}_e  \mathbf{a} \Tran{\mathbf{x}} + \mathbf{W}_{A1}, 
\end{equation}
where $\mathbf{W}_{A1}$ is \gls{awgn} with \gls{iid} $\mathcal{CN}(0,\sigma^2)$ entries.

\subsubsection*{Stage-III}
Let 
\begin{equation}
	c = \frac{N}{\Norm{\mathbf{Y}_{A1}}_F^2}. 
\end{equation}
In the third stage, \gls{ap} A transmits a scaled version of the conjugate of the signal received during the second stage to \gls{ap} B. The received signal at \gls{ap} B, $\mathbf{Y}_{B2} \in \mathbb{C}^{M_B\times N}$ can be written as
\begin{equation}
	\label{eqn:rxSignalPhaseEstimationStage3}
\begin{aligned}
	\mathbf{Y}_{B2} & =  \sqrt{c}  \frac{t^A_1}{r^A_1}  \mathbf{G}_e^{\text{T}} \mathbf{Y}_{A1}^* + \mathbf{W}_{B2} \\
	& = \sqrt{c}  \frac{t^A_1}{r^A_1} \mathbf{G}_e^{\text{T}} \Brac{\frac{t^B_1}{r^B_1} \mathbf{G}_e \mathbf{a} \Tran{\mathbf{x}} + \mathbf{W}_{A1}}^* + \mathbf{W}_{B2} ,
\end{aligned}
\end{equation}
where $\mathbf{W}_{B2}$ is \gls{awgn} with \gls{iid} $\mathcal{CN}(0,\sigma^2)$ entries. 

\subsection{Phase Offset Estimation}
Let 
\begin{equation}
\begin{aligned}
	c_1 & = \sqrt{c} \frac{\Abs{t_1^A}\Abs{t_1^B}}{\Abs{r_1^A}\Abs{r_1^B}} \\
	c_2 & = \sqrt{c} \frac{\Abs{t_1^A}}{\Abs{r_1^A}}. 
\end{aligned}
\end{equation}
Then \eqref{eqn:rxSignalPhaseEstimationStage3} can we written as 
\begin{equation}
	\begin{aligned}
		\mathbf{Y}_{B2} &= e^{j\varphi}c_1 \mathbf{G}_e^\text{T} \mathbf{G}_e^\text{*} \mathbf{a}^* \Herm{\mathbf{x}} + e^{j\varphi_A} c_2  \mathbf{G}_e^\text{T} \mathbf{W}_{A1}^\text{*} + \mathbf{W}_{B2}  \\
		& = e^{j\varphi}c_1 \mathbf{G}_e^\text{T} \mathbf{G}_e^\text{*} \mathbf{a}^* \Herm{\mathbf{x}} + \mathbf{W}^{\prime}, 
	\end{aligned}
\end{equation}
where $ \mathbf{W}^{\prime} = e^{j\varphi_A} c_2  \mathbf{G}_e^\text{T} \mathbf{W}_{A1}^\text{*} + \mathbf{W}_{B2} $. 
Conditioned on $ \mathbf{G} $ and $ c_2 $, each column of $ \mathbf{W}^{\prime} $ is distributed as $ \mathcal{CN}(0, \Brac{\mathbf{I}+c_2^2 \mathbf{G}_e^\text{T}\mathbf{G}_e^\text{*}}\sigma^2) $. 
Let 
\begin{equation}
\begin{aligned}
\mathbf{Z} & = c_1 \mathbf{G}_e^\text{T} \mathbf{G}_e^\text{*} \mathbf{a}^* \Herm{\mathbf{x}} \\
\mathbf{R} & = \mathbf{I}+c_2^2 \mathbf{G}_e^\text{T}\mathbf{G}_e^\text{*}.
\end{aligned}
\end{equation}
Then $ \mathbf{Y}_{B2} = e^{j\varphi} \mathbf{Z} + \mathbf{W}^{\prime}  $. 
Using \gls{nls} estimation, the estimate of $\varphi$ is given by 
\begin{equation}
	\label{eqn:phaseMLReduction}
\begin{aligned}
\hat{\varphi} & = \argmin_\varphi \trace\Cbrac{ \Herm{\Brac{\mathbf{Y}_{B2}-e^{j\varphi}\mathbf{Z}}} \mathbf{R}^{-1} 			 \Brac{\mathbf{Y}_{B2}-e^{j\varphi}\mathbf{Z}} } \\
			  & = \argmin_\varphi \trace\Cbrac{ \mathbf{R}^{-\frac{1}{2}} \Brac{\mathbf{Y}_{B2}-e^{j\varphi}\mathbf{Z}}  \Herm{\Brac{\mathbf{Y}_{B2}-e^{j\varphi}\mathbf{Z}}} \mathbf{R}^{-\frac{1}{2}} }   \\
%			  & = \argmin_\varphi \trace\Cbrac{ \Brac{ \mathbf{R}^{-\frac{1}{2}} \mathbf{Y}_{B2}-e^{j\varphi}\mathbf{R}^{-\frac{1}{2}} \mathbf{Z}}   \Herm{\Brac{\mathbf{R}^{-\frac{1}{2}} \mathbf{Y}_{B2}-e^{j\varphi}\mathbf{R}^{-\frac{1}{2}}\mathbf{Z}}} }  \\
			  & = \argmin_\varphi \Norm{ \mathbf{R}^{-\frac{1}{2}} \mathbf{Y}_{B2}-e^{j\varphi}\mathbf{R}^{-\frac{1}{2}} \mathbf{Z}}_\text{F}^2 \\
			  & = \argmin_\varphi \Norm{ \Vect\Brac{\mathbf{R}^{-\frac{1}{2}} \mathbf{Y}_{B2}}-e^{j\varphi}\Vect\Brac{\mathbf{R}^{-\frac{1}{2}} \mathbf{Z}}}_2^2 .
\end{aligned}
\end{equation}
From \eqref{eqn:phaseMLReduction}, the estimate of $\varphi$ is the angle between the two vectors $ \Vect\Brac{\mathbf{R}^{-\frac{1}{2}} \mathbf{Y}_{B2}} $ and $ \Vect\Brac{\mathbf{R}^{-\frac{1}{2}} \mathbf{Z}} $. Mathematically, 
\begin{equation}
	\label{eqn:phaseMLEstimate}
	\begin{aligned}
		\hat{\varphi} & = \angle\Brac{ \Vect\Brac{\mathbf{R}^{-\frac{1}{2}} \mathbf{Z}}^\text{H} \Vect\Brac{\mathbf{R}^{-\frac{1}{2}} \mathbf{Y}_{B2}} } \\
		& = \angle \Brac{\trace\Cbrac{\mathbf{Z}^\text{H}\mathbf{R}^{-1}\mathbf{Y}_{B2}}} \\
		& = \angle \Brac{\trace\Cbrac{c_1 \mathbf{x}\Tran{\mathbf{a}} \mathbf{G}_e^\text{T} \mathbf{G}_e^\text{*} \mathbf{R}^{-1}\mathbf{Y}_{B2}}} \\
		& = \angle \Brac{\Tran{\mathbf{a}} \mathbf{G}_e^\text{T} \mathbf{G}_e^\text{*} \mathbf{R}^{-1}\mathbf{Y}_{B2} \mathbf{x} } \\
		& = \angle \Brac{\Tran{\mathbf{a}} \mathbf{G}_e^\text{T} \mathbf{G}_e^\text{*} \Brac{\mathbf{I} + c_2^2 \mathbf{G}_e^\text{T} \mathbf{G}_e^\text{*} }^{-1}\mathbf{Y}_{B2} \mathbf{x} }.
	\end{aligned}
\end{equation}
%The effective channel $ \mathbf{G}_e $ is unknown at \gls{ap} B and to estimate $ \mathbf{G}_e $ the \glspl{ap} need to spend resources of $ \mathcal{O}(M_AM_B) $.
The complexity of the \gls{nls} estimator is $\mathcal{O}(M_B^3)$ due to the matrix inversion term. 
Thus, as the number of antennas in the \gls{ap} increases the complexity also increases.
However, if the \gls{snr} is high at stage-III, we can approximate $ \mathbf{R} $ as $  c_2^2 \mathbf{G}_e^\text{T} \mathbf{G}_e^\text{*} $, which is possible in practice as the \glspl{ap} are not power limited. Thus, the estimate for $\varphi$ at high \gls{snr} is given by 
\begin{equation}
	\label{eqn:phaseSubOptimalEstimate}
		\hat{\varphi} 	= \angle\Brac{\Tran{\mathbf{a}} \mathbf{Y}_{B2} \mathbf{x} } .
%		& = \arctan \left(\frac{\Im\{\mathbf{a}^\text{T} \mathbf{Y}_{B2}\mathbf{x}\}}{\Re\{\mathbf{a}^\text{T} \mathbf{Y}_{B2}\mathbf{x}\}}\right).
\end{equation}
The simple sub-optimal estimator in \eqref{eqn:phaseSubOptimalEstimate} performs the same as \gls{nls} estimator at high \gls{snr} with significantly lower complexity, and  requiring only matrix multiplications.

\subsection{Optimal Beamforming Direction}
\label{sec:OptimalBeamformingDirectionPhase}
Let 
\begin{equation}
	b = \Norm{\Brac{\mathbf{I} + c_2^2 \mathbf{G}_e^\text{T} \mathbf{G}_e^\text{*} }^{-\frac{1}{2}} \mathbf{G}_e^\text{T} \mathbf{G}_e^\text{*} \mathbf{a}^\text{*}}_2^2.
\end{equation}
Consider the test statistic in \eqref{eqn:phaseMLEstimate}
\begin{equation}
	\begin{aligned}
	y_B & =  \Tran{\mathbf{a}} \mathbf{G}_e^\text{T} \mathbf{G}_e^\text{*} \Brac{\mathbf{I} + c_2^2 \mathbf{G}_e^\text{T} \mathbf{G}_e^\text{*} }^{-1}\mathbf{Y}_{B2} \mathbf{x} \\
	& = \Tran{\mathbf{a}} \mathbf{G}_e^\text{T} \mathbf{G}_e^\text{*} \Brac{\mathbf{I} + c_2^2 \mathbf{G}_e^\text{T} \mathbf{G}_e^\text{*} }^{-1} (e^{j\varphi}c_1 \mathbf{G}_e^\text{T} \mathbf{G}_e^\text{*} \mathbf{a}^* \Herm{\mathbf{x}} + \mathbf{W}^{\prime})  \mathbf{x} \\
		& = e^{j\varphi} c_1 b \Norm{\mathbf{x}}_2^2 + w \\
		& = e^{j\varphi} c_1 N b + w , 
	\end{aligned}
	\label{eqn:optimalTestStatistic}
\end{equation}
where $ w = \Tran{\mathbf{a}} \mathbf{G}_e^\text{T} \mathbf{G}_e^\text{*}  \Brac{\mathbf{I} + c_2^2 \mathbf{G}_e^\text{T} \mathbf{G}_e^\text{*} }^{-1} \mathbf{W}^{\prime} \mathbf{x} $. 
In \eqref{eqn:optimalTestStatistic}, the columns of the matrix $\mathbf{W}'$ are colored with covariance $\mathbf{R}$. Hence, the noise power depends on the choice of phase synchronization signal $\mathbf{x}$. 
From \eqref{eqn:optimalTestStatistic}, the estimation error in $\varphi$ can be reduced by increasing the synchronization signal length $N$ as well as by maximizing the value of $b$.
Specifically, the \gls{snr} in (\ref{eqn:optimalTestStatistic}) is proportional to both $N$ and $b$. 
In this subsection, we find the optimal beamforming direction $ \mathbf{a} $ that maximizes $b$.

Let the \gls{svd} of the effective channel $\mathbf{G}_e$ be 
\begin{equation}
	\label{eqn:svdEffectiveChannel}
	\mathbf{G}_e = \mathbf{U} \mathbf{\Sigma} \mathbf{V}^\text{H},
\end{equation}
where $\mathbf{U} \in \mathbb{C}^{M_A\times M_A} $ and $\mathbf{V} \in \mathbb{C}^{M_B\times M_B} $ are unitary matrices and $\mathbf{\Sigma}\in\mathbb{R}^{M_A\times M_B}$ with singular values of $\mathbf{G}_e$ in decreasing order on its principal diagonal. Then, we have 
\begin{equation}
\mathbf{G}_e^\text{T}  \mathbf{G}_e^\text{*} = \mathbf{V}^\text{*} \mathbf{\Lambda}\mathbf{V}^\text{T},
\label{eqn:svdEffectiveChannelTranConj}
\end{equation}
where $ \mathbf{\Lambda} = \mathbf{\Sigma}^\text{T} \mathbf{\Sigma} \in \mathbb{R}^{M_B\times M_B}$ is a diagonal matrix. 
Let $ \lambda_1 \geq \lambda_2 \geq \cdots \geq \lambda_{M_B} $ be the diagonal elements of $\mathbf{\Lambda}$ in decreasing order.
Substituting \eqref{eqn:svdEffectiveChannelTranConj} for $ b $, we have
\begin{equation}
\begin{aligned}
b & =  \mathbf{a}^\text{T} \mathbf{G}_e^\text{T}  \mathbf{G}_e^\text{*}  \Brac{ \mathbf{I} + c_2^2\mathbf{G}_e^\text{T}  \mathbf{G}_e^\text{*}}^{-1}  \mathbf{G}_e^\text{T}  \mathbf{G}_e^\text{*}  \mathbf{a}^\text{*} \\
 & = \mathbf{a}^\text{T} \mathbf{V}^\text{*} \mathbf{\Lambda} \mathbf{V}^\text{T} \Brac{ \mathbf{V}^\text{*} \mathbf{V}^\text{T} + c_2^2\mathbf{V}^\text{*} \mathbf{\Lambda} \mathbf{V}^\text{T}}^{-1}  \mathbf{V}^\text{*} \mathbf{\Lambda} \mathbf{V}^\text{T}  \mathbf{a}^\text{*} \\
 & = \mathbf{a}^\text{T} \mathbf{V}^\text{*} \mathbf{\Lambda}  \Brac{ \mathbf{I} + c_2^2 \mathbf{\Lambda}  }^{-1} \mathbf{\Lambda} \mathbf{V}^\text{T}  \mathbf{a}^\text{*} \\
 & =  \mathbf{a}^\text{T} \mathbf{V}^\text{*} \mathbf{\Lambda}^{\prime} \mathbf{V}^\text{T}  \mathbf{a}^\text{*},
\end{aligned}
\label{eqn:phaseOptimalBeamformingDirection}
\end{equation}
where $ \mathbf{\Lambda}^{\prime} = \mathbf{\Lambda}  \Brac{ \mathbf{I} + c_2^2 \mathbf{\Lambda}  }^{-1} \mathbf{\Lambda} $ is a diagonal matrix.
Consider the difference between the first and the second diagonal elements of $\mathbf{\Lambda}'$, i.e.,
\begin{equation}
\begin{aligned}
\frac{\lambda_1^2}{1+c_2^2\lambda_1} - \frac{\lambda_2^2}{1+c_2^2\lambda_2} & = \frac{(\lambda_1 - \lambda_2  )( \lambda_1 + \lambda_2 +  c_2^2\lambda_1\lambda_2)}{(1+c_2^2\lambda_1)(1+c_2^2\lambda_2)} \\
& \geq 0.
\end{aligned}
\end{equation}
Thus, the diagonal entries of $\mathbf{\Lambda}'$ is also in the decreasing order. 
The value of $b$ can be maximized by choosing $ \mathbf{a} = \mathbf{v}_1 $, where the vector $\mathbf{v}_1$ is the first column of the matrix $\mathbf{V}$, which corresponds to the dominant direction of the effective channel in which the signal was received at \gls{ap} B from \gls{ap} A. Thus, by beamforming the synchronization signal $\mathbf{x}$ in the dominant direction of the effective channel, a better estimate of the phase offset $\varphi$ could be obtained.

Consider the test statistic in \eqref{eqn:phaseMLEstimate} and when we beamform the synchronization signal in the optimal direction, the statistic can be further reduced as follows: 
\begin{equation}
\begin{aligned}
\hat{\varphi}	& = \angle \Brac{\Tran{\mathbf{a}} \mathbf{G}_e^\text{T} \mathbf{G}_e^\text{*} \Brac{\mathbf{I} + c_2^2 \mathbf{G}_e^\text{T} \mathbf{G}_e^\text{*} }^{-1}\mathbf{Y}_{B2} \mathbf{x} } \\
& = \angle \Brac{\mathbf{v}_1^\text{T} \mathbf{V}^\text{*} \mathbf{\Lambda}\mathbf{V}^\text{T} \Brac{\mathbf{V}^\text{*} \mathbf{V}^\text{T} + c_2^2 \mathbf{V}^\text{*} \mathbf{\Lambda}\mathbf{V}^\text{T} }^{-1}\mathbf{Y}_{B2} \mathbf{x} } \\
& = \angle \Brac{ \frac{\lambda_1}{1+c_2^2\lambda_1}  \mathbf{v}_1^\text{T} \mathbf{Y}_{B2} \mathbf{x} } \\
& = \angle \Brac{ \mathbf{v}_1^\text{T} \mathbf{Y}_{B2} \mathbf{x} }.
\end{aligned}
\end{equation}
Thus, if we beamform in the optimal direction, the optimal test statistic in \eqref{eqn:phaseMLEstimate} becomes same as the simpler test statistic in \eqref{eqn:phaseSubOptimalEstimate}.

\subsection{Estimating the Beamforming Direction}
\label{subsec:BeamformingDirection}
From Sec. \ref{sec:OptimalBeamformingDirectionPhase}, it is evident that the optimal direction to beamform is the dominant direction of the effective channel in which \gls{ap} B receives the signal from the \gls{ap} A. 
In practice, the effective channel $\mathbf{G}_e$ is not perfectly known at either \glspl{ap}. However, as the effective channel is reciprocal, the \gls{ap} B can estimate the dominant direction from the signal received in stage-I, without the need to estimate the effective channel, by computing the \gls{svd} of the received signal $ \mathbf{Y}_{B1} $. Let
\begin{equation}
	\mathbf{Y}_{B1} = \mathbf{U}_B \mathbf{\Sigma}_B \mathbf{V}_B^\text{H}, 
\end{equation}
where $\mathbf{U}_B \in \mathbb{C}^{M_B\times M_B} $ and $\mathbf{V}_B \in \mathbb{C}^{L\times L} $ are unitary matrices and $\mathbf{\Sigma}_B\in\mathbb{R}^{M_B\times L}$ is a diagonal matrix with the singular values of $\mathbf{Y}_{B1} $ in decreasing order. 
%The columns of $\mathbf{U}_B$ correspond to the direction of the received signal ordered according to the dominance of power received in each direction. 
The dominant direction in which the signal is received is $\mathbf{u}_{B1}$, where $\mathbf{u}_{B1}$ is the first column of $\mathbf{U}_{B}$. Thus, the beamforming direction is chosen as $ \mathbf{a} = \mathbf{u}_{B1}^\text{*} $.
The BeamSync protocol to synchronize the phase difference between \glspl{ap} A and B is provided in algorithm~\ref{alg:phaseSyncAlgo}.
In terms of computational complexity, the algorithm consists of linear operations plus a computation of the dominant eigenvector of a matrix. 
The dominant eigenvector can be computed by a variety of algorithms, for example, a power iteration~\cite{epperson2021introduction}.

\floatstyle{spaceruled}
\restylefloat{algorithm}
\begin{algorithm}[!t]
	\caption{\strut Phase Synchronization Algorithm}
	\begin{algorithmic}[1]
		\label{alg:phaseSyncAlgo}
		\STATE Transmit omnidirectional pilot signal $ \boldsymbol{\Phi} $ from \gls{ap} A
		\STATE Compute the \gls{svd} of the received signal $ \mathbf{Y}_{B1} = \mathbf{U}_B\boldsymbol{\Sigma}_B\mathbf{V}_B^\text{H}$ at \gls{ap} B
		\STATE \gls{ap} B sends the phase synchronization signal $\mathbf{x}$ by beamforming in the direction $ \mathbf{a} = \mathbf{u}_{B1}^\text{*} $
		\STATE \gls{ap} A transmits the conjugate of the received signal back to \gls{ap} B
		\STATE \gls{ap} B estimates the phase difference $\varphi = \angle\Brac{\Tran{\mathbf{a}} \mathbf{Y}_{B2} \mathbf{x} }$ and precompensates its phase during transmission.		
	\end{algorithmic}
\end{algorithm}

\subsection{Phase Synchronization Among Multiple APs}\label{subsec:PhaseSyncProtocol}

\begin{figure}
	\centering
	\resizebox{0.45\textwidth}{!}{%
	\begin{tikzpicture}
		\node[rectangle, rounded corners] (master) at (0,0) [align=center, minimum height =8mm, draw] {master};
		\node[rectangle, rounded corners] (sec1) [align=center, minimum height =8mm, right=10mm of master,draw] {slave-1};
		\node[rectangle, rounded corners] (sec2) [align=center, minimum height =8mm, right=10mm of sec1,draw] {slave-2};
		\node (sec3) [align=center, minimum height =10mm, right=10mm of sec2] {$\cdots$};
		\node[rectangle, rounded corners] (seck) [align=center, minimum height =8mm, right=10mm of sec3,draw] {slave-K};

		\draw [<->, dashed, shorten <= 1mm, shorten >= 1mm] (master) to[out=80,in=100]  node [align=center,midway,below]{$\mathbf{G}_{1}$} (sec1);
		\draw [<->, dashed, shorten <= 1mm, shorten >= 1mm] (master) to[out=80,in=100]  node [align=center,midway,below]{$\mathbf{G}_{2}$} (sec2);
		\draw [<->, dashed, shorten <= 1mm, shorten >= 1mm] (master) to[out=80,in=100]  node [align=center,midway,below]{$\mathbf{G}_{K}$} (seck);
		
		%\draw[->] ([xshift=-4mm,yshift=-4mm]master.south) to[out=-90,in=90]  node [midway,rotate=90,above]{Time} +(-90:6.5cm);
		
		\draw (master.south) -- +(-90:7cm);
		\draw (sec1.south) -- +(-90:7cm);
		\draw (sec2.south) -- +(-90:7cm);
		\draw (seck.south) -- +(-90:7cm);
		
		\draw[-{Latex[round]}]  ([yshift=-10mm]master.south) to[out=0,in=180]  node [at start,anchor=south west]{Pilot $\mathbf{\Phi}$} ([yshift=-10mm]sec1.south) ;
		\draw[-{Latex[round]}]  ([yshift=-10mm]master.south) -- ([yshift=-10mm]sec2.south) ;
		\draw[-{Latex[round]}]  ([yshift=-10mm]master.south) -- ([yshift=-10mm]seck.south) ;

		\draw[{Latex[round]}-]  ([yshift=-20mm]master.south) to[out=0,in=180]  node [at start,anchor=south west]{Sync signal} ([yshift=-20mm]sec1.south) ;	
		\draw[-{Latex[round]}]  ([yshift=-20mm]master.south) to[out=-10,in=180]  node[at end, align=left,right=1pt]{Estimate $\varphi_1$}  ([yshift=-25mm]sec1.south) ;			
		
		\draw[{Latex[round]}-]  ([yshift=-35mm]master.south) to[out=0,in=180]  node [at start,anchor=south west]{Sync signal} ([yshift=-35mm]sec2.south) ;	
		\draw[-{Latex[round]}]  ([yshift=-35mm]master.south) to[out=-10,in=180]  node[at end, align=left,right=1pt]{Estimate $\varphi_2$}  ([yshift=-40mm]sec2.south) ;			
		
		\node (dotsnode) [align=center, minimum height =10mm, below=40mm of sec3] {$\cdots$};
		\node [align=center, minimum height =10mm, left=50mm of dotsnode] {$\vdots$};
		
		\draw[{Latex[round]}-]  ([yshift=-60mm]master.south) to[out=0,in=180]  node [at start,anchor=south west]{Sync signal} ([yshift=-60mm]seck.south) ;	
		\draw[-{Latex[round]}]  ([yshift=-60mm]master.south) to[out=-10,in=180]  node[at end, align=left,right=1pt]{Estimate $\varphi_K$}  ([yshift=-65mm]seck.south) ;			
		
		\node[rectangle] (process) [align=center, below left=5mm and 30mm of master, rotate=90,minimum height =8mm, fill=yellow,draw] {Master sends back\\conjugate of received\\ signal};					
		
		\draw[->,dotted]  (process.south) -- ([yshift=-20mm]master.south) ;
		\draw[->,dotted]  (process.south) -- ([yshift=-35mm]master.south) ;
		\draw[->,dotted]  (process.south) -- ([yshift=-60mm]master.south) ;

		\node[rectangle] (process1) [align=center, below right=5mm and 30mm of seck, rotate=-90,minimum height =8mm, fill=green,draw] {Slaves estimates the \\dominant direction of \\signal reception};
		
		\draw[->,green]  (process1.south) to[out=180,in=-10] ([yshift=-10mm]sec1.south) ;
		\draw[->,green]  (process1.south) to[out=180,in=-10] ([yshift=-10mm]sec2.south) ;
		\draw[->,green]  (process1.south) to[out=180,in=-10] ([yshift=-10mm]seck.south) ;

	\end{tikzpicture}
}
	\caption{BeamSync protocol for phase synchronization.}
	\label{fig:BeamSyncProtocolPhaseSync}

\end{figure}
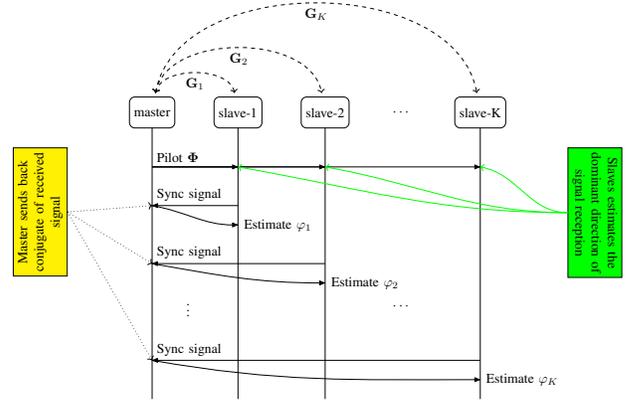

In this section, we consider how to extend the BeamSync phase synchronization algorithm to the case of multiple \glspl{ap}. 
To synchronize the \glspl{ap} to a common reference, we nominate one of the \glspl{ap} as the master \gls{ap} and the remaining \glspl{ap} as slave \glspl{ap}. 
Let there be $K$ slave \glspl{ap} in the network. 
Let $\mathbf{G}_k$ be the channel between the master \gls{ap} and the $k$th slave \gls{ap}. The BeamSync protocol for phase synchronization among multiple \glspl{ap} is illustrated in Fig.~\ref{fig:BeamSyncProtocolPhaseSync}. 
The stage-I pilot signal is broadcast from the master \gls{ap} to all slave \glspl{ap} and all the slave \glspl{ap} estimates its dominant direction of the signal received from the master \gls{ap}. 
Note that, the pilot signal from the master needs to be sent only once during the entire synchronization. 
Afterward, sequentially or multiplexed in frequency, to avoid interference, each slave \gls{ap} sends the phase synchronization signal to the master \gls{ap} (stage-II) and the master sends back the conjugate of the received signal (stage-III) to each of the \glspl{ap} for estimation of the phase offset. 
The frequency sub-bands and timeslots will not be orthogonal until the system is completely synchronized. 
For synchronization purposes, guard bands in time and frequency need to be inserted to avoid interference due to non-orthogonality.

The synchronization scheme in \cite{vieira2021reciprocity} requires all the participating \glspl{ap} to sequentially transmit beams in all directions and listen to others. 
This requires a large number of measurements to be carried out. 
Considering $M$ antennas per each \gls{ap}, the total number of measurements required is $(K+1)M^2$ and all these measurements need to be sent to the \gls{cpu} for joint estimation of the phase offsets, which requires additional fronthaul signaling. 
However, with BeamSync, the master \gls{ap} needs to send the pilot signal once, which requires only $L$ measurements. 
The slave \glspl{ap} then estimate the dominant direction of the received signal and carry out the synchronization procedure using stage-II and III sequentially, requiring a total of $2N$ measurements for each master-slave pair and hence, a total of $2KN$ measurements.

\section{BeamSync: Frequency Synchronization}
\label{sec:frequencySynchornization}
In this section, we discuss how BeamSync can be used for synchronizing the carrier frequency between two reciprocity-calibrated \glspl{ap}. 
As the local oscillators driving the RF chain of the different base station differs, there exists a carrier frequency offset between the base stations. Let $\Delta \in \mathbb{R}$ be the frequency offset between A and B.

\subsection{BeamSync for Frequency Synchronization}
The protocol consists of two stages described as follows:
\subsubsection*{Stage-I}
\Gls{ap} B transmits an orthonormal pilot sequence of length $ L_B \geq M_B $ from each of its antennas. Let the columns of the matrix $\mathbf{\Phi}_\text{B} \in \mathbb{C}^{L_B \times M_B}$, where  $\mathbf{\Phi}_\text{B}^\text{H}\mathbf{\Phi}_\text{B}=\mathbf{I}_{M_B}$,  denote the set of orthonormal pilot sequences. 
Let 
\begin{equation}
\boldsymbol{\Delta}_\tau = \Tran{\Sbrac{ e^{j2\pi \Delta T} ~ e^{j2\pi 2\Delta T} ~ \cdots ~ e^{j2\pi \tau\Delta T} }} \in \mathbb{C}^{\tau\times1},
\end{equation}
where $T$ is the symbol arrival time.
The collective signal received in $L_B$ time instants at \gls{ap} A, $\mathbf{Y}_A \in \mathbb{C}^{M_A\times L_B} $ can be written as 
\begin{equation}
			\mathbf{Y}_A  =  \frac{t^B_1}{r^B_1} \mathbf{G}_e \Tran{\mathbf{\Phi}}_\text{B} \mathbf{D}_{\boldsymbol{\Delta}_{L_B}}  + \mathbf{W}_A,
\end{equation}
where $\mathbf{W}_A \in  \mathbb{C}^{M_A\times L_B} $ is \gls{awgn} with \gls{iid} $\mathcal{CN}(0,\sigma^2)$ entries.

\subsubsection*{Stage-II}
\Gls{ap} A processes the signal $\mathbf{Y}_A$ received in stage-I and determines a unitary beamforming vector $\mathbf{a}_\text{f}~\in~\mathbb{C}^{M_A\times1}$. It then beamforms an $ N_f$-length frequency synchronization signal $\mathbf{x}_\text{f}$ towards \gls{ap} B. Moreover, we assume that $ \mathbf{x}_{\text{f}} $ is real-valued.
%to have a constant rotation of the recieved signal in $ (n+1) $th and $ n $th time slots. 
The received signal at \gls{ap} B, $\mathbf{Y}_B \in \mathbb{C}^{M_B \times N_f} $ can be written as 
\begin{equation}
	\label{eqn:SecondarySyncSignal}
	\mathbf{Y}_B = \frac{t^A_1}{r^A_1} \mathbf{G}_e^{\text{T}} \mathbf{a}_\text{f} \Tran{\mathbf{x}}_\text{f} \mathbf{D}^*_{\boldsymbol{\Delta}_{N_f}} + \mathbf{W}_B,
\end{equation}
where $\mathbf{W}_B \in \mathbb{C}^{M_B\times N_f}$ is \gls{awgn} with \gls{iid} $\mathcal{CN}(0,\sigma^2)$ entries. \Gls{ap} B needs to estimate its frequency offset $\Delta $ with respect to \gls{ap} A from (\ref{eqn:SecondarySyncSignal}). 
Let 
\begin{equation}
\mathbf{b} = \frac{t^A_1}{r^A_1} \mathbf{G}_e^{\text{T}} \mathbf{a}_\text{f}.
\end{equation}
Then (\ref{eqn:SecondarySyncSignal}) can be rewritten as 
\begin{equation}
	\label{eqn:SecondarySyncSignal_2}
	\mathbf{Y}_B = \mathbf{b} \Tran{\mathbf{x}}_\text{f} \mathbf{D}^*_{\boldsymbol{\Delta}_{N_f}}  + \mathbf{W}_B. 
\end{equation}
The joint maximum-likelihood estimates of $\mathbf{b}$ and $\Delta$ are given by 
\begin{equation}
	\label{eqn:joint_estimate}
	(\hat{\mathbf{b}},\hat{\Delta}) = \argmin_{\mathbf{b},\Delta} \lVert \mathbf{Y}_B - \mathbf{b} \Tran{\mathbf{x}}_\text{f} \mathbf{D}^*_{\boldsymbol{\Delta}_{N_f}} \rVert^2.
\end{equation}
Solving (\ref{eqn:joint_estimate}) using \gls{nls} estimation in Gaussian noise~\cite[Sec. 8.9]{kay1993fundamentals} with $\mathbf{b}$ as a nuisance parameter, the estimates of $\mathbf{b}$ and $\Delta$ are given by 
\begin{align}
	\hat{\mathbf{b}} & = \frac{\mathbf{Y}_B \mathbf{D}_{\boldsymbol{\Delta}_{N_f}} \mathbf{x}_\text{f}}{ \Norm{\mathbf{x}_\text{f}}^2  }  \\
	\hat{\Delta} & = \argmax_\Delta \lVert \mathbf{Y}_B \mathbf{D}_{\boldsymbol{\Delta}_{N_f}} \mathbf{x}_\text{f} \rVert^2.
\end{align}
In the next subsection, we discuss the optimal beamforming direction and the parameters affecting the frequency offset estimation performance.

\subsection{Optimal Beamforming Direction}
\label{sec:OptimalBeamforming}
Here, we derive the optimal beamforming direction that minimizes the frequency offset estimation error. We look at the conditions for which the \gls{crb} on the estimate of $\Delta$, is minimized. We have $\mathbf{b} = \mathbf{b}_\text{R} + j \mathbf{b}_\text{I}$. Let 
\begin{equation}
	\boldsymbol{\theta} = [ \mathbf{b}_\text{R}^\text{T} \ \mathbf{b}_\text{I}^\text{T} \ \Delta]^\text{T}
\end{equation}
be the unknown vector parameter at \gls{ap} B. From (\ref{eqn:SecondarySyncSignal_2}), the signal received at the $ n^\text{th} $ time instant, $\mathbf{y}_B(n) \in \mathbb{C}^{M_B\times1}$ is given by 
\begin{equation}
	\mathbf{y}_B (n) = \mathbf{b} x_{\text{f}} (n) e^{-j2\pi n \Delta} + \mathbf{w}_B(n) ,
\end{equation}
where $ x_{\text{f}}(n) $ is the $ n $th component of $ \mathbf{x}_{\text{f}} $ and $\mathbf{w}_B(n)$ is the $ n $th column of $\mathbf{W}_B$. Also $ \mathbf{y}_B(n) $ is 
distributed as $\mathcal{CN}(\mathbf{b} x_\text{f}(n) e^{-j 2 \pi n \Delta }, \sigma^2\mathbf{I}_{M_B})$. 
Let $ \mathbf{y}_{B\text{R}}(n) $ and $ \mathbf{y}_{B\text{I}}(n) $ be the real and imaginary parts of $\mathbf{y}_B(n) $.
Thus, $\bar{\mathbf{y}}_B(n)~=~[\mathbf{y}_{B\text{R}}^\text{T}(n) \ \mathbf{y}_{B\text{I}}^\text{T}(n)]^\text{T}~\in~\mathbb{R}^{2M_B\times 1}$ is distributed as $\mathcal{N}(\boldsymbol{\mu}_n(\boldsymbol{\theta}),\mathbf{C}(\boldsymbol{\theta}))$, where $ \boldsymbol{\mu}_n $ and $\mathbf{C}$ are mean vector and covariance matrix, respectively, of $\bar{\mathbf{y}}_B$ parameterized by $ \boldsymbol{\theta} $, and are given by
\begin{align}\label{eqn:mean}
	\boldsymbol{\mu}_n(\boldsymbol{\theta}) & = x_\text{f}(n) \begin{bmatrix*}[r]
		\mathbf{b}_\text{R} \cos(2\pi n \Delta) + \mathbf{b}_\text{I} \sin(2\pi n \Delta) \\
		-\mathbf{b}_\text{R} \sin(2\pi n \Delta) + \mathbf{b}_\text{I} \cos(2\pi n \Delta) \\
	\end{bmatrix*} \\
	\label{eqn:covariance}
	\mathbf{C}(\boldsymbol{\theta}) &  = \frac{\sigma^2}{2}\mathbf{I}_{2M_B}. 
\end{align}

Using the Slepian Bang theorem \cite[Sec. 3.9]{kay1993fundamentals}, each element of the \gls{fim} of $\boldsymbol{\theta}$ at $ n^\text{th} $ time instant, $ \mathbf{J}_n(\boldsymbol{\theta})~\in~\mathbb{R}^{(2M_B+1) \times (2M_B+1)} $, can be computed as 
\begin{equation}
	\begin{aligned}
	[\mathbf{J}_n(\boldsymbol{\theta})]_{k,l} & = 
\Tran{\Sbrac{\frac{\partial \boldsymbol{\mu}_n(\boldsymbol{\theta})}{\partial \boldsymbol{\theta}_k}}}  \mathbf{C}^{-1}(\boldsymbol{\theta})  
\Sbrac{\frac{\partial \boldsymbol{\mu}_n(\boldsymbol{\theta})}{\partial \boldsymbol{\theta}_l} } \\
& + \frac{1}{2} \trace\left[ \mathbf{C}^{-1}(\boldsymbol{\theta}) \frac{\partial \mathbf{C}(\boldsymbol{\theta})}{\partial \boldsymbol{\theta}_k}\mathbf{C}^{-1}(\boldsymbol{\theta}) \frac{\partial \mathbf{C}(\boldsymbol{\theta})}{\partial \boldsymbol{\theta}_l}  \right] .
	\end{aligned}
\end{equation}
By computing the partial derivatives of \eqref{eqn:mean} and \eqref{eqn:covariance}, we obtain
\begin{equation}\label{eqn:FIM_n}
	\mathbf{J}_n(\boldsymbol{\theta}) = \frac{2 x_\text{f}^2(n) }{\sigma^2}
	\begin{bmatrix*}[c]
		\mathbf{I}_{M_B} & \mathbf{0} & 2\pi n \mathbf{b}_{\text{I}}\\
		\mathbf{0} &  \mathbf{I}_{M_B} & -2\pi n  \mathbf{b}_{\text{R}} \\
		2\pi n  \mathbf{b}_{\text{I}}^{\text{T}} & -2\pi n \mathbf{b}_{\text{R}}^{\text{T}} & 4\pi^2 n^2  \lVert \mathbf{b} \rVert^2
	\end{bmatrix*}.
\end{equation}
The received signal $\mathbf{y}_B(n)$ is independent for different time instants. Thus, using the additive property of FIM, the overall FIM of $\boldsymbol{\theta}$, $ \mathbf{J}(\boldsymbol{\theta}) $, is given by 
\begin{equation}\label{eqn:FIM}
	\mathbf{J}(\boldsymbol{\theta}) = \sum_{n=1}^{N_f}    \mathbf{J}_n(\boldsymbol{\theta}).
\end{equation}
The \gls{crb} of $\hat{\Delta}$ can be computed from $ \text{J} (\boldsymbol{\theta}) $ as 
\begin{equation}
	\text{CRB}(\hat{\Delta}) = \left[\mathbf{J}^{-1}(\boldsymbol{\theta}) \right]_{2M_B+1,2M_B+1}, 
\end{equation}
which is the lower right corner element of $ \mathbf{J}^{-1}(\boldsymbol{\theta}) $. Using the inverse of a block partitioned matrix~\cite[Sec. 0.7.3]{horn2012matrix}, the \gls{crb} of $\hat{\Delta}$ is given by 
\begin{equation}\label{eqn:CRB_Delta}
	\text{CRB}(\hat{\Delta}) = \frac{\sigma^2}{ 8 \pi^2  \Norm{\mathbf{b}}^2 \left( \sum_{n=1}^{N_f} n^2x_\text{f}^2(n) - \frac{(\sum_{n=1}^{N_f} n x_\text{f}^2(n))^2}{\sum_{n=1}^{N_f} x_\text{f}^2(n)}\right)  }.
\end{equation}
From (\ref{eqn:CRB_Delta}), the \gls{crb} of $\hat{\Delta}$ will be minimized when 
\begin{equation}
\Norm{\mathbf{b}}^2~=~  \Brac{\frac{\Abs{t^A_1}}{\Abs{r^A_1}}}^2 \Norm{  \mathbf{G}_e^{\text{T}} \mathbf{a}_\text{f}}^2 
\end{equation}
is maximized. 
From \eqref{eqn:svdEffectiveChannel}, we have 
\begin{equation}
		\label{eqn:OptimalBeamForming}
		\Norm{\mathbf{G}_e^{\text{T}} \mathbf{a}_\text{f}}^2  = \mathbf{a}_\text{f}^\text{H}\mathbf{U}^\text{*}\mathbf{\Sigma} \mathbf{\Sigma}^\text{T} \mathbf{U}^\text{T} \mathbf{a}_\text{f}, 
\end{equation}
and can be maximized by choosing $ \mathbf{a}_\text{f}=\mathbf{u}_{1}^*$. The vector $\mathbf{u}_{1}$, is the first column of matrix $\mathbf{U}$. Hence, the optimal beamforming direction $\mathbf{a}_\text{f}$ corresponds to the dominant direction of the effective channel in which the signal is received at \gls{ap} A from \gls{ap} B. Similar to the phase synchronization, the beamforming vector can be obtained by \gls{ap} A from $\mathbf{Y}_A$. Let the \gls{svd} of $\mathbf{Y}_A$ be
\begin{equation}
\mathbf{Y}_A = \mathbf{U}_A \mathbf{\Sigma}_A \mathbf{V}_A^\text{H}.
\end{equation}
Then the beamforming vector is chosen as $ \mathbf{a}_\text{f} = \mathbf{u}_{A1}^\text{*} $, where $ \mathbf{u}_{A1} $ is the first column of matrix $\mathbf{U}_A$. 

From (\ref{eqn:CRB_Delta}), for estimating $\Delta$, the synchronization signal length $ N_f $ should be at least 2. 
Also, as $N_f$ increases, the denominator  in \eqref{eqn:CRB_Delta} increases. Thus, increasing the synchronization sequence length $ N_f $  improves the estimation quality. 
Moreover, the \gls{crb} of the estimate of $\Delta$ can be reduced by increasing the \gls{snr}, which in turn improves the estimation quality.
The frequency synchronization algorithm is provided in algorithm~\ref{alg:frequencySyncAlgo}.

\floatstyle{spaceruled}
\restylefloat{algorithm}
\begin{algorithm}[!t]
	\caption{\strut Frequency Synchronization Algorithm}
	\begin{algorithmic}[1]
		\label{alg:frequencySyncAlgo}
		\STATE Transmit omnidirectional pilot signal $ \boldsymbol{\Phi}_\text{B} $ from \gls{ap} B
		\STATE Compute \gls{svd} of received signal $ \mathbf{Y}_{A} = \mathbf{U}_A\boldsymbol{\Sigma}_A\mathbf{V}_A^\text{H}$ at \gls{ap} A
		\STATE \gls{ap} A sends the frequency synchronization signal $\mathbf{x}_\text{f}$ by beamforming in the direction $ \mathbf{a}_\text{f} = \mathbf{u}_{A1}^\text{*} $
		\STATE \gls{ap} B estimates the carrier frequency offset $ \hat{\Delta} $ and precompensate its frequency during transmission.		
	\end{algorithmic}
\end{algorithm}

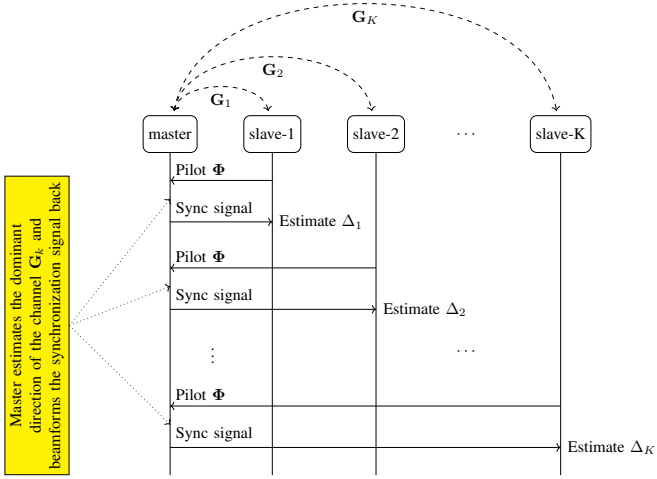
\begin{figure}
	\centering
	\resizebox{0.49\textwidth}{!}{
	\begin{tikzpicture}
		\node[rectangle, rounded corners] (master) at (0,0) [align=center, minimum height =8mm, draw] {master};
		\node[rectangle, rounded corners] (sec1) [align=center, minimum height =8mm, right=10mm of master,draw] {slave-1};
		\node[rectangle, rounded corners] (sec2) [align=center, minimum height =8mm, right=10mm of sec1,draw] {slave-2};
		\node (sec3) [align=center, minimum height =10mm, right=10mm of sec2] {$\cdots$};
		\node[rectangle, rounded corners] (seck) [align=center, minimum height =8mm, right=10mm of sec3,draw] {slave-K};

		\draw [<->, dashed, shorten <= 1mm, shorten >= 1mm] (master) to[out=80,in=100]  node [align=center,midway,below]{$\mathbf{G}_{1}$} (sec1);
		\draw [<->, dashed, shorten <= 1mm, shorten >= 1mm] (master) to[out=80,in=100]  node [align=center,midway,below]{$\mathbf{G}_{2}$} (sec2);
		\draw [<->, dashed, shorten <= 1mm, shorten >= 1mm] (master) to[out=80,in=100]  node [align=center,midway,below]{$\mathbf{G}_{K}$} (seck);
		
		%\draw[->] ([xshift=-4mm,yshift=-4mm]master.south) to[out=-90,in=90]  node [midway,rotate=90,above]{Time} +(-90:6.5cm);
		
		\draw (master.south) -- +(-90:7cm);
		\draw (sec1.south) -- +(-90:7cm);
		\draw (sec2.south) -- +(-90:7cm);
		\draw (seck.south) -- +(-90:7cm);
		
		\draw[<-]  ([yshift=-6mm]master.south) to[out=0,in=180]  node [at start,anchor=south west]{Pilot $\mathbf{\Phi}$} ([yshift=-6mm]sec1.south) ;
		\draw[->]  ([yshift=-15mm]master.south) to[out=0,in=180]  node [at start,anchor=south west]{Sync signal} node[at end, align=left,right=1pt]{Estimate $\Delta_1$}  ([yshift=-15mm]sec1.south) ;		
		
		\draw[<-]  ([yshift=-25mm]master.south) to[out=0,in=180]  node [at start,anchor=south west]{Pilot $\mathbf{\Phi}$} ([yshift=-25mm]sec2.south) ;
		\draw[->]  ([yshift=-34mm]master.south) to[out=0,in=180]  node [at start,anchor=south west]{Sync signal} node[at end, align=left,right=1pt]{Estimate $\Delta_2$}  ([yshift=-34mm]sec2.south) ;				
		
		\node (dotsnode) [align=center, minimum height =10mm, below=37mm of sec3] {$\cdots$};
		\node [align=center, minimum height =10mm, left=50mm of dotsnode] {$\vdots$};
		
		\draw[<-]  ([yshift=-55mm]master.south) to[out=0,in=180]  node [at start,anchor=south west]{Pilot $\mathbf{\Phi}$} ([yshift=-55mm]seck.south) ;
		\draw[->]  ([yshift=-64mm]master.south) to[out=0,in=180]  node [at start,anchor=south west]{Sync signal} node[at end, align=left,right=1pt]{Estimate $\Delta_K$}  ([yshift=-64mm]seck.south) ;	
		
		\node[rectangle] (process) [align=center, below left=5mm and 30mm of master, rotate=90,minimum height =8mm, fill=yellow,draw] {Master estimates the dominant\\direction of the channel $\mathbf{G}_k$ and \\beamforms the synchronization signal back};					
		
		\draw[->,dotted]  (process.south) -- ([yshift=-10mm]master.south) ; ;
		\draw[->,dotted]  (process.south) -- ([yshift=-29mm]master.south) ;
		\draw[->,dotted]  (process.south) -- ([yshift=-59mm]master.south) ;
	\end{tikzpicture} 
}
	\caption{BeamSync protocol for frequency synchronization.}
	\label{fig:BeamSyncProtocolFrequencySync}
\end{figure}

\subsection{Frequency Synchronization Among Multiple APs}\label{subsec:FrequencySyncProtocol}
	In this section, we extend the BeamSync frequency synchronization algorithm to the case of multiple \glspl{ap}. We consider the same scenario as in Sec.~\ref{subsec:PhaseSyncProtocol}. The BeamSync protocol for frequency synchronization among multiple \glspl{ap} is illustrated in Fig.~\ref{fig:BeamSyncProtocolFrequencySync}. The slave \glspl{ap} synchronize with the master \gls{ap} in a sequential fashion. In each synchronization slot, the slave \gls{ap} sends the pilot to the master \gls{ap} for the estimation of the dominant direction of the received signal  (stage-I). Then the master sends the frequency synchronization signal to the slave \gls{ap} for the estimation of the frequency offset.

\subsection{BeamSync Protocol} \label{subsec:BeamsyncProtocol}
\begin{figure}[!t]
	\centering
	\resizebox{0.48\textwidth}{!}{
	\begin{tikzpicture}
		\node (a) at (0,0)  [align=center,minimum height=10mm,minimum width=5mm,draw,fill=yellow] {};
		\node (b) [align=center,minimum height=10mm,minimum width=5mm,right=0 of a,draw,fill=green]{};
		\node (c) [align=center,minimum height=10mm,minimum width=10mm,right=0 of b,draw,fill=red]{Data};
		\node (d) [align=center,minimum height=10mm,minimum width=5mm,right=0 of c,draw,fill=green]{};
		\node (e) [align=center,minimum height=10mm,minimum width=10mm,right=0 of d,draw,fill=red]{Data};
		\node (f) [align=center,minimum height=10mm,minimum width=10mm,right=0 of e]{$ \cdots $};
		\node (g) [align=center,minimum height=10mm,minimum width=5mm,right=0 of f,draw,fill=green]{};
		\node (h) [align=center,minimum height=10mm,minimum width=10mm,right=0 of g,draw,fill=red]{Data};
%		\node (i) [align=center,minimum height=10mm,minimum width=5mm,right=0 of h,draw,fill=yellow] {};
		\node (j) [align=center,minimum height=10mm,minimum width=5mm,right=0 of h,draw,fill=green]{};
		\node (k) [align=center,minimum height=10mm,minimum width=10mm,right=0 of j,draw,fill=red]{Data};
		\node (l) [align=center,minimum height=10mm,minimum width=10mm,right=0 of k]{$ \cdots $};
	
		\draw[->,thick] (-2.5mm,-5mm)--(80mm,-5mm) node[below]{time};
		
		\node (fs)  [align=center,above=10mm of a,draw,fill=yellow]{Frequency\\Synchronization};
		\draw[->] (fs.south) -- (a.north);
%		\draw[->] (fs.south) -- (i.north);
		
		\node (ps)  [align=center,right=10mm of fs,draw,fill=green]{Phase\\Synchronization};
		\draw[->] (ps.south) -- (b.north);
		\draw[->] (ps.south) -- (d.north);
		\draw[->] (ps.south) -- (g.north);
		\draw[->] (ps.south) -- (j.north);
		
		\draw[->,thick] (-2.5mm,-10mm)--(-2.5mm,-5mm) node[align=center,below right]{cold start};
		
	\end{tikzpicture}
 }
	\caption{Synchronization procedure.}
	\label{fig:coldStartSynchronization}
\end{figure}
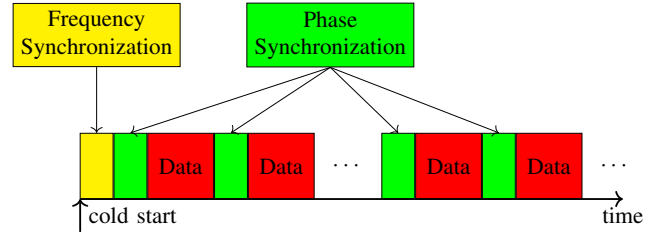

In a distributed system, the \glspl{ap} are likely to be out of synchronization during the cold start and require a synchronization stage before users can communicate with the \glspl{ap}. This synchronization stage requires dedicated resources for synchronization of the \glspl{ap}, which need to be allocated prior to transmission to/from the users. Therefore, the current architecture needs to be redesigned to accommodate this requirement. %During the synchronization stage, the resource grid needs to be dedicated to distributed AP synchronization. 
%The BeamSync protocol can synchronize the \glspl{ap} faster without the need for channel estimation.
%During the cold start or initialization of the entire system, all the distributed transceivers will be out of sync. 
After the initial power-up, all the \glspl{ap} will synchronize the frequency and the phase with a master \gls{ap} in a sequential fashion using the BeamSync protocol. 
After the synchronization stage, regular data transmission stage with users take place, which include channel estimation, uplink and downlink data transmission.
%Afterward, the distributed system enters the data transmission phase with \glspl{ue} which includes channel estimation. 
The synchronization procedure needs to be done when the \glspl{ap} go out of synchronization. 
Due to the presence of phase noise in the transceivers, phase synchronization needs to be done more often compared to frequency synchronization.
The synchronization procedure dispersed over time is represented in Fig.~\ref{fig:coldStartSynchronization}. 

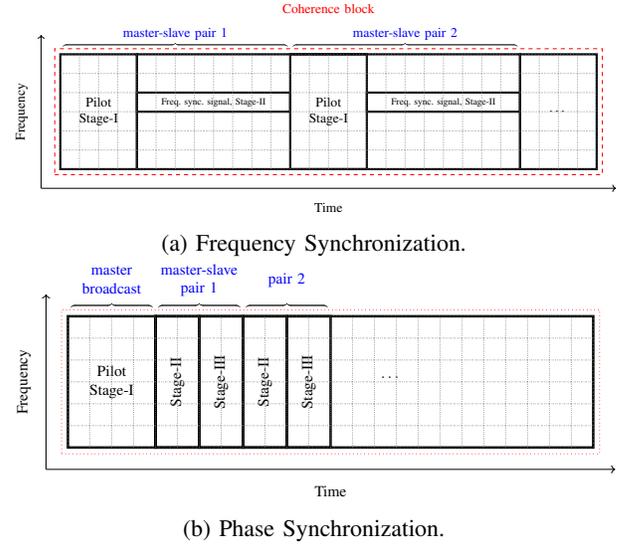
\begin{figure}
	\begin{subfigure}[t]{0.5\textwidth}
		\centering
			\resizebox{0.9\textwidth}{!}{
			\begin{tikzpicture}
			\node (a1) at (0,0)  [align=center,minimum height=30mm,minimum width=20mm,draw,ultra thick] {Pilot\\Stage-I};
			\node (a2) at (60mm,0) [align=center,minimum height=30mm,minimum width=100mm,draw,ultra thick] {};
			\node (a3) at (30mm,2.5mm) [align=center,minimum height=5mm,minimum width=40mm,draw,ultra thick] {\scriptsize Freq. sync. signal, Stage-II};
			\node (a4) at (60mm,0) [align=center,minimum height=30mm,minimum width=20mm,draw,ultra thick] {Pilot\\Stage-I};
			\node (a5) at (90mm,2.5mm) [align=center,minimum height=5mm,minimum width=40mm,draw,ultra thick] {\scriptsize Freq. sync. signal, Stage-II};
			\node (a6) at (120mm,0) [align=center,minimum height=30mm,minimum width=20mm,draw,ultra thick] {$\cdots$};
				
			\foreach \i in {0,...,27}
			{ \foreach \j in {0,...,5}
				{
					\node at (-7.5mm+\i*5mm,-12.5mm++\j*5mm) [draw,minimum height=5mm,minimum width=5mm,color=gray,dotted] {} ;
				}
			}

			\draw [thick,decorate, decoration = {calligraphic brace,raise=5pt},color=blue] ([xshift=1mm]a1.north west) --  ([xshift=(39mm)]a1.north east) node[align=center,midway,yshift=5mm,color=blue] {master-slave pair 1};	
			
			\draw [thick,decorate, decoration = {calligraphic brace,raise=5pt},color=blue] ([xshift=1mm]a4.north west) --  ([xshift=(39mm)]a4.north east) node[align=center,midway,yshift=5mm,color=blue] {master-slave pair 2};	
				
			\node (cohblk) [draw,dashed,fit=(a1) (a2) (a6),color=red] {};
			\node [align=center,above=8mm of cohblk,color=red] {Coherence block};

			\draw[->] (-15mm,-20mm) --( -15mm,20mm) node[align=center,midway,xshift=-5mm]{\rotatebox{90}{Frequency}};
			\draw[->] (-15mm,-20mm) --(135mm ,-20mm) node[align=center,midway,yshift=-5mm]{Time};
				
			\end{tikzpicture} }
			\caption{Frequency Synchronization.}
			\label{fig:FreqSyncOfdmResourceGrid}
	\end{subfigure}
	\hfill
	% right subfigure
	\begin{subfigure}[t]{0.5\textwidth}
		\centering
			\resizebox{0.9\textwidth}{!}{
		\begin{tikzpicture}
			\node (a1) at (0,0)  [align=center,minimum height=30mm,minimum width=20mm,draw,ultra thick] {Pilot\\Stage-I};
			\node (a2) at (60mm,0) [align=center,minimum height=30mm,minimum width=100mm,draw,ultra thick] {};
			\node (a3) at (15mm,0) [align=center,minimum height=30mm,minimum width=10mm,draw,ultra thick] {\rotatebox{90}{Stage-II}};
			\node (a4) at (25mm,0) [align=center,minimum height=30mm,minimum width=10mm,draw,ultra thick] {\rotatebox{90}{Stage-III}};
			\node (a5) at (35mm,0) [align=center,minimum height=30mm,minimum width=10mm,draw,ultra thick] {\rotatebox{90}{Stage-II}};
			\node (a6) at (45mm,0) [align=center,minimum height=30mm,minimum width=10mm,draw,ultra thick] {\rotatebox{90}{Stage-III}};			
			\node (a7) [align=center,right=10mm of a6,yshift=1mm] {$\cdots$};		
			
			\foreach \i in {0,...,23}
			{ \foreach \j in {0,...,5}
				{
					\node at (-7.5mm+\i*5mm,-12.5mm++\j*5mm) [draw,minimum height=5mm,minimum width=5mm,color=gray,dotted] {} ;
				}
			}

			\draw [thick,decorate, decoration = {calligraphic brace,raise=5pt},color=blue] ([xshift=1mm]a1.north west) --  ([xshift=-1mm]a1.north east) node[align=center,midway,yshift=8mm,color=blue] {master\\broadcast};
			
			\draw [thick,decorate, decoration = {calligraphic brace,raise=5pt},color=blue] ([xshift=1mm]a3.north west) --  ([xshift=-1mm]a4.north east) node[align=center,midway,yshift=8mm,color=blue] {master-slave\\pair 1};	
			
			\draw [thick,decorate, decoration = {calligraphic brace,raise=5pt},color=blue] ([xshift=1mm]a5.north west) --  ([xshift=-1mm]a6.north east) node[align=center,midway,yshift=8mm,color=blue] {pair 2};							
						
			\node (cohblk) [draw,dotted,fit=(a1) (a2),color=red] {};
%			\node [align=center,right=5mm of cohblk,color=red] {Coherence\\block};
			
			\draw[->] (-15mm,-20mm) --( -15mm,20mm) node[align=center,midway,xshift=-5mm]{\rotatebox{90}{Frequency}};
			\draw[->] (-15mm,-20mm) --(115mm ,-20mm) node[align=center,midway,yshift=-5mm]{Time};
			
		\end{tikzpicture}  }
		\caption{Phase Synchronization.}
		\label{fig:PhaseSyncOfdmResourceGrid}
	\end{subfigure}

	\caption{An example of resource allocation in OFDM resource grid for synchronization.}
	\label{fig:OfdmResourceAllocation}
\end{figure}

Fig.~\ref{fig:OfdmResourceAllocation} shows an example of resource allocation for synchronization in an OFDM resource grid. For frequency synchronization, the synchronization signal is sent over different time intervals to capture the phase shifts occurring over time due to the frequency offset. Resources for pilot signaling and frequency synchronization signal are allocated for each master-slave pair in a coherence block for frequency synchronization. In the case of phase synchronization, the resources allocated for pilot can be used by all the slave \glspl{ap} who are synchronizing in the same coherence interval. Afterward, resources for stage-II and stage-III are dedicated for each master-slave pair.

As the number of \glspl{ap} increases in a distributed system, the time required for synchronization also increases. 
Additionally, if the \glspl{ap} are spread over a wide geographic area, some of them may not be able to communicate with each other which limits the current algorithm for large area synchronization.
In this situation, one possible approach could be to consider a clustering approach, where the \glspl{ap} in each cluster form an independent distributed system, but we leave this for potential future work.
%In this situation, it would be advantageous to consider a clustering approach where the APs in each cluster form an independent distributed system. Please note that clustering based distributed network synchronization and data transmission is an area that requires further exploration and is not within the scope of this work.

\section{Simulations}
\label{sec:Simulations}
In this section, we discuss the performance of the proposed BeamSync protocol for synchronization in distributed massive \gls{mimo} systems. 
The BeamSync algorithm does not rely on any particular channel model: the algorithm is agnostic to the channel model and therefore to the array topology.
All processing is done within single coherence intervals over which the channel is static.
For simplicity we consider an independent Rayleigh block fading channel, i.e., each element of the channel matrix $\mathbf{G}$ is distributed as $\mathcal{CN}(0,1)$, thus, not modeling any particular antenna array configuration or topology. 
%However, if we impose specific structure to the array with limited beamforming directions, then the directions will be estimated from the received signal which is the dominant direction of signal reception for that antenna array.
We consider a carrier frequency of $ 3 $ GHz, a signal bandwidth of $ 20 $ MHz and a symbol timing of $T=\frac{1}{14}$~ms for the simulations. 
Furthermore, we draw the transmit and receive channel gains at random from a $\mathcal{CN}(0,1)$	distribution. This makes the  phase offsets    uniformly random over  $[0,2\pi)$.
We use  $ 10^5 $ Monte Carlo trials for the simulations.
The downlink rates from the \glspl{ap} to the users depend on the magnitude of the alignment error, but not on which specific synchronization method that was used for the alignment.
%If the alignment between the APs is same, the downlink rates from the APs to the users will be the same irrespective of the synchronization method used.
Hence, when comparing   different schemes, we chose the alignment error or the \gls{rmse} of the estimates as the performance metric.

In the system setup, we assume that all the antennas are driven by their own \gls{rf} chain thereby enabling fully digital beamforming. 
Hence, BeamSync is a fully digital beamforming-based synchronization technique, which allows us to beamform the signal in any direction in a 3-dimensional environment. 
%Synchronization with BeamSync neither require the \glspl{ap} to estimate the channel nor the \glspl{ap} to send measurement datas to the \gls{cpu} making BeamSync a highly practical option.
In this work, we consider \gls{fgb} synchronization as the main benchmark scheme. 
In \gls{fgb}, the transmitter performs transmit beamforming and the receiver performs receive beamforming from a fixed set of available beamforming directions. 
\Gls{fgb} is implemented with a single \gls{rf} chain driving all antennas, and thus, is an analog beamforming technique. 
In our simulations, we consider the columns of a \gls{dft} matrix as the   set of orthogonal beams. Let $ \{ \mathbf{f}_{A,k} \in\mathbb{C}^{M_A\times 1}, ~k=1,2,\cdots,M_A  \} $ be the fixed set of beams available at the \gls{ap} A. The transmit and receive beamforming vectors are chosen such that the received signal power is maximized. Let 
	\begin{equation}
		k = \argmax_{ k'} \ \lVert \mathbf{f}_{A,k'}^\text{H}\mathbf{Y}_A \rVert ^2, \ l = \argmax_{ l'} \ \lVert \mathbf{f}_{A,l'}^\text{H}\mathbf{Y}_A \rVert ^2.
		\label{eqn:FGB_Description}
	\end{equation}
Then, the transmit beamforming vector is chosen to be  $\mathbf{a}_t~=~\mathbf{f}_{A,k}^* $, and the receive beamforming vector is $\mathbf{a}_r~=~\mathbf{f}_{A,l}$ for \gls{ap} A. 
Synchronization with the \gls{fgb} scheme does not require estimation of the channel between the \glspl{ap},
and does not require sending  measurement data to the \gls{cpu}. Thus, we use the \gls{fgb} scheme for comparison with BeamSync.
Note that \gls{fgb} results in the reception of a scalar quantity for each beam in the grid, rather than a vector quantity (measurements from all antennas) as for BeamSync. 
Unless there is phase coherency between different beams in the analog beamforming (which we do not assume), the analog beamforming would not allow for fully phase-coherent digital processing of the received pilots.
Thus, with FGB, $ M_A\times M_B $ measurements are needed to determine the best transmit and receive beamforming vectors.
In contrast, with BeamSync, the beamforming vector is obtained from $L$ measurements (length of the pilot signal) and computation of the dominant eigenvector of the measurement matrix.

\subsection{Phase Synchronization}

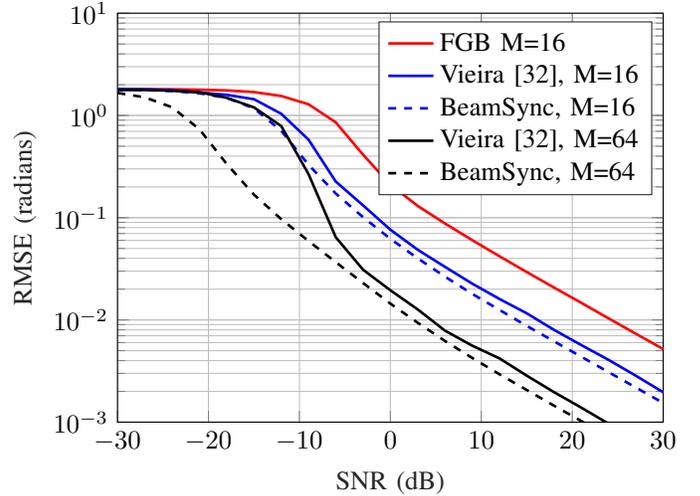
\begin{figure}[!t]
	\centering
	\input{phaseEstimateMethodsComparison.tex} 
	\caption{Comparison of BeamSync with other reciprocity calibration methods. For this plot, the parameters considered are $ M= M_A = M_B$, $ L=M $, and $ N=\frac{M}{2} $.}
	\label{fig:phaseEstimatePerfromanceMethodsComparison}
\end{figure}

For simulations, we have experimented with different phase synchronization signals, viz. (i) a complex Gaussian signal, (ii) an all-one signal, and (iii) a sinusoidal signal, all with the same norm $\Norm{\mathbf{x}}^2 = N$. 
The Monte-Carlo performance of BeamSync with the above signals were indistinguishable. 
Thus, the changes in noise covariance had more or less no effect on the performance. 
For the plots, we have used the complex Gaussian signal as the phase synchronization signal. 
In Fig.~\ref{fig:phaseEstimatePerfromanceMethodsComparison}, we show the superiority of the BeamSync over the calibration method for distributed massive \gls{mimo} proposed in~\cite{vieira2021reciprocity} (which to our knowledge	is state-of-the-art) and over the FGB scheme. 
For the plot, we consider $ M=M_A=M_B $. 
We consider a modified version of~\cite{vieira2021reciprocity} such that all the antennas at \gls{ap}-A and \gls{ap}-B are equipped with \gls{rf} chains. 
Thus, the algorithm in~\cite{vieira2021reciprocity} requires $ 2M $ measurements with the modified scheme ($ M $ measurements in each direction). 
Note that, the original paper~\cite{vieira2021reciprocity} considers analog beamforming  with one \gls{rf} chain per \gls{ap} and hence, requires $2M^2$ measurements. 
For a fair comparison, in BeamSync we consider $L=M$ and $N=\frac{M}{2}$, thereby requiring a total of $2M$ measurements. 
We keep the total energy and resources spent for transmission the same in all schemes. 
From the figure, it can be seen that BeamSync performs $3$~dB better. 
Also, BeamSync performs much better at low \gls{snr}  compared to the FGB and \cite{vieira2021reciprocity} schemes.
Note that the approach in~\cite{vieira2021reciprocity} is centralized, and requires the measurement data from \glspl{ap} to be sent to the \gls{cpu} through fronthaul. 
In contrast, BeamSync does not require any measurement data to be sent to the \gls{cpu} and the \glspl{ap} can estimate their phase offsets independently.

\begin{figure}[!t]
	\centering
	\input{phaseEstimatePlot.tex}
	\caption{BeamSync performance for phase estimates with $ M_A = M_B=16 $, $ L=16 $, and $ N=100 $.}
	\label{fig:phaseEstimatePerfromanceRayleigh}
\end{figure}
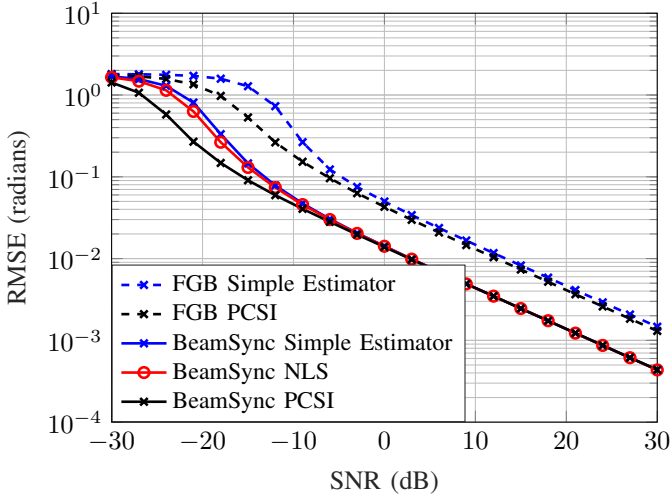

Fig.~\ref{fig:phaseEstimatePerfromanceRayleigh} shows the performance of BeamSync protocol for phase synchronization in distributed massive \gls{mimo} systems. 
It can be seen from the figure that, the simple estimator in \eqref{eqn:phaseSubOptimalEstimate} performs equally well as the optimal estimator at high \gls{snr}. 
Moreover, at high \gls{snr}, the performance of the estimators approaches the performance of the estimator if we had \gls{pcsi}. 
Thus, by using BeamSync with the simple estimator in \eqref{eqn:phaseSubOptimalEstimate}, the synchronization can be done without estimating the channel between the \glspl{ap}.
For a fixed RMSE requirement, the BeamSync protocol has an \gls{snr} gain of $ 10 $ dB when compared with the \gls{fgb} scheme.

\begin{figure}[!t]
	\centering
	\input{phaseEstimatePerformanceNumAntennas.tex}
	\caption{BeamSync performance for phase estimates for different number of antennas, $ M_A=M_B=M $, $ L=M $, and $ N=100 $.}
	\label{fig:phaseEstimatePerfromanceNumAntennas}
\end{figure}
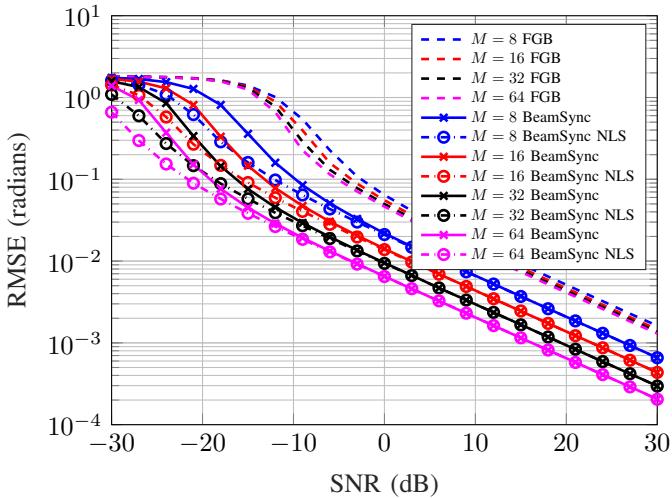

Fig.~\ref{fig:phaseEstimatePerfromanceNumAntennas} shows the phase estimation performance of BeamSync for different numbers of antennas with NLS and with the simple estimator. 
It can be seen from the plot that the BeamSync performance improves with the number of antennas at the \gls{ap}. 
The performance of the simple estimator matches with the NLS and hence, BeamSync can perform well without   knowledge of the channel between the \glspl{ap}.
For every doubling of the number of antennas, the performance of BeamSync improves by $ 3 $~dB. 
This is due to the increase in the \gls{snr} by beamforming and spatial processing gain. The improvement in the \gls{fgb} scheme is negligible. 

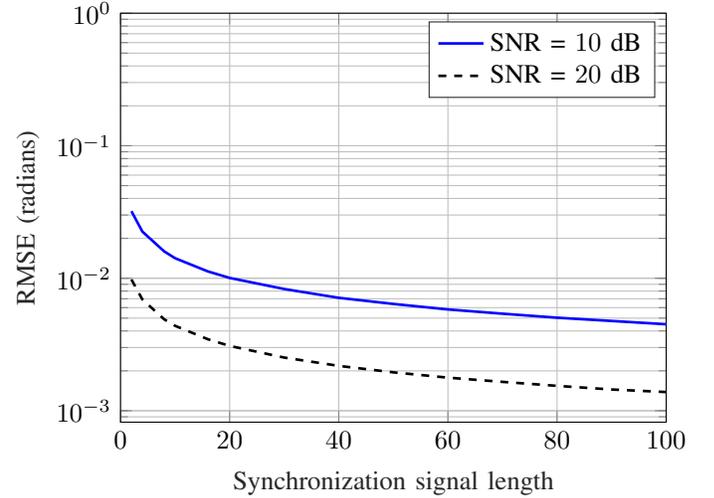
\begin{figure}[!t]
	\centering
	\input{BeamSync_SyncSignalLength_Performance.tex}
	\caption{BeamSync performance with the synchronization signal length for different \gls{snr} levels. $M_A=M_B=16$ and $L=16$.}
	\label{fig:BeamSyncSignalLegnthPerformance}
\end{figure}

The estimation performance of BeamSync for different synchronization signal sequence lengths is provided in Fig.~\ref{fig:BeamSyncSignalLegnthPerformance}. 
The  performance improves with the synchronization signal length.

\subsection{Phase Noise}
\label{sec:phaseNoise}
A practical \gls{lo} driving the \gls{rf} chain will experience phase fluctuations creating non-negligible distortions in the transmit signal.
This is referred to as \textit{phase noise} and is widely studied in~\cite{demir2000phase,mehrotra2002noise,bittner2008tutorial,pitarokoilis2014uplink}. 
We consider all the \gls{rf} chains in an \gls{ap} to be driven by a single \gls{lo} and thus, operate in synchronous mode. 
Let $\omega_A$ be the phase noise process at \gls{ap} A.
We use the widely accepted discrete Wiener phase noise model \cite{bittner2008tutorial,pitarokoilis2014uplink} given by 
\begin{equation}
	\omega_A[n+1] = \omega_A[n] + \nu_A[n] ,
	\label{eqn:discreteWienerNoiseModel}
\end{equation}
where the increments $\nu_A[n]$ are \gls{iid} zero mean Gaussian random variables, i.e., $\nu_A[n] \sim \mathcal{N}(0,\sigma_\nu^2)$. The variance $\sigma_\nu^2 = 4\pi^2f_c^2c_{vco} T_s$, where $ f_c $ is the carrier frequency, $ T_s $ is the sampling interval and $ c_{vco} $ is a constant which determines the quality of a \gls{lo}. 
Let $ \omega_B $ be the phase noise process at \gls{ap} B. Without loss of generality we can assume $\omega_A[1] = \omega_B[1] = 0$, as it could be captured into $ t_1^A $ and $ t_1^B $, respectively. 
Let $\omega[n] = \omega_A[n] + \omega_B[n]$ such that the increments of $\omega$ are distributed as $\mathcal{N}(0,2\sigma_\nu^2)$. Let 
\begin{equation}
	\boldsymbol{\omega}_p^q = \Tran{\Sbrac{ e^{j\omega[p]} ~ e^{j\omega[p+1]} ~\cdots ~ e^{j\omega[q]}}}.
\end{equation}
The phase noise effects at the receiving \glspl{ap} can be modeled into our signal model as follows:
\begin{align}
	\mathbf{Y}_{B1} & = \sqrt{L} \frac{t^A_1}{r^A_1} \mathbf{G}_e^{\text{T}} \Herm{\mathbf{\Phi}} \mathbf{D}_{\boldsymbol{\omega}_1^L} + \mathbf{W}_{B1} \\
	\mathbf{Y}_{A1} & = \frac{t^B_1}{r^B_1}\mathbf{G}_e \mathbf{a} \Tran{\mathbf{x}} \mathbf{D}_{\boldsymbol{\omega}_{L+1}^{L+N}}^\text{H} + \mathbf{W}_{A1} \\
	\mathbf{Y}_{B2} & =  \sqrt{c} \frac{t^A_1}{r^A_1}  \mathbf{G}_e^{\text{T}} \mathbf{Y}_{A1}^* \mathbf{D}_{\boldsymbol{\omega}_{L+N+1}^{L+2N}} + \mathbf{W}_{B2}.
\end{align}

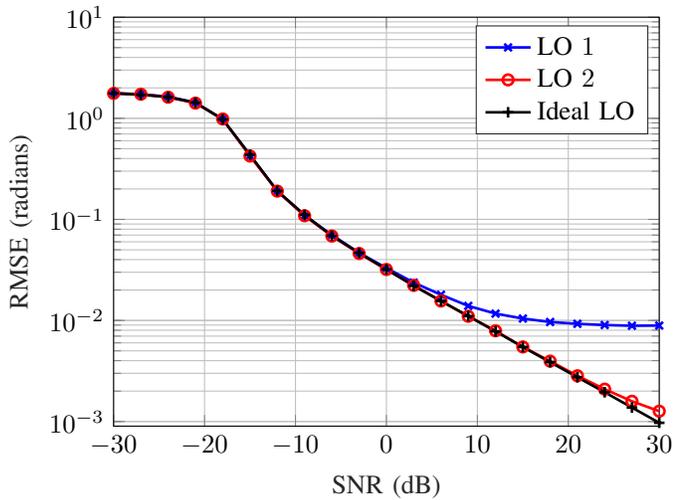
\begin{figure}[!t]
	\centering
	\input{phaseEstimatePerformanceWithPhaseNoise.tex}
	\caption{BeamSync performance for phase estimates for different LOs with $ M_A = M_B=16 $, $ L=16 $, and $ N=20 $.}
	\label{fig:phaseEstimatePerfromanceWithPhaseNoise}
\end{figure}

Fig.~\ref{fig:phaseEstimatePerfromanceWithPhaseNoise} shows the phase estimation performance of the BeamSync in the presence of phase noise at the estimating \glspl{ap}.
We consider two crystal oscillators: \gls{lo}~$1$ with $ c_{vco} = 1.7610\times10^{-19}~(\text{rad Hz})^{-1}$, and \gls{lo}~$2$ with $ c_{vco} = 1.4647\times10^{-21}~(\text{rad Hz})^{-1}$, where the latter is a high-performance \gls{lo}.
We compare the performance of the above two \glspl{lo} with an ideal \gls{lo} with no phase-noise.
%We consider the oscillator parameters for a sub-carrier spacing of $ 100 $ KHz in the simulations.
From the plot, BeamSync can perform well in the presence of phase noise. 
By using a high performing \gls{lo}, the performance can be improved further. 
Note that a free-running \gls{vco} can experience unbounded phase noise, which can cause significant synchronization issues in a distributed system. 
However, in practical systems, \glspl{pll} are typically used in conjunction with the \gls{vco} to stabilize the phase noise. 
As a result, the phase offset between the \glspl{ap} will drift   slowly over time, and the amount of drift will depend on the operating environment of the \glspl{ap}. 
%This means that the \glspl{ap} will not require synchronization as frequently as they would with a free-running \gls{vco}.

\subsection{Impact of Reciprocity Calibration Errors}
\label{subsec:ImpactReciprocityErrors}

\begin{figure}[!t]
	\centering
	\input{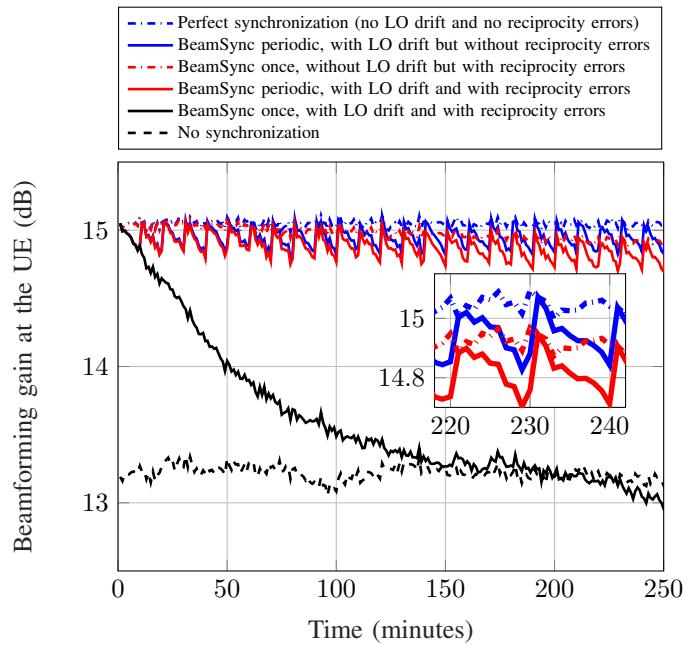}
	\caption{Time evolution of beamforming gain at the \gls{ue} from \glspl{ap} A \& B. Here $ M_A = M_B=16 $, $ L=16 $, $ N=20 $, and \gls{snr}=~$20$~dB.}
	\label{fig:reciprocityCalibrationErrorOverTime}
\end{figure}

	Cellular massive \gls{mimo} with a single \gls{ap} relies on the  assumption of channel reciprocity and  reciprocity calibration suffices to achieve this~\cite{marzetta2016fundamentals,shepard2012argos}. 
	A per-\gls{ap} reciprocity calibration method  was provided in~\cite{shepard2012argos}, wherein it was also demonstrated 
	experimentally that the reciprocity calibration can be performed rather infrequently (on the time-scale of hours).
	In contrast, in distributed \gls{mimo} systems, if the \glspl{ap} are driven by independent \glspl{lo}, then in addition to the per-\gls{ap} reciprocity calibration errors, the system will also experience phase drifts between the \glspl{ap} (on a much faster time-scale).
	Our BeamSync algorithm is a procedure to synchronize the phase between independently operating \glspl{ap}, but not a method to achieve per-\gls{ap} reciprocity calibration. 
	In fact, BeamSync relies on the \glspl{ap} being individually reciprocity-calibrated. 
	
	There are two consequences of  per-\gls{ap} reciprocity errors: first, that some beamforming gain is lost when serving \glspl{ue}; second, as a  quantitatively much smaller effect, that the beamforming between the \glspl{ap} inside of the BeamSync algorithm  loses some efficiency which in turn impacts, although very little,  the phase synchronization performance -- which in turn affects the  beamforming gain when serving  \glspl{ue}. 
	
	To give a numerical illustration of the above phenomena,  we model the per-\gls{ap} reciprocity calibration errors as additive Gaussian with zero mean and variance of $\sigma_{\epsilon}^2$. 
	We take, from the experimental results in \cite{shepard2012argos},  $\sigma_{\epsilon}^2= 10^{-9.6}$, yielding a mean angle deviation of $2.7\%$ over $4$ hours of time. 
	We ignore the variations in amplitude as the measurements in~\cite{shepard2012argos} show only tiny variations ($0.7\%$) in the amplitude over the same time duration.
	This means that the per-\gls{ap} reciprocity calibration coefficients $\{\frac{t_m^A}{r_m^A} \}$ for \gls{ap}~A and $\{\frac{t_m^B}{r_m^B} \}$ for \gls{ap}~B vary very slowly over time.	
	However, the ratio $\frac{t_1^A}{r_1^A}/\frac{t_1^B}{r_1^B}$ changes must faster, due to the \gls{lo} drifts~\cite{nissel2022correctly}. 
	For the simulations, we consider that both the \glspl{ap} are equipped with a high-quality \gls{lo}, specifically \gls{lo} $2$ 
	(see its performance figures in Section~\ref{sec:phaseNoise}). 
	Practical deployments may use \glspl{lo} of less quality, which entails much more frequent re-calibration; we use \gls{lo} $2$ here only to improve readability of the plots.
	
	To study the consequences of per-\gls{ap} reciprocity errors and \gls{lo} drift, we consider how the beamforming gain achieved at the \gls{ue} evolves over time. 
	We refer to Fig.~\ref{fig:distributedMassiveMimo} for the definition of all quantities being referred to in the following.
	For simplicity, we consider that both \glspl{ap} are at the same distance from the \gls{ue}, and that the channels $\mathbf{g}_A$ and $\mathbf{g}_B$ are Rayleigh fading with i.i.d. $\mathcal{CN}(0,1)$ entries. 
	Fig.~\ref{fig:reciprocityCalibrationErrorOverTime} shows the beamforming gain at the \gls{ue} over time. 
	In the simulation, BeamSync calibration is done once (at time $0$) for some of the curves, and  periodically thereafter for some of curves. 
	We observe the following: 
	\begin{enumerate}
		\item For reference (blue dash-dotted curve), we consider the scenario where we have perfect phase synchronization between the \glspl{ap} (this is equivalent to no  \gls{lo} drift or noise) as well as no per-\gls{ap} reciprocity calibration errors. This  achieves the maximum possible beamforming gain, $M_A + M_B$.
		
		\item In the presence of \gls{lo} drift but with BeamSync performed only once (no periodic re-synchronization between the \glspl{ap}), the beamforming gain drops rapidly over time (black solid curve). 
		For a lower-quality \glspl{lo}, e.g., \gls{lo} $1$, this drop in the gain would be much faster.
		This illustrates that  periodic phase re-synchronization is required.
		
		\item With \gls{lo} drift and BeamSync phase calibration performed periodically every $10$ minutes, but \emph{without} per-\gls{ap} reciprocity calibration errors (blue solid curve): 
		In this case, BeamSync periodic re-calibration helps to maintain a good beamforming gain.
		The re-calibration recovers the losses from the phase drift between the \glspl{ap}.
		The zig-zag behaviour in the plot is due to the loss of  synchronization during the $10$-minute interval.
		%Again, the BeamSync re-calibration would need to be done much more often with lower-quality  \glspl{lo}.
		In practice, the periodicity of BeamSync re-calibration depends on the quality of the \gls{lo} used.
		
		\item With \gls{lo} drift and BeamSync phase calibration  performed periodically every $10$ minutes,
		and \emph{with} per-\gls{ap} reciprocity calibration errors (red solid curve): 
		BeamSync is able to recover the losses from the \gls{lo} drift; however, it is unable to recover the long-term losses from per-\gls{ap} reciprocity calibration errors.
		
		\item Also for reference, we consider a hypothetical situation where we have no \gls{lo} drift, but we do have per-\gls{ap} reciprocity calibration errors (red dash-dotted curve). 
		Refer to the inset for better clarity.
		
	\end{enumerate}

\begin{figure}[!t]
	\centering
	\input{calibrationErrorPlot}
	\caption{Time evolution of the mean of the absolute phase deviation between the \glspl{ap} observed at the \gls{ue}. Here $ M_A = M_B=16 $, $ L=16 $, $ N=20 $, SNR$=20$~dB.}
	\label{fig:calibrationErrorPlot}
\end{figure}
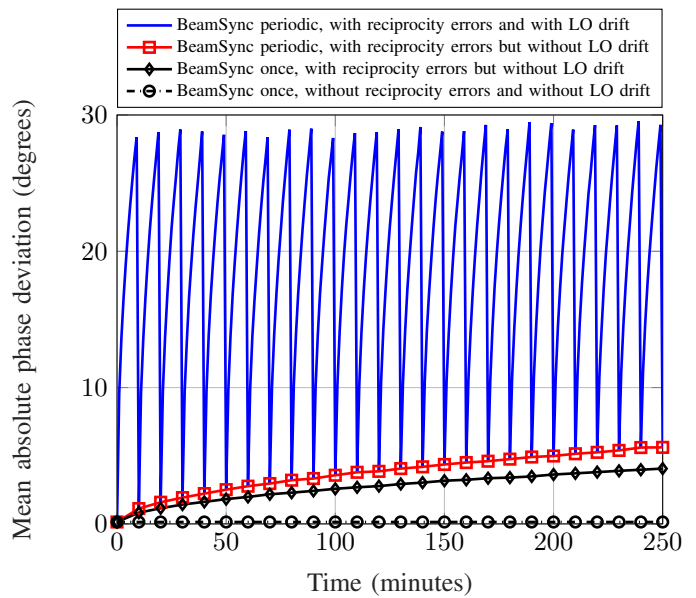

When executing the BeamSync procedure to phase-synchronize two \glspl{ap}, the per-\gls{ap} reciprocity error will impact the synchronization performance  (as already noted above). 
To illustrate this effect, Fig.~\ref{fig:calibrationErrorPlot} shows the time evolution of the mean absolute value of the phase difference between the \glspl{ap} seen at the \gls{ue} after phase correction is applied at \gls{ap}~B. 
%two \glspl{ap}, when BeamSync is used to re-calibrate periodically (every $10$ minutes).
%Here the deviation is measured as the phase difference . 
%(Only phase deviations right after running BeamSync are shown; therefore this plot does not show the zig-zag behavior as in Figure~\ref{fig:reciprocityCalibrationErrorOverTime}.)
Per-\gls{ap} reciprocity calibration is done once, at time $0$.
%For reference, we consider a hypothetical scenario where we have perfect synchronization at time $0$ and have only reciprocity calibrations errors at each \gls{ap} over time and no re-synchronization (black solid curve with diamond markers).
%With blue and red solid curves, we consider BeamSync re-calibration every $10$ minutes.
The gap between the red solid curve with square markers and black solid curve with diamond markers shows the angle deviation between the \glspl{ap} caused by the per-\gls{ap} reciprocity calibration errors inside the BeamSync algorithm.
It can be seen from the plot that BeamSync causes a tiny deviation of $\phi \approx 1.5^\circ$ after $4$ hours due to the reciprocity errors inside the BeamSync algorithm. 
This leads to a beamforming loss by a factor of $\frac{M_A+M_B}{\Abs{M_A + e^{j\phi}M_B}}$, which for the example considered in Fig.~\ref{fig:calibrationErrorPlot} is $0.0004$~dB. 
This loss should be compared to the beamforming loss in Fig.~\ref{fig:reciprocityCalibrationErrorOverTime} (the $0.13$~dB gap between blue and red dash-dotted curves at $4$~hours). 
The $0.13$~dB gap in Fig.~\ref{fig:reciprocityCalibrationErrorOverTime} is due to the loss of beamforming gain due to reciprocity errors \emph{and} the loss of BeamSync performance due to reciprocity errors; the latter amounts to only $0.0004$~dB of the $0.13$~dB in total (this is the second consequence of the reciprocity calibration errors).

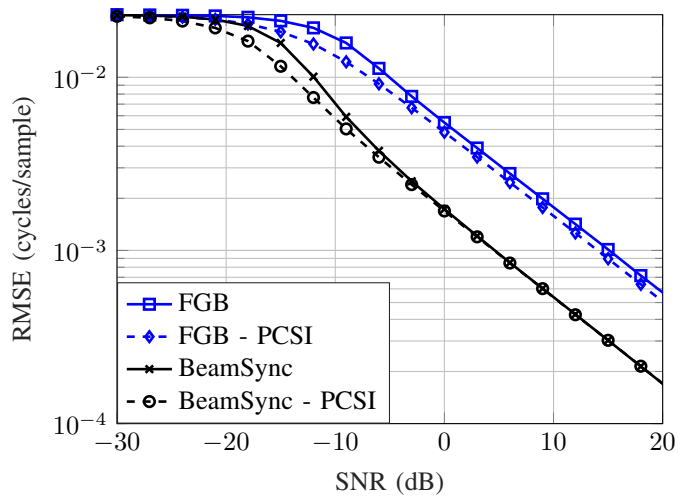
\begin{figure}[!t]
	\centering
	\input{offsetEstimationPlot.tex}
	\caption{BeamSync performance for carrier frequency offset estimates with $ M_A = M_B = 16 $, $ L=16 $, and $ N_f=10 $.}
	\label{fig:offsetEstimatePerfromanceRayleigh}
\end{figure}

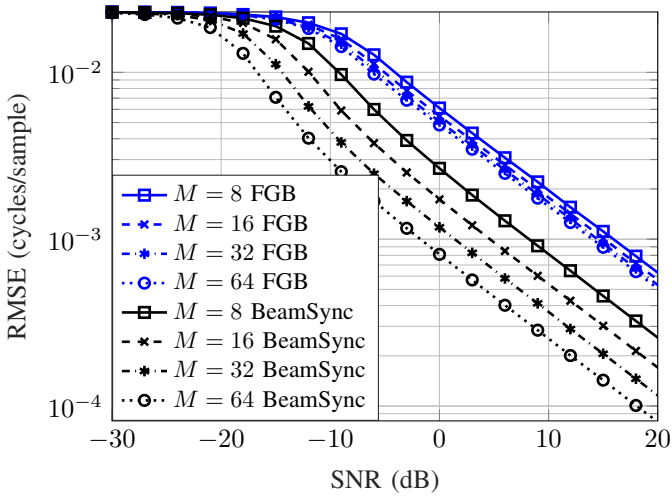
\begin{figure}[!t]
	\centering
	\input{offsetEstimationPerformanceNumAntennas.tex}
	\caption{BeamSync performance for carrier frequency offset estimates with $ M_A = M_B = M $, $ L_B = M_B $ and $ N_f = 10 $.}
	\label{fig:offsetEstimatePerfromanceNumAntennas}
\end{figure}

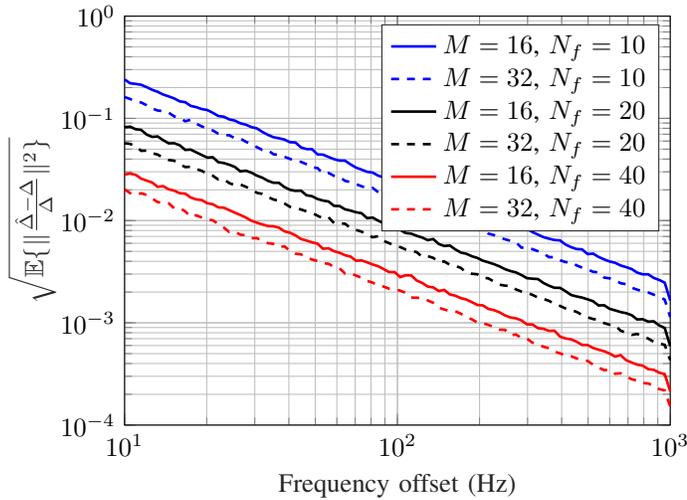
\begin{figure}[!t]
	\centering
	\input{frequencyOffsetRangePlot.tex}
	\caption{BeamSync performance with frequency offset range. $M_A=M_B=M$ and $L=M$ and \gls{snr} $=20$~dB.}
	\label{fig:frequencyOffsetRange}
\end{figure}

\subsection{Frequency Synchronization}
In this subsection, we consider the frequency synchronization performance of the BeamSync. 
As per the $3$GPP TS $38.104$, the carrier frequency of the \gls{ap} shall be accurate within $\pm 0.05$~ppm. 
Thus, for $3$~GHz carrier frequency, we assume that the simulated frequency offset $\Delta$ ranges uniformly from $-300$~Hz to $300$~Hz. 
%We consider the symbol timing to be $T=\frac{1}{14}$~ms.
Here we consider the \gls{rmse} $=\sqrt{\Exp{\Norm{\Delta T -\hat{\Delta}T}^2}}$ (cycles/sample) as the performance metric.
We have considered a sinusoidal signal  as the frequency synchronization signal. 
The frequency synchronization performances of BeamSync and the fixed grid of beams are shown in Fig.~\ref{fig:offsetEstimatePerfromanceRayleigh} and Fig.~\ref{fig:offsetEstimatePerfromanceNumAntennas}. 
The gains achieved for frequency synchronization are similar to the gains in phase synchronization. For example, for a fixed \gls{rmse} requirement, the \gls{snr} gain is approximately $ 10 $~dB for BeamSync compared to the \gls{fgb} scheme. Also, it can be seen that for a fixed \gls{rmse} requirement, the \gls{snr} requirement reduces by $ 3 $~dB when the number of antennas is doubled at the \glspl{ap} for the BeamSync protocol. 

Fig.~\ref{fig:frequencyOffsetRange} shows the performance of BeamSync with respect to the frequency offset. 
For a coarse synchronization, we need less synchronization signal length $N_f$, while for fine synchronization we need larger $N_f$, or a larger number of antennas, or a higher \gls{snr}. 
By doubling $N_f$, the \gls{snr} improves by $6$~dB.

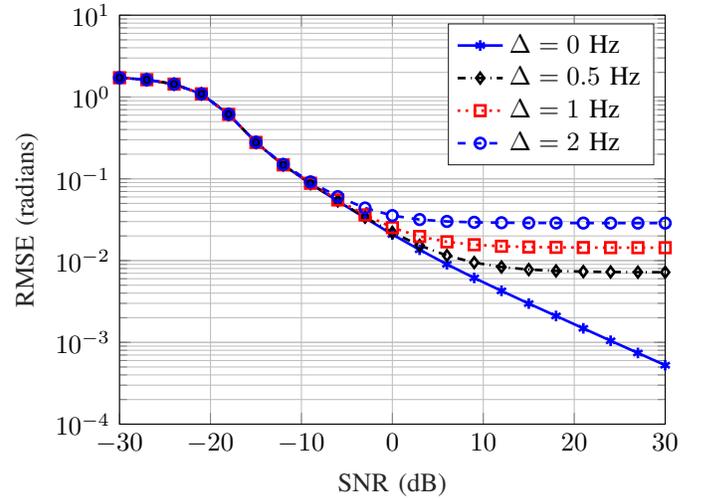
\begin{figure}[!t]
	\centering
	\input{BeamSyncPhasePerformanceWithFrequencyOffset.tex}
	\caption{BeamSync phase synchronization performance with imperfect frequency offset. $M_A=M_B=32$, $L=32$ and $N=32$.}
	\label{fig:PhasePerformanceWithFrequencyOffset}
\end{figure}

Fig.~\ref{fig:PhasePerformanceWithFrequencyOffset} shows the phase estimation performance  of BeamSync when we have a residual frequency offset~$\Delta$ between two \glspl{ap}. 
A residual frequency error between \glspl{ap} can lead to a significant change in phase with time, which  impacts the phase estimation. 
Hence, it is necessary to correct the phases soon after frequency synchronization. 
With perfect frequency synchronization, i.e., $\Delta=0$, the phase estimation \gls{rmse} decreases as the \gls{snr} increases. However, with residual frequency errors, the \gls{rmse} reaches a constant floor independent of the \gls{snr}. 
Furthermore, this constant value increases as the residual frequency error increases. 
To synchronize faster, different \gls{ap} pairs can use \gls{fdma} to mitigate the issue of imperfect frequency offsets. 
If the number of \gls{ap} pairs to be synchronized is large, we need to interlace frequency and phase synchronization between different pairs.

\section{Conclusion}
\label{sec:Conclusion}
In this paper, we studied the synchronization requirements in a distributed massive \gls{mimo} from a reciprocity perspective. 
We proposed two novel over-the-air synchronization protocols, BeamSync, based on digital beamforming to synchronize the carrier frequency offsets and phase offsets among distributed \glspl{ap}. 
We analytically derived the optimal frequency offset estimator and phase offset estimators. 
We perform a phase calibration of reciprocity-calibrated \glspl{ap} with one \gls{ap} to enable coherent transmission and also propose a simpler phase offset estimator which performs close to the optimal estimator at high \gls{snr}. 
We analytically showed that the synchronization signals need to be beamformed in the dominant direction of the effective channel in which the signal is received. 
BeamSync can achieve a $ 3 $ dB gain for every doubling of the number of antennas at the \glspl{ap}. 
This performance is due to the improved \gls{snr} by beamforming and spatial processing gain. 

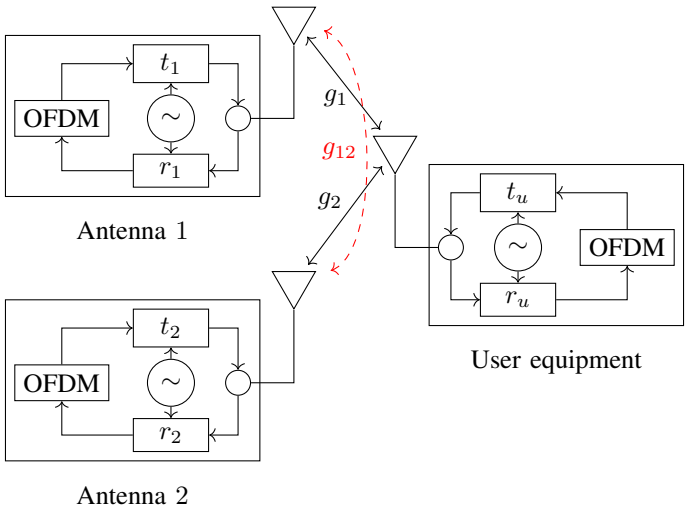
\begin{figure}
	\centering
	\begin{tikzpicture}
		%% BS 1	
		\node (ap_tx) at (0,0)  [align=center,minimum width=10mm,draw] {$ t_1 $};
		\node (ap_lo) [circle,align=center,below=1.5mm of ap_tx,draw]{$\sim$};
		\node (ap_rx) [align=center,below=1.5mm of ap_lo,minimum width=10mm,draw] {$ r_1 $};
		\node (ap_ofdm) [align=center,left=5mm of ap_lo,draw] {OFDM};
		\node (ap_l1) [circle,align=center,right=4mm of ap_lo,draw]{};	
		\node (ap_ant) [regular polygon, regular polygon sides=3,rotate=180,above right=5mm and 13mm of ap_tx,draw]{};
		\draw[->] (ap_lo) -- (ap_tx);
		\draw[->] (ap_lo) -- (ap_rx);
		
		\draw[->] (ap_tx) -| (ap_l1);
		\draw[<-] (ap_rx) -| (ap_l1);
		
		\draw[-] (ap_l1) -| (ap_ant);
		
		\draw[->] (ap_ofdm) |- (ap_tx);
		\draw[<-] (ap_ofdm) |- (ap_rx);
		
		\node (ap) [draw,fit=(ap_tx) (ap_lo) (ap_rx) (ap_l1) (ap_ofdm)] {};	
		\node [align=center,below=2mm of ap]{Antenna $ 1 $};
		
		%% BS2
		\node (ap2_tx) [align=center,minimum width=10mm,below=30mm of ap_tx,draw] {$ t_2 $};
		\node (ap2_lo) [circle,align=center,below=1.5mm of ap2_tx,draw]{$\sim$};
		\node (ap2_rx) [align=center,below=1.5mm of ap2_lo,minimum width=10mm,draw] {$ r_2 $};
		\node (ap2_ofdm) [align=center,left=5mm of ap2_lo,draw] {OFDM};
		\node (ap2_l1) [circle,align=center,right=4mm of ap2_lo,draw]{};	
		\node (ap2_ant) [regular polygon, regular polygon sides=3,rotate=180,above right=5mm and 13mm of ap2_tx,draw]{};
		\draw[->] (ap2_lo) -- (ap2_tx);
		\draw[->] (ap2_lo) -- (ap2_rx);
		
		\draw[->] (ap2_tx) -| (ap2_l1);
		\draw[<-] (ap2_rx) -| (ap2_l1);
		
		\draw[-] (ap2_l1) -| (ap2_ant);
		
		\draw[->] (ap2_ofdm) |- (ap2_tx);
		\draw[<-] (ap2_ofdm) |- (ap2_rx);
		
		\node (ap2) [draw,fit=(ap2_tx) (ap2_lo) (ap2_rx) (ap2_l1) (ap2_ofdm)] {};	
		\node [align=center,below=2mm of ap2]{Antenna $ 2 $};
		
		%% UE 
		\node (ue_tx) [align=center,below right=12mm and 36mm of ap_tx,minimum width=10mm,draw] {$t_u$};
		\node (ue_lo) [circle,align=center,below=1.5mm of ue_tx,draw]{$\sim$};
		\node (ue_rx) [align=center,below=1.5mm of ue_lo,minimum width=10mm,draw] {$ r_u $};
		\node (ue_ofdm) [align=center,right=5mm of ue_lo,draw] {OFDM};	
		\node (ue_l1) [circle,align=center,left=4mm of ue_lo,draw]{};	
		\node (ue_ant) [regular polygon, regular polygon sides=3,rotate=180,above left=5mm and 13mm of ue_tx,draw]{};
		\node (ue) [draw,fit=(ue_tx) (ue_lo) (ue_rx) (ue_l1) (ue_ofdm)] {};	
		\node [align=center,below=2mm of ue]{User equipment};
		
		\draw[->] (ue_lo) -- (ue_tx);
		\draw[->] (ue_lo) -- (ue_rx);
		
		\draw[->] (ue_tx) -| (ue_l1);
		\draw[<-] (ue_rx) -| (ue_l1);
		
		\draw[-] (ue_l1) -| (ue_ant);
		
		\draw[->] (ue_ofdm) |- (ue_tx);
		\draw[<-] (ue_ofdm) |- (ue_rx);

		\draw[<->,shorten <= 1mm, shorten >= 1mm] (ap_ant) -- (ue_ant) node [align=center,midway,xshift=-1mm,yshift=-2mm] {$g_1$};	
		\draw[<->,shorten <= 1mm, shorten >= 1mm] (ap2_ant) -- (ue_ant) node [align=center,midway,xshift=-2mm,yshift=2mm] {$g_2$};	
		
		\draw [red, <->, dashed, shorten <= 0.25cm, shorten >= 0.25cm] (ap_ant) to[out=-20,in=20]  node [align=center,midway,xshift=-3mm,yshift=0mm]{$g_{12}$} (ap2_ant);
	\end{tikzpicture}
	\caption{System model with $ 2 $ antennas serving a UE.}
	\label{fig:multiAntennaSystemModel}
\end{figure}

\appendix
\section{Reciprocity Calibration of Multi-Antenna AP}
\label{app:reciprocityCalibrationTechnique}
In this appendix, we briefly layout how reciprocity calibration is obtained at a multi-antenna \gls{ap}~\cite{shepard2012argos}.
Consider two antennas jointly serving a \gls{ue} as shown in Fig.~\ref{fig:multiAntennaSystemModel}. Let $ g_1 $ be the reciprocal channel between the \gls{ue} and antenna~$ 1 $, and similarly $ g_2 $ be the reciprocal channel between the \gls{ue} and antenna~$ 2 $. Let $ t_1,~t_2 $ be the \gls{rf} transmit chain gains at antennas $ 1 $ and $ 2 $, respectively. Also, in a similar way, we define $ r_1,~r_2 $ as the \gls{rf} receive chain gains at antennas $ 1 $ and $ 2 $, respectively. The two uplink channels are given by 
\begin{align}
	g_{u1}^{\text{UL}} & = t_u ~ g_1 ~ r_1 \\
	g_{u2}^{\text{UL}} & = t_u ~ g_2 ~ r_2.
\end{align}
Similarly, the downlink channels are given by 
\begin{align}
	g_{1u}^{\text{DL}} & = t_1 ~ g_1 ~ r_u \\
	g_{2u}^{\text{DL}} & = t_2 ~ g_2 ~ r_u.
\end{align}
With pilot signaling from the \gls{ue}, antennas $ 1 $ and $ 2 $ can estimate the uplink channels $ g_{u1}^{\text{UL}} $ and $ g_{u2}^{\text{UL}} $, respectively. On the downlink, both the antennas $ 1 $ and $ 2 $ applies \textit{conjugate beamforming }~\cite{marzetta2016fundamentals} and the total channel gains seen by the \gls{ue} from both antennas are given by
\begin{align}
	h_1 & = \Conj{\Brac{g_{u1}^{\text{UL}}}} g_{1u}^{\text{DL}} = t_u^*r_1^* \Abs{g_1}^2 t_1 r_u \\
	h_2 & = \Conj{\Brac{g_{u2}^{\text{UL}}}} g_{2u}^{\text{DL}} = t_u^*r_2^* \Abs{g_2}^2 t_2 r_u . 
\end{align}
The channel gains $ h_1 $ and $ h_2 $ can be written in terms of amplitude and phase differences of \gls{rf} chains as follows
\begin{equation}
	\label{eqn:twoTerminalChannels}
	\begin{aligned}
		h_1 & = \Abs{t_u} \Abs{r_1} \Abs{g_1}^2 \Abs{t_1} \Abs{r_u} e^{j\Brac{(\angle{t_1} - \angle{r_1}) - (\angle{t_u} - \angle{r_u} )} } \\
		h_2 & = \Abs{t_u} \Abs{r_2} \Abs{g_2}^2 \Abs{t_2} \Abs{r_u} e^{j\Brac{(\angle{t_2} - \angle{r_2}) - (\angle{t_u} - \angle{r_u})} }.
	\end{aligned}
\end{equation}
Note that in \eqref{eqn:twoTerminalChannels} the phase shifts caused by the reciprocal channels $ g_1 $ and $ g_2 $ are removed through conjugate beamforming. For the signals from antennas $ 1 $ and $ 2 $ to add up coherently at \gls{ue}, the phases of $ h_1 $ and $ h_2 $ need to be precompensated at the antennas. 
%However, an aboslute precompensation is unneccessary. 
A relative compensation of antenna $ 2 $ with antenna $ 1 $ suffices to add up the signals at the \gls{ue} coherently. Mathematically, what matters is the difference between the phases of two antennas, i.e., $ \Brac{(\angle{t_1} - \angle{r_1}) - (\angle{t_2} - \angle{r_2})} $. If we have many antennas, then we need to relatively precompensate the phases of all antennas with antenna $ 1 $, i.e., compensate $ \Brac{(\angle{t_1} - \angle{r_1}) - (\angle{t_i} - \angle{r_i})} $ for $ i=2,3,\cdots $.

%\subsubsection{Weak Calibration}
Now, consider two antennas $ 1 $ and $ 2 $, and let $ g_{12} \in \mathbb{C}$ be the reciprocal channel gain between them. Using a bidirectional measurement between antennas $ 1 $ and $ 2 $, we obtain 
\begin{align}
	z_{12} & = t_1 g_{12} r_2 = \Abs{t_1} \Abs{g_{12}} \Abs{r_2} e^{j\Brac{\angle{t_1} + \angle{g_{12}} + \angle{r_2}}} \\
	z_{21} & = t_2 g_{12} r_1 = \Abs{t_2} \Abs{g_{12}} \Abs{r_1} e^{j\Brac{\angle{t_2} + \angle{g_{12}} + \angle{r_1}}}.  
\end{align}
Dividing these two measurements we get  
\begin{equation}
	\frac{z_{12}}{z_{21}} = \frac{\Abs{t_1} \Abs{r_2}}{\Abs{t_2} \Abs{r_1}} e^{j\Brac{(\angle{t_1} - \angle{r_1}) - (\angle{t_2} - \angle{r_2}) }}. 
	\label{eqn:weakCalibrationMeasurement}
\end{equation}
Thus, with \eqref{eqn:weakCalibrationMeasurement}, antenna $ 2 $ can precompensate the phase, so as to have coherent combining of the signals at the \gls{ue} and we obtain the overall channel gain from antenna $ 2 $ as 
\begin{equation}
	\begin{aligned}
		\frac{z_{12}}{z_{21}} h_2 & = \frac{\Abs{t_1} \Abs{r_2}}{\Abs{t_2} \Abs{r_1}} e^{j\Brac{(\angle{t_1} - \angle{r_1}) - (\angle{t_2} - \angle{r_2}) }}  \\
			& \qquad \times \Abs{t_u} \Abs{r_2} \Abs{g_2}^2 \Abs{t_2} \Abs{r_u} e^{j\Brac{(\angle{t_2} - \angle{r_2}) - (\angle{t_u} - \angle{r_u} )} } \\
		& = \frac{\Abs{t_1} \Abs{r_u}}{\Abs{r_1} \Abs{t_u}} \Abs{t_u}^2 \Abs{g_2}^2 \Abs{r_2}^2 e^{j((\angle{t_1} - \angle{r_1}) - (\angle{t_u} - \angle{r_u}))}.	
	\end{aligned}
	\label{eqn:calibration2Antennas}
\end{equation}

Thus, for many antennas, we could perform bidirectional measurements of all antennas ($ 2,3,\cdots $) with antenna $ 1 $ and obtain the calibration coefficients of all the antennas.
For a collocated \gls{ap}, where all the antennas are at the same unit, the channels between the antennas remain same for a long duration of time. 
Moreover, as all the antennas in the \gls{ap} share the clocks, the internal calibration coefficients obtained are stable for longer periods of time. 
Hence, calibrating the antennas in a collocated setup can be done very infrequently \cite{shepard2012argos}.

\bibliographystyle{IEEEtran}
\bibliography{IEEEabrv,references}

\vfill

\begin{IEEEbiography}[{\includegraphics[width=1in,height=1.25in,clip,keepaspectratio]{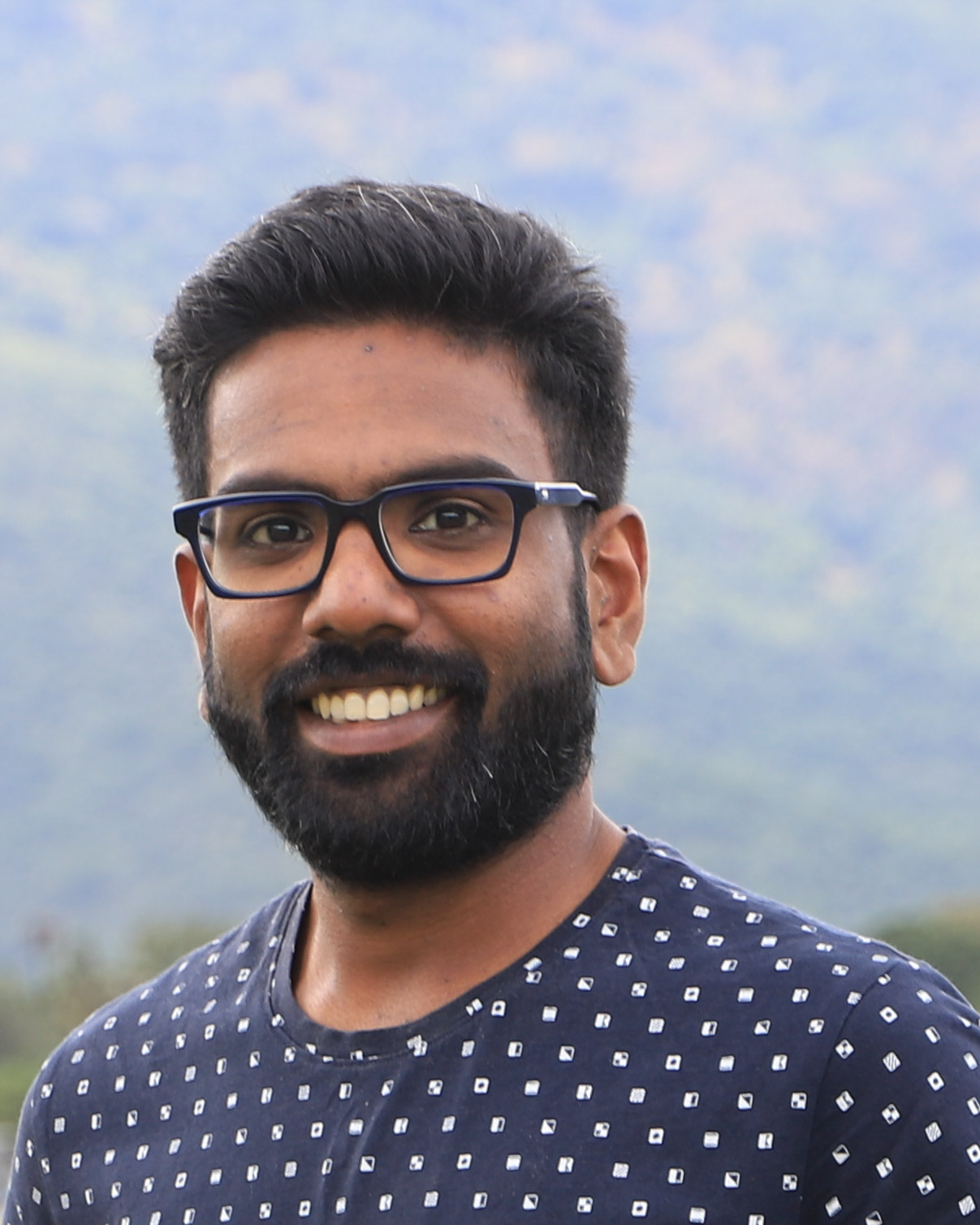}}]{Unnikrishnan Kunnath Ganesan (S'20)} received the Bachelor of Technology degree in Electronics and Communication Engineering from University of Calicut, Kerala, India in 2011 and the Masters in Engineering degree in Telecommunication Engineering from Indian Institute of Science, Bangalore, India in 2014. 
From 2014 to 2017, he worked as modem systems engineer with Qualcomm India Private Limited, Bangalore and from 2017 to 2019 he worked as senior firmware engineer with Intel. 
He is currently pursuing the Ph.D. degree with the Department of Electrical Engineering (ISY), Link\"oping University, Sweden. 
His primary research interests includes MIMO wireless communications, signal processing, and information theory.
\end{IEEEbiography}

\vfill

\begin{IEEEbiography}[{\includegraphics[width=1.1in,height=2in,clip,keepaspectratio]{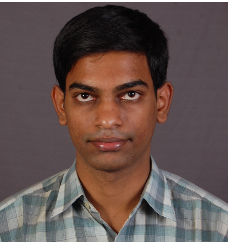}}]{Rimalapudi Sarvendranth (S'12-M'21)} received his Bachelor of Technology degree in Electrical and Electronics Engineering from the National Institute of Technology Karnataka, Surathkal in 2009. 
He received his Master of Engineering and Ph.D. degrees from the Department of Electrical Communication Engineering, Indian Institute of Science, Bangalore in 2012 and 2020, respectively. 
He is currently working as an assistant professor in the Department of Electrical Engineering at the Indian Institute of Technology Tirupati. 
In 2021, he was a postdoctoral researcher in the Department of Electrical Engineering, at Linköping University, Sweden. 
He worked as an assistant professor in the  Electronics and Electrical Engineering department of the Indian Institute of Technology Guwahati from January 2022 to June 2023. 
From 2012 to 2016, he was with Broadcom Communications Technologies, Bangalore, India, where he worked on the development and implementation of algorithms for LTE and IEEE 802.11ac wireless standards.  
His research interests include machine learning for wireless communication, multiple antenna techniques, spectrum sharing, and next-generation wireless standards.
	
\end{IEEEbiography}

\begin{IEEEbiography}[{\includegraphics[width=1.1in,height=1.35in,clip,keepaspectratio]{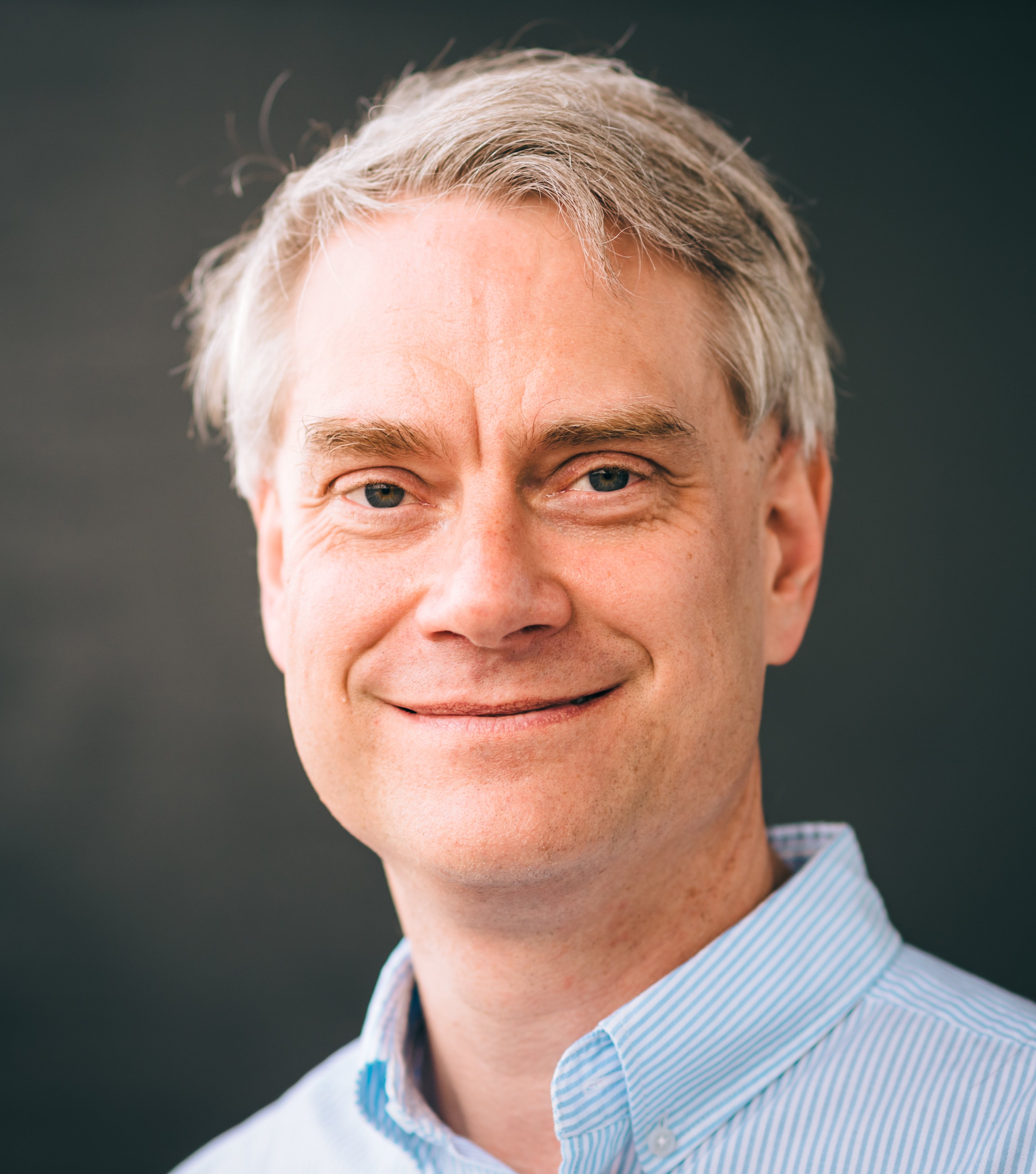}}]{Erik G. Larsson (S'99--M'03--SM'10--F'16)} 
received the Ph.D. degree from Uppsala University, Uppsala, Sweden, in 2002.  
He is currently Professor of Communication Systems at Link\"oping University (LiU) in Link\"oping, Sweden. 
He was with the KTH Royal Institute of Technology in Stockholm, Sweden, the George Washington University, USA, the University of Florida, USA, and Ericsson Research, Sweden.  
His main professional interests are within the areas of wireless communications and signal processing. 
He co-authored \emph{Space-Time Block Coding for  Wireless Communications} (Cambridge University Press, 2003) and \emph{Fundamentals of Massive MIMO} (Cambridge University Press, 2016). 

He served as  chair  of the IEEE Signal Processing Society SPCOM technical committee (2015--2016), chair of  the \emph{IEEE Wireless  Communications Letters} steering committee (2014--2015), member of the  \emph{IEEE Transactions on Wireless Communications} steering committee (2019-2022), General and Technical Chair of the Asilomar SSC conference (2015, 2012), 
technical co-chair of the IEEE Communication Theory Workshop (2019), and   member of the  IEEE Signal Processing Society Awards Board (2017--2019). 
He was Associate Editor for, among others, the \emph{IEEE Transactions on Communications} (2010-2014), the \emph{IEEE Transactions on Signal Processing} (2006-2010),and the \emph{IEEE Signal  Processing Magazine} (2018-2022).

He received the IEEE Signal Processing Magazine Best Column Award twice, in 2012 and 2014, the IEEE ComSoc Stephen O. Rice Prize in
Communications Theory in 2015, the IEEE ComSoc Leonard G. Abraham Prize in 2017, the IEEE ComSoc Best Tutorial Paper Award in 2018, and the IEEE ComSoc Fred W. Ellersick Prize in 2019.
He is a member of the Swedish Royal Academy of Sciences (KVA), and Highly Cited according to ISI Web of Science.
	
\end{IEEEbiography}

\end{document}

%% file: phaseEstimateMethodsComparison.tex
\begin{tikzpicture}

\begin{axis}[%
%width=0.55\textwidth,
%height=0.4\textwidth,
width=0.4\textwidth,
height=0.3\textwidth,
at={(1.083in,0.808in)},
scale only axis,
xmin=-30,
xmax=30,
xlabel style={font=\color{white!15!black}},
xlabel={SNR (dB)},
ymode=log,
ymin=0.001,
ymax=10,
yminorticks=true,
ylabel style={font=\color{white!15!black}},
ylabel={RMSE (radians)},
axis background/.style={fill=white},
xmajorgrids,
ymajorgrids,
yminorgrids,
legend style={nodes={scale=1, transform shape},legend cell align=left, align=right, draw=white!15!black}
]
\addplot [color=red, line width=1.0pt]
table[row sep=crcr]{%
	-30	1.81001000409932\\
	-27	1.80632985919353\\
	-24	1.80442663601759\\
	-21	1.78561494909131\\
	-18	1.75787856195224\\
	-15	1.6933675952393\\
	-12	1.55041169016791\\
	-9	1.286882013535\\
	-6	0.855425372894017\\
	-3	0.41304054037942\\
	0	0.210571401486011\\
	3	0.130461651815512\\
	6	0.0878043091574807\\
	9	0.0603095302378284\\
	12	0.0419530362791463\\
	15	0.0293864062556573\\
	18	0.0207323179466313\\
	21	0.0146866662702139\\
	24	0.0103573611133648\\
	27	0.00731345505501753\\
	30	0.00519234213599701\\
};
\addlegendentry{FGB M=16}

\addplot [color=blue, line width=1.0pt]
table[row sep=crcr]{%
	-30	1.81510389959825\\
	-27	1.80347350375235\\
	-24	1.76188488272128\\
	-21	1.67418119863928\\
	-18	1.59102537306051\\
	-15	1.44207141168141\\
	-12	1.0390264535507\\
	-9	0.580096408032933\\
	-6	0.225532341180083\\
	-3	0.131930863485958\\
	0	0.0763869352736541\\
	3	0.0486357520555312\\
	6	0.0330596343292018\\
	9	0.0226483278828872\\
	12	0.0160331029417841\\
	15	0.0116015505974038\\
	18	0.00799395765040354\\
	21	0.00567216222845568\\
	24	0.0040643574539416\\
	27	0.00284318274140454\\
	30	0.00196153830662327\\
};
\addlegendentry{Vieira \cite{vieira2021reciprocity}, M=16}

%\addplot [color=blue, dashdotted, line width=1.0pt, mark=diamond, mark options={solid, blue}]
%table[row sep=crcr]{%
%	-30	1.77788027380234\\
%	-27	1.74574725835592\\
%	-24	1.67835164633497\\
%	-21	1.54184753074044\\
%	-18	1.28828917948372\\
%	-15	0.889769226068822\\
%	-12	0.476890860008855\\
%	-9	0.245025233195167\\
%	-6	0.148323682593157\\
%	-3	0.0973851914972845\\
%	0	0.0669339247308312\\
%	3	0.046396197986235\\
%	6	0.032662588866518\\
%	9	0.0229801289645212\\
%	12	0.0161851722263173\\
%	15	0.0114758064431301\\
%	18	0.00813434776940711\\
%	21	0.00575830703295645\\
%	24	0.00406665821208293\\
%	27	0.0028823507090748\\
%	30	0.00204278157488338\\
%};
%\addlegendentry{BeamSync Random M=16}

\addplot [color=blue, dashed, line width=1.0pt]
table[row sep=crcr]{%
	-30	1.79218554675561\\
	-27	1.77075431307759\\
	-24	1.73402487582252\\
	-21	1.64860326609067\\
	-18	1.48497189823752\\
	-15	1.17561010366622\\
	-12	0.708744662827941\\
	-9	0.329264249561466\\
	-6	0.172298886242753\\
	-3	0.101108444382159\\
	0	0.0621531801152479\\
	3	0.0401036606964569\\
	6	0.0265097335687498\\
	9	0.0179888161147714\\
	12	0.0125222445056914\\
	15	0.008739774474968\\
	18	0.00617248778553452\\
	21	0.00436407843066703\\
	24	0.00308348640004004\\
	27	0.0021807301456067\\
	30	0.00154015899568334\\
};
\addlegendentry{BeamSync, M=16}

%\addplot [color=blue, dotted, line width=1.0pt, mark=star, mark options={solid, blue}]
%table[row sep=crcr]{%
%	-30	1.77716305244143\\
%	-27	1.74368348816628\\
%	-24	1.6735962269081\\
%	-21	1.53392088733872\\
%	-18	1.2731015180696\\
%	-15	0.873026580789822\\
%	-12	0.462925796838701\\
%	-9	0.241390415885353\\
%	-6	0.143841927496101\\
%	-3	0.0903134668846655\\
%	0	0.0567844503471131\\
%	3	0.0359236921423849\\
%	6	0.0230202074303393\\
%	9	0.0149391045151724\\
%	12	0.0100330950885337\\
%	15	0.00691433193198878\\
%	18	0.00478307865061221\\
%	21	0.00337746869270408\\
%	24	0.002362874312176\\
%	27	0.0016794037694773\\
%	30	0.00117841659402376\\
%};
%\addlegendentry{BeamSync $\alpha=0.3$, M=16}

\addplot [color=black, line width=1.0pt]
table[row sep=crcr]{%
	-30	1.77007728291866\\
	-27	1.77435249953252\\
	-24	1.74693105945668\\
	-21	1.68842043860463\\
	-18	1.48963485465258\\
	-15	1.19977196596148\\
	-12	0.787774252759524\\
	-9	0.267312111008167\\
	-6	0.0643937078369497\\
	-3	0.0309886171091989\\
	0	0.0195058475075836\\
	3	0.0126878827736341\\
	6	0.00786011328414567\\
	9	0.00560088993098003\\
	12	0.00420396341085961\\
	15	0.00284974758760835\\
	18	0.00196559037051878\\
	21	0.00138638512304956\\
	24	0.000965490672725636\\
	27	0.000688039288184907\\
	30	0.000486765244740381\\
};
\addlegendentry{Vieira \cite{vieira2021reciprocity}, M=64}

%\addplot [color=black, dashdotted, line width=1.0pt, mark=diamond, mark options={solid, black}]
%table[row sep=crcr]{%
%	-30	1.5543659983117\\
%	-27	1.29446542644203\\
%	-24	0.870371060496766\\
%	-21	0.412371358433325\\
%	-18	0.201190201841622\\
%	-15	0.113867261864972\\
%	-12	0.0711888882330265\\
%	-9	0.047256487150628\\
%	-6	0.0324246154277118\\
%	-3	0.0226485079843467\\
%	0	0.015858781086385\\
%	3	0.011185608238784\\
%	6	0.00790139108201727\\
%	9	0.00556742411333159\\
%	12	0.00395919548392885\\
%	15	0.00280387313588536\\
%	18	0.00197682875254401\\
%	21	0.00139892638887582\\
%	24	0.000992687870064825\\
%	27	0.000703624731850595\\
%	30	0.000498637662563495\\
%};
%\addlegendentry{BeamSync Random M=64}

\addplot [color=black, dashed, line width=1.0pt]
table[row sep=crcr]{%
	-30	1.66220279495309\\
	-27	1.50314293961073\\
	-24	1.20206734959463\\
	-21	0.733906909368105\\
	-18	0.33556118061562\\
	-15	0.168134345462372\\
	-12	0.0967231943809227\\
	-9	0.059163284645779\\
	-6	0.0367402513080777\\
	-3	0.0229519869637858\\
	0	0.0144480863512541\\
	3	0.00939220434181049\\
	6	0.00623085918761468\\
	9	0.00424870201629087\\
	12	0.00294142537984863\\
	15	0.00206383736631642\\
	18	0.00144932029097005\\
	21	0.00102484660653487\\
	24	0.000719647418570436\\
	27	0.000510725057191553\\
	30	0.000361939479134401\\
};
\addlegendentry{BeamSync, M=64}

%\addplot [color=black, dotted, line width=1.0pt, mark=star, mark options={solid, black}]
%table[row sep=crcr]{%
%	-30	1.53939798336857\\
%	-27	1.2636540405994\\
%	-24	0.825618094040427\\
%	-21	0.388498060238583\\
%	-18	0.194921565994765\\
%	-15	0.113779781048577\\
%	-12	0.0718803226544328\\
%	-9	0.0477161835121299\\
%	-6	0.0319980427001975\\
%	-3	0.0211120948985257\\
%	0	0.0138238540617443\\
%	3	0.00880159040390604\\
%	6	0.00559229828005714\\
%	9	0.00358089971084691\\
%	12	0.00237000092158813\\
%	15	0.00160323922734295\\
%	18	0.00110373347562079\\
%	21	0.000770773176319107\\
%	24	0.000543646777780423\\
%	27	0.000384025701351007\\
%	30	0.000271066616940035\\
%};
%\addlegendentry{BeamSync $\alpha=0.2$, M=64}

\end{axis}
\end{tikzpicture}%

%% file: phaseEstimatePlot.tex
\begin{tikzpicture}

\begin{axis}[%
%width=0.55\textwidth,
%height=0.4\textwidth,
width=0.4\textwidth,
height=0.3\textwidth,
at={(1.083in,0.797in)},
scale only axis,
xmin=-30,
xmax=30,
xlabel style={font=\color{white!15!black}},
xlabel={SNR (dB)},
ymode=log,
ymin=0.0001,
ymax=10,
yminorticks=true,
ylabel style={font=\color{white!15!black}},
ylabel={RMSE (radians)},
axis background/.style={fill=white},
xmajorgrids,
ymajorgrids,
yminorgrids,
legend style={nodes={scale=0.9, transform shape}, at={(0,0)}, anchor=south west, legend cell align=left, align=left, draw=white!15!black}
]
\addplot [color=blue, dashed, line width=1.0pt, mark=x, mark options={solid, blue}]
table[row sep=crcr]{%
	-30	1.80223428427646\\
	-27	1.79206130848495\\
	-24	1.76664138533848\\
	-21	1.71519075536454\\
	-18	1.58238591849146\\
	-15	1.28068644592681\\
	-12	0.730018696743736\\
	-9	0.265361525619414\\
	-6	0.123003405886324\\
	-3	0.0751830132669807\\
	0	0.0498561478192219\\
	3	0.0340250986956466\\
	6	0.0236722353585625\\
	9	0.0166125172156128\\
	12	0.011670423222127\\
	15	0.00823927634499967\\
	18	0.00584924814634694\\
	21	0.00412308891914215\\
	24	0.00291504989492436\\
	27	0.00206960588391262\\
	30	0.00146748214348882\\
};
\addlegendentry{FGB Simple Estimator}

\addplot [color=black, dashed, line width=1.0pt, mark=x, mark options={solid, black}]
table[row sep=crcr]{%
	-30	1.74857397428301\\
	-27	1.69189921918086\\
	-24	1.57209945695946\\
	-21	1.35079379854288\\
	-18	0.976139057610415\\
	-15	0.531711191499785\\
	-12	0.263609411673463\\
	-9	0.15274629288389\\
	-6	0.0961503732967792\\
	-3	0.0630042857716697\\
	0	0.0430161264281249\\
	3	0.0297691825430363\\
	6	0.0208749531164295\\
	9	0.0146245288788425\\
	12	0.0103880768822721\\
	15	0.0073445883335749\\
	18	0.00520962544235919\\
	21	0.00366130284983225\\
	24	0.00259693867225795\\
	27	0.00183864876301653\\
	30	0.00130379354856879\\
};
\addlegendentry{FGB PCSI}

\addplot [color=blue, line width=1.0pt, mark=x, mark options={solid, blue}]
table[row sep=crcr]{%
	-30	1.69211708123126\\
	-27	1.56541667045662\\
	-24	1.29631115289207\\
	-21	0.808445988306759\\
	-18	0.332408487804599\\
	-15	0.145920844288228\\
	-12	0.0784065563382547\\
	-9	0.0472028690613885\\
	-6	0.030490639536768\\
	-3	0.0204779762737628\\
	0	0.0142042960334994\\
	3	0.00986342125917406\\
	6	0.00695664231565165\\
	9	0.0049005788855416\\
	12	0.00347126014341212\\
	15	0.00245211406805259\\
	18	0.00173352910062541\\
	21	0.00122630505685803\\
	24	0.000868764304495489\\
	27	0.000613737973058972\\
	30	0.000436997838183999\\
};
\addlegendentry{BeamSync Simple Estimator}

\addplot [color=red, line width=1.0pt, mark=o, mark options={solid, red}]
table[row sep=crcr]{%
	-30	1.64551523624946\\
	-27	1.47560492001868\\
	-24	1.1410708605862\\
	-21	0.629740393382674\\
	-18	0.264496576120632\\
	-15	0.1300828858613\\
	-12	0.0738585145208951\\
	-9	0.0456913085381908\\
	-6	0.0300077525513121\\
	-3	0.0203122640817738\\
	0	0.0141429226800815\\
	3	0.00984216485540364\\
	6	0.00694835318986816\\
	9	0.00489838407304311\\
	12	0.00346988785555346\\
	15	0.00245174024131076\\
	18	0.00174004453367803\\
	21	0.00122540089307971\\
	24	0.000867079781604926\\
	27	0.000613730485396313\\
	30	0.00043501780115497\\
};
\addlegendentry{BeamSync NLS}

\addplot [color=black, line width=1.0pt, mark=x, mark options={solid, black}]
table[row sep=crcr]{%
	-30	1.42278229648361\\
	-27	1.06611969202464\\
	-24	0.577724284210583\\
	-21	0.268623924654616\\
	-18	0.147704367642352\\
	-15	0.0906579690666521\\
	-12	0.059605325801631\\
	-9	0.0405520118881969\\
	-6	0.0281622214491666\\
	-3	0.0196641078903966\\
	0	0.0138670930668519\\
	3	0.00974099375953853\\
	6	0.00689294013705124\\
	9	0.00490693220149492\\
	12	0.00347570955614677\\
	15	0.00244756302154075\\
	18	0.0017344029520544\\
	21	0.00122804930950461\\
	24	0.00086887903836859\\
	27	0.00061455742792348\\
	30	0.000436503113895072\\
};
\addlegendentry{BeamSync PCSI}

\end{axis}

\end{tikzpicture}%

%% file: phaseEstimatePerformanceNumAntennas.tex
\definecolor{mycolor1}{rgb}{1.00000,0.00000,1.00000}%
\begin{tikzpicture}

\begin{axis}[%
%width=0.55\textwidth,
%height=0.4\textwidth,
width=0.4\textwidth,
height=0.3\textwidth,
at={(1.083in,0.797in)},
scale only axis,
xmin=-30,
xmax=30,
xlabel style={font=\color{white!15!black}},
xlabel={SNR (dB)},
ymode=log,
ymin=0.0001,
ymax=10,
yminorticks=true,
ylabel style={font=\color{white!15!black}},
ylabel={RMSE (radians)},
axis background/.style={fill=white},
xmajorgrids,
ymajorgrids,
yminorgrids,
legend style={nodes={scale=0.6, transform shape}, legend cell align=left, align=left, draw=white!15!black}
]
\addplot [color=blue, dashed, line width=1.0pt]
  table[row sep=crcr]{%
-30	1.80264686942972\\
-27	1.788118962219\\
-24	1.76887256118538\\
-21	1.72700473056583\\
-18	1.62061749818554\\
-15	1.40132340912537\\
-12	1.01448837372981\\
-9	0.541900289849804\\
-6	0.226592995179056\\
-3	0.108007943602401\\
0	0.0643343949256654\\
3	0.0413583913300577\\
6	0.0274966941171493\\
9	0.0188734751439425\\
12	0.0130641835727554\\
15	0.00917512117064892\\
18	0.00647608413173635\\
21	0.00456802676308261\\
24	0.0032359554549641\\
27	0.00227619995178313\\
30	0.00161994849275565\\
};
\addlegendentry{$M=8$ FGB}

\addplot [color=red, dashed, line width=1.0pt]
  table[row sep=crcr]{%
-30	1.80204716385096\\
-27	1.79137710683968\\
-24	1.77184045015808\\
-21	1.71584107140124\\
-18	1.61078189540286\\
-15	1.35556542465714\\
-12	0.894717248454308\\
-9	0.396067122488787\\
-6	0.158295500269191\\
-3	0.0884353896436173\\
0	0.0555343657972028\\
3	0.0364636874500327\\
6	0.0246536638905822\\
9	0.0169501860011337\\
12	0.0118177782524361\\
15	0.00832710100360921\\
18	0.00585036361684318\\
21	0.00415559805964683\\
24	0.0029237977360381\\
27	0.00206537289280922\\
30	0.00146865328949188\\
};
\addlegendentry{$M=16$ FGB}

\addplot [color=black, dashed, line width=1.0pt]
  table[row sep=crcr]{%
-30	1.79752200214495\\
-27	1.79314277873631\\
-24	1.77295628819153\\
-21	1.71496038076091\\
-18	1.59806123119395\\
-15	1.30724946777797\\
-12	0.775840663105337\\
-9	0.288283645129738\\
-6	0.130656752336212\\
-3	0.0780257326944006\\
0	0.0499193889281149\\
3	0.0334223078573775\\
6	0.0227946941196952\\
9	0.015729751400542\\
12	0.011000775777566\\
15	0.00771253369325246\\
18	0.00545965878609961\\
21	0.00386109764205013\\
24	0.00273039323298427\\
27	0.00193647441007391\\
30	0.00136566921563338\\
};
\addlegendentry{$M=32$ FGB}

\addplot [color=mycolor1, dashed, line width=1.0pt]
  table[row sep=crcr]{%
-30	1.80359053119965\\
-27	1.79337692963814\\
-24	1.76495645445154\\
-21	1.71246685542309\\
-18	1.57765624523247\\
-15	1.24657427425625\\
-12	0.664041667451346\\
-9	0.225708626955278\\
-6	0.114590322732046\\
-3	0.070991844138906\\
0	0.046436764197389\\
3	0.031104555325433\\
6	0.021325431049936\\
9	0.0148803474679652\\
12	0.0104206756426983\\
15	0.00736727196186816\\
18	0.00518536690312098\\
21	0.00366913060469619\\
24	0.0025979817052215\\
27	0.00183155507169699\\
30	0.00129966243839421\\
};
\addlegendentry{$M=64$ FGB}

\addplot [color=blue, line width=1.0pt, mark=x, mark options={solid, blue}]
  table[row sep=crcr]{%
-30	1.74490575324902\\
-27	1.68454585931934\\
-24	1.54277002058336\\
-21	1.27419074497532\\
-18	0.811731162222901\\
-15	0.360059739563409\\
-12	0.15938986014236\\
-9	0.0847220565096077\\
-6	0.0507879433303973\\
-3	0.0328007588359525\\
0	0.0218385121162621\\
3	0.0151093464840372\\
6	0.0105836749402561\\
9	0.00745987414183778\\
12	0.00525647057288693\\
15	0.00370804149096303\\
18	0.00262117278294492\\
21	0.00185592752573631\\
24	0.00131682732599717\\
27	0.000930448561022707\\
30	0.000659222680220079\\
};
\addlegendentry{$M=8$ BeamSync}

\addplot [color=blue, dashdotted, line width=1.0pt, mark=o, mark options={solid, blue}]
  table[row sep=crcr]{%
-30	1.6164730241133\\
-27	1.43170074444937\\
-24	1.09115097696068\\
-21	0.618497436327968\\
-18	0.287152018264917\\
-15	0.160461739794476\\
-12	0.0980436076545754\\
-9	0.0642969318976659\\
-6	0.0433702432279855\\
-3	0.030012494950039\\
0	0.0210932345176592\\
3	0.0147754372349663\\
6	0.0104273311376541\\
9	0.00738883322258807\\
12	0.00523924043612812\\
15	0.00369478070733764\\
18	0.00261476506460084\\
21	0.00185594293660869\\
24	0.00130904675207477\\
27	0.00092973708661273\\
30	0.000658050137250022\\
};
\addlegendentry{$M=8$ BeamSync NLS}

\addplot [color=red, line width=1.0pt, mark=x, mark options={solid, red}]
  table[row sep=crcr]{%
-30	1.69025239012902\\
-27	1.5641195977205\\
-24	1.29430506433406\\
-21	0.810608877240708\\
-18	0.332126831168304\\
-15	0.146372375192206\\
-12	0.078176473863569\\
-9	0.0472970663001617\\
-6	0.0303800202497493\\
-3	0.0205489220120222\\
0	0.0141790830473086\\
3	0.00988442204140499\\
6	0.00695048717201133\\
9	0.00490778216747686\\
12	0.0034551837754956\\
15	0.00244487700724954\\
18	0.00172878851454977\\
21	0.0012270951719715\\
24	0.000870634235871306\\
27	0.000615637914846048\\
30	0.000433768395232318\\
};
\addlegendentry{$M=16$ BeamSync}

\addplot [color=red, dashdotted, line width=1.0pt, mark=o, mark options={solid, red}]
  table[row sep=crcr]{%
-30	1.42278229648361\\
-27	1.06611969202464\\
-24	0.577724284210583\\
-21	0.268623924654616\\
-18	0.147921552399311\\
-15	0.0913125429534357\\
-12	0.0595003131965857\\
-9	0.0405028295111508\\
-6	0.0281457076265184\\
-3	0.0196944584220971\\
0	0.0138068159224453\\
3	0.00974099375953853\\
6	0.00691233767249654\\
9	0.00489096896778992\\
12	0.00346724660210969\\
15	0.00245545730807231\\
18	0.00172977286961448\\
21	0.00122395168662631\\
24	0.000870673999783473\\
27	0.00061329907157054\\
30	0.000435348584739648\\
};
\addlegendentry{$M=16$ BeamSync NLS}

\addplot [color=black, line width=1.0pt, mark=x, mark options={solid, black}]
  table[row sep=crcr]{%
-30	1.58465771123285\\
-27	1.33976767849992\\
-24	0.85490952422428\\
-21	0.337839721228337\\
-18	0.144548263681463\\
-15	0.0760435461541792\\
-12	0.0454857202604712\\
-9	0.0292776012938255\\
-6	0.0197454404331748\\
-3	0.0136059842301034\\
0	0.00952506084657318\\
3	0.0066989981432604\\
6	0.00468698885713894\\
9	0.00333146318380806\\
12	0.002348554391152\\
15	0.00167119321771425\\
18	0.0011758581365806\\
21	0.000833548448756065\\
24	0.000590046813431774\\
27	0.000418513228975635\\
30	0.000297538426087277\\
};
\addlegendentry{$M=32$ BeamSync}

\addplot [color=black, dashdotted, line width=1.0pt, mark=o, mark options={solid, black}]
  table[row sep=crcr]{%
-30	1.09028907868021\\
-27	0.597907514663122\\
-24	0.273531982351667\\
-21	0.146806755608207\\
-18	0.0884985441724216\\
-15	0.0576575000789848\\
-12	0.0389685197795311\\
-9	0.0270599036796657\\
-6	0.0188612172531671\\
-3	0.0133065198124922\\
0	0.00936955410011862\\
3	0.00664538630361776\\
6	0.00469862705526938\\
9	0.00333413311489992\\
12	0.00236283114427647\\
15	0.00166219502416274\\
18	0.00117998995024784\\
21	0.000836763746721697\\
24	0.000590378512100911\\
27	0.00041946819014997\\
30	0.000296472865071471\\
};
\addlegendentry{$M=32$ BeamSync NLS}

\addplot [color=mycolor1, line width=1.0pt, mark=x, mark options={solid, mycolor1}]
  table[row sep=crcr]{%
-30	1.38934910395261\\
-27	0.932403451837722\\
-24	0.371513749378332\\
-21	0.151560843636064\\
-18	0.0766638385143692\\
-15	0.0449466649766651\\
-12	0.0288111138843175\\
-9	0.0192513017081494\\
-6	0.0132507165517407\\
-3	0.00928245740075405\\
0	0.00652418963974265\\
3	0.00458543495487982\\
6	0.00324941870126264\\
9	0.00229637685296058\\
12	0.00162819825077283\\
15	0.00114932666797039\\
18	0.000814212409411747\\
21	0.000578620516771141\\
24	0.000409047890597844\\
27	0.0002905567689958\\
30	0.000205744275150023\\
};
\addlegendentry{$M=64$ BeamSync}

\addplot [color=mycolor1, dashdotted, line width=1.0pt, mark=o, mark options={solid, mycolor1}]
  table[row sep=crcr]{%
-30	0.662195600302976\\
-27	0.297801701298096\\
-24	0.153839309054312\\
-21	0.0896237183666835\\
-18	0.0574485126285227\\
-15	0.03841619093652\\
-12	0.0264344534918364\\
-9	0.0184495947749463\\
-6	0.0129578996724441\\
-3	0.00917852363313061\\
0	0.00647561610665643\\
3	0.00459723938717531\\
6	0.00326015226478966\\
9	0.00230650207427264\\
12	0.00162736904282261\\
15	0.00115271167507506\\
18	0.000817853733729979\\
21	0.00057631563942741\\
24	0.000409141692975628\\
27	0.000289575860261279\\
30	0.00020440315413046\\
};
\addlegendentry{$M=64$ BeamSync NLS}

\end{axis}
\end{tikzpicture}%

%% file: BeamSync_SyncSignalLength_Performance.tex
\begin{tikzpicture}

\begin{axis}[%
%width=0.55\textwidth,
%height=0.4\textwidth,
width=0.4\textwidth,
height=0.3\textwidth,
at={(1.083in,0.808in)},
scale only axis,
xmin=0,
xmax=100,
xlabel style={font=\color{white!15!black}},
xlabel={Synchronization signal length},
ymode=log,
ymin=0.000820205816884309,
ymax=1,
yminorticks=true,
ylabel style={font=\color{white!15!black}},
ylabel={RMSE (radians)},
axis background/.style={fill=white},
xmajorgrids,
ymajorgrids,
yminorgrids,
legend style={legend cell align=left, align=left, draw=white!15!black}
]
\addplot [color=blue, line width=1.0pt]
  table[row sep=crcr]{%
2	0.03205845202783\\
4	0.0225057219254303\\
8	0.0159522683189166\\
10	0.0141659273838387\\
16	0.0112611478296804\\
20	0.0100532678131115\\
30	0.00829234450896739\\
40	0.00711663380199763\\
50	0.00639794376375745\\
60	0.0058144009770649\\
70	0.00540067468521559\\
80	0.00503711626003712\\
90	0.00475974070699704\\
100	0.00449581990572682\\
};
\addlegendentry{SNR = $10$ dB}

\addplot [color=black, dashed, line width=1.0pt]
  table[row sep=crcr]{%
2	0.00979057636088956\\
4	0.00692438866278426\\
8	0.00489998608234564\\
10	0.00437510427187919\\
16	0.00346650441499943\\
20	0.00308497879634254\\
30	0.00251990072698084\\
40	0.00218083752777056\\
50	0.00194947084931883\\
60	0.00177751613839527\\
70	0.00165192439841532\\
80	0.00154146145108276\\
90	0.00144695008279963\\
100	0.00138418680660854\\
};
\addlegendentry{ SNR = $20$ dB}

\end{axis}
\end{tikzpicture}%

%% file: phaseEstimatePerformanceWithPhaseNoise.tex

\begin{tikzpicture}

\begin{axis}[%
%width=0.55\textwidth,
%height=0.4\textwidth,
width=0.4\textwidth,
height=0.3\textwidth,
at={(1.083in,0.808in)},
scale only axis,
xmin=-30,
xmax=30,
xlabel style={font=\color{white!15!black}},
xlabel={SNR (dB)},
ymode=log,
ymin=0.0009,
ymax=10,
yminorticks=true,
ylabel style={font=\color{white!15!black}},
ylabel={RMSE (radians)},
axis background/.style={fill=white},
xmajorgrids,
ymajorgrids,
yminorgrids,
legend style={legend cell align=left, align=left, draw=white!15!black}
]
\addplot [color=blue, line width=1.0pt, mark=x, mark options={solid, blue}]
table[row sep=crcr]{%
	-30	1.76978186696359\\
	-27	1.73231328139391\\
	-24	1.63641145973069\\
	-21	1.43157245926709\\
	-18	0.977207126475685\\
	-15	0.429794084309211\\
	-12	0.19032940713365\\
	-9	0.10909357953799\\
	-6	0.070130207277799\\
	-3	0.0465829251024552\\
	0	0.0330364439331758\\
	3	0.0237009523058377\\
	6	0.0179657372735448\\
	9	0.0139265239174246\\
	12	0.0116851300401479\\
	15	0.0104427001929787\\
	18	0.00965151862606665\\
	21	0.0092619102887655\\
	24	0.00900787012122946\\
	27	0.00886302511042938\\
	30	0.00889983424831889\\
};
\addlegendentry{LO $1$}

\addplot [color=red, line width=1.0pt, mark=o, mark options={solid, red}]
table[row sep=crcr]{%
	-30	1.76346546073102\\
	-27	1.73181773953163\\
	-24	1.61627614303599\\
	-21	1.41335784701761\\
	-18	0.984480231485164\\
	-15	0.423192725602032\\
	-12	0.190326989863663\\
	-9	0.108570174051971\\
	-6	0.0684081126746177\\
	-3	0.0459564337067482\\
	0	0.0319097988375849\\
	3	0.0222436492088742\\
	6	0.0155650946611141\\
	9	0.0109981529488089\\
	12	0.00790019623532391\\
	15	0.00546945206035373\\
	18	0.00395973670099913\\
	21	0.0028358706405027\\
	24	0.00208966899200943\\
	27	0.00158688175353001\\
	30	0.00126567471321411\\
};
\addlegendentry{LO $2$}

\addplot [color=black, line width=1.0pt, mark=+, mark options={solid, black}]
table[row sep=crcr]{%
	-30	1.76710972661241\\
	-27	1.7112851110519\\
	-24	1.62158228558616\\
	-21	1.42795229623162\\
	-18	0.9760080323902\\
	-15	0.435201213273816\\
	-12	0.191127922445404\\
	-9	0.110766146046731\\
	-6	0.0689423731624022\\
	-3	0.0465603516540082\\
	0	0.0320583984350463\\
	3	0.0220117347888619\\
	6	0.0155896054132689\\
	9	0.0110435747841797\\
	12	0.00779394390303564\\
	15	0.00546139114417613\\
	18	0.00386709554014543\\
	21	0.00274314602122205\\
	24	0.00194973474765969\\
	27	0.00137516885512645\\
	30	0.000971516541615483\\
};
\addlegendentry{Ideal LO}

\end{axis}

\end{tikzpicture}%

%% file: calibrationErrorPlot.tex
\begin{tikzpicture}
		\pgfplotsset{
		every axis legend/.append style={ at={(0,1.26)}, nodes={scale=0.675},anchor=north west, legend cell align=left, align=left, legend columns = 1}}
	\begin{axis}[%
		%width=0.55\textwidth,
		%height=0.4\textwidth,
		width=0.4\textwidth,
		height=0.3\textwidth,
		at={(0.758in,0.481in)},
		scale only axis,
		xmin=0,
		xmax=250,
		xlabel style={font=\color{white!15!black}},
		xlabel={Time (minutes)},
		ymin=0,
		ymax=30,
		ylabel style={font=\color{white!15!black}},
		ylabel={Mean absolute phase deviation (degrees)},
		axis background/.style={fill=white},
		xmajorgrids,
		ymajorgrids,
		]
\addplot [color=blue, line width=1.0pt]
table[row sep=crcr]{%
	0	0.140119544151472\\
	1	9.55049309173667\\
	2	13.6222119497165\\
	3	16.5704125433554\\
	4	19.0765868997552\\
	5	21.3267121754967\\
	6	23.3144779821825\\
	7	25.1763144184997\\
	8	26.9110472727966\\
	9	28.3800666496834\\
	10	1.12437614994714\\
	11	9.63891817102832\\
	12	13.488181341353\\
	13	16.5467100262581\\
	14	19.1630973002884\\
	15	21.4991026334238\\
	16	23.5109540400244\\
	17	25.3417963900066\\
	18	27.054983439235\\
	19	28.7493913949693\\
	20	1.5886821019458\\
	21	9.67735724840093\\
	22	13.6374109215612\\
	23	16.6483011295896\\
	24	19.2907767470107\\
	25	21.72297903273\\
	26	23.6394405286634\\
	27	25.443546542857\\
	28	27.3611834445092\\
	29	28.9392373278602\\
	30	1.92758987287384\\
	31	9.72343833120831\\
	32	13.454869878437\\
	33	16.6188625195335\\
	34	19.0186686252018\\
	35	21.2660297553369\\
	36	23.2611521398421\\
	37	25.0359768133269\\
	38	26.8896757428694\\
	39	28.7885643681298\\
	40	2.22184554138603\\
	41	9.78203860200575\\
	42	13.8091427797011\\
	43	16.8257675352777\\
	44	19.2702079059099\\
	45	21.4871383299476\\
	46	23.4984643130353\\
	47	25.3260918601314\\
	48	26.9094879776627\\
	49	28.5483468403804\\
	50	2.51110709922626\\
	51	9.83789882694973\\
	52	13.7976475859486\\
	53	16.8794239882041\\
	54	19.3313289503636\\
	55	21.5761194901339\\
	56	23.6366051402734\\
	57	25.5331908234824\\
	58	27.3313357553845\\
	59	28.8205646589744\\
	60	2.7663578291474\\
	61	9.89682378010506\\
	62	13.7472198382042\\
	63	16.4541169136293\\
	64	18.8167619964083\\
	65	21.132287225265\\
	66	23.167272710204\\
	67	25.0693552790392\\
	68	26.6758138417899\\
	69	28.3764864460002\\
	70	2.95763925679374\\
	71	10.0489314477815\\
	72	13.8679920708435\\
	73	16.7876759733475\\
	74	19.3291676364793\\
	75	21.4570092290499\\
	76	23.6405171441149\\
	77	25.4796314007589\\
	78	27.2880252364697\\
	79	28.9237751143123\\
	80	3.20525273497832\\
	81	10.1975530042061\\
	82	13.9390426059337\\
	83	16.8950644517688\\
	84	19.2987283722059\\
	85	21.6114701446915\\
	86	23.652393599941\\
	87	25.6098291040364\\
	88	27.284644378478\\
	89	28.9973254594883\\
	90	3.32814655297284\\
	91	10.1266473909544\\
	92	13.8760269424925\\
	93	16.8747962901605\\
	94	19.176141213442\\
	95	21.3723452584777\\
	96	23.3758443028922\\
	97	25.3618330151772\\
	98	26.8179894782262\\
	99	28.3149081836811\\
	100	3.56973534086631\\
	101	10.2904154054555\\
	102	14.1142625098691\\
	103	17.1704933831724\\
	104	19.7360515304047\\
	105	21.884750416848\\
	106	23.6691811799242\\
	107	25.5313756196099\\
	108	27.2149789301792\\
	109	28.6750752103238\\
	110	3.78692081605589\\
	111	10.2004964688175\\
	112	14.0862172465114\\
	113	16.9872070123128\\
	114	19.4784659603696\\
	115	21.6412031279426\\
	116	23.6068387679855\\
	117	25.3845781239754\\
	118	27.143064436102\\
	119	28.7308335198794\\
	120	3.85586494139463\\
	121	10.4467460712156\\
	122	14.2139681653878\\
	123	17.0771833524339\\
	124	19.6490233928301\\
	125	21.8161555784524\\
	126	23.7339463032545\\
	127	25.5654733253864\\
	128	27.2832170847497\\
	129	28.9639207264064\\
	130	4.05950274943789\\
	131	10.2957310831866\\
	132	14.1131113443074\\
	133	17.1293151757853\\
	134	19.5233415882518\\
	135	21.6990038067641\\
	136	23.7561504292082\\
	137	25.7525455447304\\
	138	27.2463328542141\\
	139	29.1200636671499\\
	140	4.18572653002796\\
	141	10.4322659988315\\
	142	14.2993861631348\\
	143	17.0182564815299\\
	144	19.4568335253525\\
	145	21.7579553202786\\
	146	23.821470233484\\
	147	25.5955570055073\\
	148	27.1662168159344\\
	149	28.782021230297\\
	150	4.36045455640437\\
	151	10.5056542873783\\
	152	14.0119848574796\\
	153	17.0185587018783\\
	154	19.5686649576274\\
	155	21.8692559813114\\
	156	23.6558431670806\\
	157	25.4695588122021\\
	158	27.051298473843\\
	159	28.8115780206976\\
	160	4.50040370817764\\
	161	10.5885264163282\\
	162	14.1392019622348\\
	163	17.0653586377259\\
	164	19.6116534970569\\
	165	21.8913754137226\\
	166	23.8531571505107\\
	167	25.8819099588668\\
	168	27.6370550908698\\
	169	29.2725252526932\\
	170	4.60036169519233\\
	171	10.4601059422263\\
	172	14.2894151140383\\
	173	17.3911205509562\\
	174	19.8447983670413\\
	175	22.0069759598274\\
	176	23.9395076945021\\
	177	25.7649166797954\\
	178	27.5277273818532\\
	179	28.9480388298342\\
	180	4.75209749603683\\
	181	10.6490681871306\\
	182	14.3110389276904\\
	183	17.2477555160371\\
	184	19.8377560861796\\
	185	22.0506492804787\\
	186	24.0935594857286\\
	187	25.9156087044554\\
	188	27.6593659048543\\
	189	29.4662721849625\\
	190	4.90770187671691\\
	191	10.7819012926374\\
	192	14.4088256164742\\
	193	17.4318037857849\\
	194	20.0914146021062\\
	195	22.2798462246784\\
	196	24.3434048001293\\
	197	26.0416757084639\\
	198	27.8023700590357\\
	199	29.3982643010381\\
	200	4.98235861289706\\
	201	10.7717288314263\\
	202	14.4418594565428\\
	203	17.2713243879355\\
	204	19.7818714038811\\
	205	21.80593910766\\
	206	23.7871386988829\\
	207	25.6854709732444\\
	208	27.3411127115161\\
	209	28.9276377711848\\
	210	5.14617014951014\\
	211	10.9881334683797\\
	212	14.7397220735804\\
	213	17.5688139067578\\
	214	19.9098745751801\\
	215	22.0538582008369\\
	216	24.1934605050565\\
	217	25.7444407302609\\
	218	27.6089336064153\\
	219	29.2569253344324\\
	220	5.25027055158622\\
	221	10.8005390129758\\
	222	14.2685017825684\\
	223	17.2710903566438\\
	224	19.8407049597303\\
	225	21.8584597789014\\
	226	23.9208257062389\\
	227	25.8200018134199\\
	228	27.5437109468766\\
	229	29.2334432909051\\
	230	5.3907544501125\\
	231	11.071778920538\\
	232	14.5670354550459\\
	233	17.3973818590197\\
	234	19.9386362129902\\
	235	22.0779837882707\\
	236	24.2610350287728\\
	237	26.1337076088942\\
	238	27.9721754160155\\
	239	29.5506155163824\\
	240	5.60168392403289\\
	241	11.1289918860426\\
	242	14.7118885702595\\
	243	17.4948579010642\\
	244	19.9071973587723\\
	245	22.0066235786322\\
	246	24.0348609947097\\
	247	25.9838393716856\\
	248	27.6733691650019\\
	249	29.2780885354948\\
	250	5.61435525601757\\
};
\addlegendentry{BeamSync periodic, with reciprocity errors and with LO drift}

\addplot [color=red, line width=1.0pt, mark=square, mark options={solid, red}]
table[row sep=crcr]{%
	0	0.140119544151472\\
	10	1.12437614994714\\
	20	1.5886821019458\\
	30	1.92758987287384\\
	40	2.22184554138603\\
	50	2.51110709922626\\
	60	2.7663578291474\\
	70	2.95763925679374\\
	80	3.20525273497832\\
	90	3.32814655297284\\
	100	3.56973534086631\\
	110	3.78692081605589\\
	120	3.85586494139463\\
	130	4.05950274943789\\
	140	4.18572653002796\\
	150	4.36045455640437\\
	160	4.50040370817764\\
	170	4.60036169519233\\
	180	4.75209749603683\\
	190	4.90770187671691\\
	200	4.98235861289706\\
	210	5.14617014951014\\
	220	5.25027055158622\\
	230	5.3907544501125\\
	240	5.60168392403289\\
	250	5.61435525601757\\
};
\addlegendentry{BeamSync periodic, with reciprocity errors but without LO drift}

\addplot [color=black,  line width=1.0pt, mark=diamond, mark options={solid, black}]
table[row sep=crcr]{%
	0	0.140119544151472\\
	10	0.825816384125895\\
	20	1.14822736513919\\
	30	1.41933173122091\\
	40	1.6157701943148\\
	50	1.81769436789308\\
	60	1.97954661192199\\
	70	2.17339583299325\\
	80	2.29084868504505\\
	90	2.42219184253086\\
	100	2.57615232472604\\
	110	2.68540238092298\\
	120	2.77198093549559\\
	130	2.92651757535836\\
	140	3.02896994088058\\
	150	3.16793866435514\\
	160	3.23466399200971\\
	170	3.35107427082852\\
	180	3.397975141123\\
	190	3.48324626047557\\
	200	3.62248757234146\\
	210	3.70866031857502\\
	220	3.79845157103478\\
	230	3.89229814990658\\
	240	3.97475118020746\\
	250	4.05690367476749\\
};
\addlegendentry{BeamSync once, with reciprocity errors but without LO drift}

\addplot [color=black,dashdotted ,line width=1.0pt, mark=o, mark options={solid, black}]
table[row sep=crcr]{%
	0	0.140119544151472\\
	10	0.140119544151472\\
	20	0.140119544151472\\
	30	0.140119544151472\\
	40	0.140119544151472\\
	50	0.140119544151472\\
	60	0.140119544151472\\
	70	0.140119544151472\\
	80	0.140119544151472\\
	90	0.140119544151473\\
	100	0.140119544151472\\
	110	0.140119544151472\\
	120	0.140119544151472\\
	130	0.140119544151472\\
	140	0.140119544151472\\
	150	0.140119544151472\\
	160	0.140119544151472\\
	170	0.140119544151472\\
	180	0.140119544151472\\
	190	0.140119544151472\\
	200	0.140119544151472\\
	210	0.140119544151472\\
	220	0.140119544151472\\
	230	0.140119544151472\\
	240	0.140119544151472\\
	250	0.140119544151473\\
};
\addlegendentry{BeamSync once, without reciprocity errors and without LO drift}
		
	\end{axis}

\end{tikzpicture}%

%% file: offsetEstimationPlot.tex
\begin{tikzpicture}

\begin{axis}[%
%width=0.55\textwidth,
%height=0.4\textwidth,
width=0.4\textwidth,
height=0.3\textwidth,
at={(1.083in,0.808in)},
scale only axis,
xmin=-30,
xmax=20,
xlabel style={font=\color{white!15!black}},
xlabel={SNR (dB)},
ymode=log,
ymin=0.0001,
ymax=0.0229955511799563,
yminorticks=true,
ylabel style={font=\color{white!15!black}},
ylabel={RMSE (cycles/sample)},
axis background/.style={fill=white},
xmajorgrids,
ymajorgrids,
yminorgrids,
legend style={nodes={scale=1, transform shape}, at={(0,0)}, anchor=south west, legend cell align=left, align=left, draw=white!15!black}
]
\addplot [color=blue, line width=1.0pt, mark=square, mark options={solid, blue}]
  table[row sep=crcr]{%
-30	0.0229955511799563\\
-27	0.0229037738135514\\
-24	0.0228030173378104\\
-21	0.0226150265766007\\
-18	0.0222059729228878\\
-15	0.021182222013058\\
-12	0.0192998574661651\\
-9	0.015789271317268\\
-6	0.0112591368806374\\
-3	0.00775987833629158\\
0	0.00547387971563796\\
3	0.00390433598203077\\
6	0.00278157905655294\\
9	0.00198726352366709\\
12	0.00141938512450833\\
15	0.00101277357296229\\
18	0.000715696217230723\\
21	0.000510561571278594\\
24	0.000360995172041204\\
27	0.000254900278160573\\
30	0.000181099048625218\\
};
\addlegendentry{FGB}

\addplot [color=blue, dashed, line width=1.0pt, mark=diamond, mark options={solid, blue}]
  table[row sep=crcr]{%
-30	0.0228277069243354\\
-27	0.0226269919659595\\
-24	0.0222516207349539\\
-21	0.0216030921165592\\
-18	0.0203122884396009\\
-15	0.0183185013405467\\
-12	0.0155690375396524\\
-9	0.0122858075065867\\
-6	0.00916803506135625\\
-3	0.00667221031944465\\
0	0.00481766332001041\\
3	0.00346318900616889\\
6	0.00246904586382704\\
9	0.00177145086856991\\
12	0.00126258824752757\\
15	0.000895339703303364\\
18	0.000637431715987862\\
21	0.000451603836646304\\
24	0.000319736690281996\\
27	0.00022667417859447\\
30	0.000160730508632214\\
};
\addlegendentry{FGB - PCSI}

\addplot [color=black, line width=1.0pt, mark=x, mark options={solid, black}]
  table[row sep=crcr]{%
-30	0.0228759907408387\\
-27	0.0227290556429086\\
-24	0.0223846284640361\\
-21	0.0215649881160793\\
-18	0.0197879826839584\\
-15	0.0157849654956666\\
-12	0.0100835073608687\\
-9	0.00590684772804445\\
-6	0.0037621973466344\\
-3	0.00250992287212311\\
0	0.00173269259994445\\
3	0.00121143141975661\\
6	0.000849776533941595\\
9	0.000603055222474534\\
12	0.000427522244270884\\
15	0.00030262984400091\\
18	0.000213967442419456\\
21	0.000151628885056808\\
24	0.000107432296408523\\
27	7.5845347919378e-05\\
30	5.35640540749541e-05\\
};
\addlegendentry{BeamSync}

\addplot [color=black, dashed, line width=1.0pt, mark=o, mark options={solid, black}]
  table[row sep=crcr]{%
-30	0.0225162173379508\\
-27	0.0220564312611401\\
-24	0.0211213796609134\\
-21	0.0192897491785656\\
-18	0.016179823524472\\
-15	0.0115791568860102\\
-12	0.0076400617073558\\
-9	0.00503413468326795\\
-6	0.00345589782142386\\
-3	0.00240010362015167\\
0	0.00169068178726513\\
3	0.00119866899023176\\
6	0.000845918271345721\\
9	0.000601876356260598\\
12	0.000425303447966816\\
15	0.000302770510154243\\
18	0.000214521813855667\\
21	0.000151441307268899\\
24	0.000107322359714827\\
27	7.60143299169542e-05\\
30	5.36646839091446e-05\\
};
\addlegendentry{BeamSync - PCSI}

\end{axis}
\end{tikzpicture}%

%% file: offsetEstimationPerformanceNumAntennas.tex
\begin{tikzpicture}

\begin{axis}[%
%width=0.55\textwidth,
%height=0.4\textwidth,
width=0.4\textwidth,
height=0.3\textwidth,
at={(1.083in,0.808in)},
scale only axis,
xmin=-30,
xmax=20,
xlabel style={font=\color{white!15!black}},
xlabel={SNR (dB)},
ymode=log,
ymin=8.14382442220741e-05,
ymax=0.0230095829935606,
yminorticks=true,
ylabel style={font=\color{white!15!black}},
ylabel={RMSE (cycles/sample)},
axis background/.style={fill=white},
xmajorgrids,
ymajorgrids,
yminorgrids,
legend style={nodes={scale=0.9, transform shape}, at={(0,0)}, anchor=south west, legend cell align=left, align=left, draw=white!15!black}
]
\addplot [color=blue, line width=1.0pt, mark=square, mark options={solid, blue}]
  table[row sep=crcr]{%
-30	0.0230095829935606\\
-27	0.0229668668161296\\
-24	0.0228799786664205\\
-21	0.0226874323841992\\
-18	0.0222118495027348\\
-15	0.0214825217685645\\
-12	0.0198258407893291\\
-9	0.0170186312436554\\
-6	0.0127360752200218\\
-3	0.00873426132728589\\
0	0.00610911651963951\\
3	0.00432313273901665\\
6	0.00307610725542441\\
9	0.00220474361427366\\
12	0.001554585696797\\
15	0.00111324696812979\\
18	0.00079107916522915\\
21	0.00056326447968863\\
24	0.000398267513481881\\
27	0.000281351723923451\\
30	0.00020022778440045\\
};
\addlegendentry{$M=8$ FGB}

\addplot [color=blue, dashed, line width=1.0pt, mark=x, mark options={solid, blue}]
  table[row sep=crcr]{%
-30	0.0229955511799563\\
-27	0.0229037738135514\\
-24	0.0228030173378104\\
-21	0.0226150265766007\\
-18	0.0222059729228878\\
-15	0.021182222013058\\
-12	0.0192998574661651\\
-9	0.015789271317268\\
-6	0.0112591368806374\\
-3	0.00775987833629158\\
0	0.00547387971563796\\
3	0.00390433598203077\\
6	0.00278157905655294\\
9	0.00198726352366709\\
12	0.00141938512450833\\
15	0.00101277357296229\\
18	0.000715696217230723\\
21	0.000510561571278594\\
24	0.000360995172041204\\
27	0.000254900278160573\\
30	0.000181099048625218\\
};
\addlegendentry{$M=16$ FGB}

\addplot [color=blue, dashdotted, line width=1.0pt, mark=asterisk, mark options={solid, blue}]
  table[row sep=crcr]{%
-30	0.0229841630207206\\
-27	0.0229287226494468\\
-24	0.0227690158399624\\
-21	0.0225967264850492\\
-18	0.0219875749226295\\
-15	0.0208982808282388\\
-12	0.0186499873973464\\
-9	0.0148486298589839\\
-6	0.0104439872827455\\
-3	0.0071750987739755\\
0	0.00507592529406149\\
3	0.00364059170669155\\
6	0.00259911245863455\\
9	0.00185617526754236\\
12	0.00132632545228919\\
15	0.000941403005388447\\
18	0.000668377896967333\\
21	0.000473260038946052\\
24	0.000337358378768589\\
27	0.000238725542346846\\
30	0.000168745402557057\\
};
\addlegendentry{$M=32$ FGB}

\addplot [color=blue, dotted, line width=1.0pt, mark=o, mark options={solid, blue}]
  table[row sep=crcr]{%
-30	0.022988819561275\\
-27	0.0229095753758912\\
-24	0.0227809945384338\\
-21	0.0224686417197077\\
-18	0.0219312107401175\\
-15	0.0207328582916312\\
-12	0.0182991506320721\\
-9	0.0142650630998978\\
-6	0.00975959979175895\\
-3	0.00680303748938095\\
0	0.00483346553269473\\
3	0.00345967805801846\\
6	0.0024799997892823\\
9	0.00176153150276381\\
12	0.00125917806576752\\
15	0.000894184831885822\\
18	0.000638621644365014\\
21	0.000449543232300409\\
24	0.000319468083634367\\
27	0.000225510162638544\\
30	0.000160650077904439\\
};
\addlegendentry{$M=64$ FGB}

\addplot [color=black, line width=1.0pt, mark=square, mark options={solid, black}]
  table[row sep=crcr]{%
-30	0.0229385530108334\\
-27	0.0228408480939543\\
-24	0.0226032302091732\\
-21	0.0221397876903283\\
-18	0.0211136368198018\\
-15	0.0189207898088329\\
-12	0.0148938564027738\\
-9	0.00970127457048587\\
-6	0.00599182911111971\\
-3	0.00390660285600095\\
0	0.00265835149746855\\
3	0.00183759529867254\\
6	0.00128884402779878\\
9	0.000912483511778538\\
12	0.000647001142668155\\
15	0.000456583393581002\\
18	0.000323411241246294\\
21	0.000228956991966956\\
24	0.000162126640689183\\
27	0.000115077441166329\\
30	8.13394750515948e-05\\
};
\addlegendentry{$M=8$ BeamSync}

\addplot [color=black, dashed, line width=1.0pt, mark=x, mark options={solid, black}]
  table[row sep=crcr]{%
-30	0.0228759907408387\\
-27	0.0227290556429086\\
-24	0.0223846284640361\\
-21	0.0215649881160793\\
-18	0.0197879826839584\\
-15	0.0157849654956666\\
-12	0.0100835073608687\\
-9	0.00590684772804445\\
-6	0.0037621973466344\\
-3	0.00250992287212311\\
0	0.00173269259994445\\
3	0.00121143141975661\\
6	0.000849776533941595\\
9	0.000603055222474534\\
12	0.000427522244270884\\
15	0.00030262984400091\\
18	0.000213967442419456\\
21	0.000151628885056808\\
24	0.000107432296408523\\
27	7.5845347919378e-05\\
30	5.35640540749541e-05\\
};
\addlegendentry{$M=16$ BeamSync}

\addplot [color=black, dashdotted, line width=1.0pt, mark=asterisk, mark options={solid, black}]
  table[row sep=crcr]{%
-30	0.0228856223948215\\
-27	0.0225552209110532\\
-24	0.0219848853825775\\
-21	0.0205592751989466\\
-18	0.017086413635542\\
-15	0.0111958235695158\\
-12	0.0062517954068447\\
-9	0.00379958767115718\\
-6	0.00246734271001202\\
-3	0.00169132676769839\\
0	0.00117827743492453\\
3	0.000827870605006949\\
6	0.000581587019274752\\
9	0.000412587437146647\\
12	0.000289372765560796\\
15	0.0002056146501887\\
18	0.000145654459632552\\
21	0.000103112807703934\\
24	7.32875785668642e-05\\
27	5.20338237068706e-05\\
30	3.68517735516034e-05\\
};
\addlegendentry{$M=32$ BeamSync}

\addplot [color=black, dotted, line width=1.0pt, mark=o, mark options={solid, black}]
  table[row sep=crcr]{%
-30	0.0227271955364965\\
-27	0.0222092339683699\\
-24	0.0211405782121112\\
-21	0.0185468209199605\\
-18	0.0129929259442317\\
-15	0.00708457954095307\\
-12	0.00403239060619084\\
-9	0.00253914403004126\\
-6	0.00169142055265163\\
-3	0.00116185995227629\\
0	0.000810931999090541\\
3	0.000569718527533819\\
6	0.000401671913815601\\
9	0.000284824790381946\\
12	0.000200652151799761\\
15	0.000142826283860709\\
18	0.000100724758669532\\
21	7.17949869983449e-05\\
24	5.05699367042818e-05\\
27	3.59094833623695e-05\\
30	2.54594387489725e-05\\
};
\addlegendentry{$M=64$ BeamSync}

\end{axis}
\end{tikzpicture}%

%% file: frequencyOffsetRangePlot.tex
\begin{tikzpicture}

\begin{axis}[%
%width=0.55\textwidth,
%height=0.4\textwidth,
width=0.4\textwidth,
height=0.3\textwidth,
at={(1.083in,0.808in)},
scale only axis,
xmode=log,
xmin=10,
xmax=1000,
xminorticks=true,
xlabel style={font=\color{white!15!black}},
xlabel={Frequency offset (Hz)},
ymode=log,
ymin=0.0001,
ymax=1,
yminorticks=true,
ylabel style={font=\color{white!15!black}},
ylabel={$\sqrt{\mathbb{E}\{\Vert\frac{\hat{\Delta} - \Delta}{\Delta}\Vert^2 \}}$},
axis background/.style={fill=white},
xmajorgrids,
xminorgrids,
ymajorgrids,
yminorgrids,
legend style={legend cell align=left, align=left, draw=white!15!black}
]
\addplot [color=blue, line width=1.0pt]
  table[row sep=crcr]{%
10	0.24033331021729\\
10.4761575278966	0.220288616741993\\
10.9749876549306	0.217275959948811\\
11.4975699539774	0.213033400626021\\
12.0450354025878	0.201976881559923\\
12.6185688306602	0.190155943930011\\
13.2194114846603	0.181896886551727\\
13.8488637139387	0.171024008890479\\
14.5082877849594	0.164031881020523\\
15.1991108295293	0.156697395677869\\
15.9228279334109	0.146678193130131\\
16.6810053720006	0.143953122955919\\
17.4752840000768	0.132500228202039\\
18.3073828029537	0.128702631685162\\
19.1791026167249	0.123352915070927\\
20.0923300256505	0.120705800078723\\
21.0490414451202	0.113431105209669\\
22.0513073990305	0.105510339552245\\
23.1012970008316	0.103757817266565\\
24.2012826479438	0.0976991827984426\\
25.3536449397011	0.0950935596425558\\
26.5608778294669	0.0894600317360579\\
27.8255940220712	0.0867353360933286\\
29.1505306282518	0.0822062475052677\\
30.5385550883342	0.0798150823680501\\
31.9926713779738	0.0736780987428901\\
33.5160265093884	0.0690499411820302\\
35.1119173421513	0.0654206000416685\\
36.7837977182863	0.063895280314217\\
38.5352859371053	0.0620630099984169\\
40.3701725859655	0.0584339078574352\\
42.292428743895	0.0581029224710863\\
44.3062145758388	0.0519954352570864\\
46.4158883361278	0.0500770849516459\\
48.6260158006535	0.0499867513841118\\
50.9413801481638	0.0444452955607781\\
53.3669923120631	0.0441519909015477\\
55.9081018251222	0.0419216759896065\\
58.5702081805667	0.039871239033212\\
61.3590727341317	0.0386349475101652\\
64.2807311728432	0.0385757592051932\\
67.3415065775082	0.0354575811861029\\
70.5480231071865	0.032958846954104\\
73.9072203352578	0.031516550800459\\
77.4263682681127	0.0300059033353798\\
81.1130830789687	0.0294887996872467\\
84.9753435908644	0.0275468958794176\\
89.0215085445038	0.0264439031917461\\
93.260334688322	0.0259085609162732\\
97.7009957299226	0.0244769934377242\\
102.353102189903	0.0240539283812558\\
107.226722201032	0.0217502905770154\\
112.332403297803	0.0204445516507774\\
117.6811952435	0.0210286779853\\
123.284673944207	0.0192773832243277\\
129.154966501488	0.0181846879256602\\
135.304777457981	0.0178387208214085\\
141.74741629268	0.0173599567968163\\
148.496826225447	0.0162926064737972\\
155.567614393047	0.0151615694160177\\
162.975083462064	0.0146035099052236\\
170.735264747069	0.0137046388604723\\
178.864952905744	0.0129678817909872\\
187.381742286038	0.0123921267739334\\
196.304065004027	0.0119082492570727\\
205.651230834865	0.0110691045776255\\
215.443469003188	0.0112109680028783\\
225.701971963392	0.0107914930346067\\
236.448941264541	0.0101807592468256\\
247.707635599171	0.00949754589886214\\
259.502421139974	0.00917615017819924\\
271.858824273294	0.00829959827195627\\
284.80358684358	0.00825324367681054\\
298.364724028334	0.00831346079504071\\
312.571584968824	0.00772310497769997\\
327.454916287773	0.00728952921646206\\
343.046928631492	0.00689058086555439\\
359.381366380463	0.00665022111900514\\
376.493580679247	0.00646088847277959\\
394.420605943766	0.00640874103299127\\
413.201240011533	0.00583194741698228\\
432.876128108306	0.0055022509911804\\
453.487850812858	0.00527271031761425\\
475.081016210279	0.00523060675082717\\
497.702356433211	0.00470447660901076\\
521.400828799968	0.00475352457666643\\
546.227721768434	0.00441182219803609\\
572.236765935022	0.00417496003565835\\
599.484250318941	0.0039811142355195\\
628.029144183425	0.00386598188782268\\
657.933224657568	0.0036391757998013\\
689.261210434969	0.00350626951042421\\
722.080901838546	0.00335957636879769\\
756.463327554629	0.00310023828024287\\
792.482898353917	0.00298398355528997\\
830.217568131974	0.00298249939120437\\
869.749002617783	0.0027512345625263\\
911.162756115489	0.00262637626080689\\
954.548456661835	0.00247231816115485\\
1000	0.00165526432934438\\
};
\addlegendentry{$M=16$, $N_f=10$}

\addplot [color=blue, dashed, line width=1.0pt]
  table[row sep=crcr]{%
10	0.161492724294316\\
10.4761575278966	0.154849784847678\\
10.9749876549306	0.145606159210202\\
11.4975699539774	0.143138472010381\\
12.0450354025878	0.134496157885101\\
12.6185688306602	0.124804086239275\\
13.2194114846603	0.119411067943567\\
13.8488637139387	0.117452259324808\\
14.5082877849594	0.113763879555391\\
15.1991108295293	0.104575426223128\\
15.9228279334109	0.10120669619988\\
16.6810053720006	0.0937275809844805\\
17.4752840000768	0.0968691128923372\\
18.3073828029537	0.0887218790838845\\
19.1791026167249	0.0834831881057832\\
20.0923300256505	0.0794238911044041\\
21.0490414451202	0.0761834234194051\\
22.0513073990305	0.0708358776621778\\
23.1012970008316	0.0686622397537863\\
24.2012826479438	0.0683385397614478\\
25.3536449397011	0.0638915963453027\\
26.5608778294669	0.0634093025375981\\
27.8255940220712	0.0590902563436859\\
29.1505306282518	0.055916981573148\\
30.5385550883342	0.0523170380343257\\
31.9926713779738	0.0490572518103364\\
33.5160265093884	0.0488765268989661\\
35.1119173421513	0.0445813474200633\\
36.7837977182863	0.0438857799518058\\
38.5352859371053	0.0420877385963829\\
40.3701725859655	0.0401765636356156\\
42.292428743895	0.0381391020121923\\
44.3062145758388	0.0375664972553062\\
46.4158883361278	0.0355775122268151\\
48.6260158006535	0.033249860687763\\
50.9413801481638	0.0326477418242079\\
53.3669923120631	0.0289105136065637\\
55.9081018251222	0.0293009097809115\\
58.5702081805667	0.0274044262092453\\
61.3590727341317	0.0266806542784736\\
64.2807311728432	0.0244768552761618\\
67.3415065775082	0.0237802593715312\\
70.5480231071865	0.0226917198263028\\
73.9072203352578	0.0216778190190705\\
77.4263682681127	0.0207276167941993\\
81.1130830789687	0.0209000336852362\\
84.9753435908644	0.0189949905512185\\
89.0215085445038	0.0176799959425065\\
93.260334688322	0.0171832527344131\\
97.7009957299226	0.0162169891025702\\
102.353102189903	0.0153282294610032\\
107.226722201032	0.0149214712630265\\
112.332403297803	0.0142326017322905\\
117.6811952435	0.0136981880987221\\
123.284673944207	0.0129759435545946\\
129.154966501488	0.0124002999699692\\
135.304777457981	0.0117157183554683\\
141.74741629268	0.0113125560333839\\
148.496826225447	0.0112170608112517\\
155.567614393047	0.0100606220567051\\
162.975083462064	0.0100834951562642\\
170.735264747069	0.009469803905343\\
178.864952905744	0.00931885066250981\\
187.381742286038	0.00864852086489723\\
196.304065004027	0.00819890142218522\\
205.651230834865	0.00821434087755723\\
215.443469003188	0.0074663354263978\\
225.701971963392	0.00721922650862629\\
236.448941264541	0.00684373992739331\\
247.707635599171	0.00657313050624272\\
259.502421139974	0.00634271515442077\\
271.858824273294	0.00614592427679372\\
284.80358684358	0.00543516286873371\\
298.364724028334	0.00550133889485582\\
312.571584968824	0.00524968924577176\\
327.454916287773	0.00501975911816598\\
343.046928631492	0.00478015441181028\\
359.381366380463	0.00438716820230467\\
376.493580679247	0.00439419360826616\\
394.420605943766	0.00410166153757061\\
413.201240011533	0.00392588499242061\\
432.876128108306	0.00392399235908662\\
453.487850812858	0.00364814643757294\\
475.081016210279	0.00346695281413141\\
497.702356433211	0.00332381534801999\\
521.400828799968	0.00315718478939586\\
546.227721768434	0.00311757257662689\\
572.236765935022	0.00277144705282317\\
599.484250318941	0.00272531101826469\\
628.029144183425	0.00260450631153318\\
657.933224657568	0.0024394897915949\\
689.261210434969	0.00238907470679474\\
722.080901838546	0.00220875935268623\\
756.463327554629	0.00211120226770158\\
792.482898353917	0.00203397913264589\\
830.217568131974	0.0019818325090309\\
869.749002617783	0.0018856341727433\\
911.162756115489	0.00180490090121556\\
954.548456661835	0.0016859006779014\\
1000	0.00113786642449802\\
};
\addlegendentry{$M=32$, $N_f=10$}

\addplot [color=black, line width=1.0pt]
  table[row sep=crcr]{%
10	0.0820408434866439\\
10.4761575278966	0.0829754839702284\\
10.9749876549306	0.0783025241751489\\
11.4975699539774	0.077199034504535\\
12.0450354025878	0.069412977686127\\
12.6185688306602	0.0658805196050549\\
13.2194114846603	0.0665473954321679\\
13.8488637139387	0.0611727976563724\\
14.5082877849594	0.0569744409290507\\
15.1991108295293	0.0571203154459344\\
15.9228279334109	0.0545585898903251\\
16.6810053720006	0.0491591234487888\\
17.4752840000768	0.0494778091219014\\
18.3073828029537	0.0465578602706422\\
19.1791026167249	0.043841235964851\\
20.0923300256505	0.0413200945973439\\
21.0490414451202	0.0416718829709264\\
22.0513073990305	0.0372775034957458\\
23.1012970008316	0.0365926973823448\\
24.2012826479438	0.0351819492942242\\
25.3536449397011	0.033514238085789\\
26.5608778294669	0.0303124671722211\\
27.8255940220712	0.0307705640259263\\
29.1505306282518	0.0294684847061053\\
30.5385550883342	0.0275387105980293\\
31.9926713779738	0.0262709982780946\\
33.5160265093884	0.0245197371918848\\
35.1119173421513	0.0234600581088689\\
36.7837977182863	0.0234004127481022\\
38.5352859371053	0.0212791860604625\\
40.3701725859655	0.0200043121831004\\
42.292428743895	0.019539075372638\\
44.3062145758388	0.0196485184247774\\
46.4158883361278	0.0176983380003479\\
48.6260158006535	0.0173289277174374\\
50.9413801481638	0.0162746217308277\\
53.3669923120631	0.0156934956777667\\
55.9081018251222	0.015389409947203\\
58.5702081805667	0.0139422602959638\\
61.3590727341317	0.0138730287490237\\
64.2807311728432	0.0128936852456062\\
67.3415065775082	0.0125275202401987\\
70.5480231071865	0.0120698401688626\\
73.9072203352578	0.011808752235304\\
77.4263682681127	0.0105643895114608\\
81.1130830789687	0.0107104100279959\\
84.9753435908644	0.00998999922211861\\
89.0215085445038	0.00942108820738954\\
93.260334688322	0.0090971362255645\\
97.7009957299226	0.00877293917924186\\
102.353102189903	0.0081117748360576\\
107.226722201032	0.00755825912188182\\
112.332403297803	0.00745978527095174\\
117.6811952435	0.00709231028530732\\
123.284673944207	0.00682740811377922\\
129.154966501488	0.00644272066342862\\
135.304777457981	0.00660303916273674\\
141.74741629268	0.0059062434987855\\
148.496826225447	0.00564464304033817\\
155.567614393047	0.00571124930109839\\
162.975083462064	0.00523138797656379\\
170.735264747069	0.00486792279940615\\
178.864952905744	0.00473806819649274\\
187.381742286038	0.00448125675499786\\
196.304065004027	0.00431294723506305\\
205.651230834865	0.00405515129847531\\
215.443469003188	0.00390640017614788\\
225.701971963392	0.00375100294522101\\
236.448941264541	0.00357120672268963\\
247.707635599171	0.00335167677667681\\
259.502421139974	0.00321175878123514\\
271.858824273294	0.00311304726060458\\
284.80358684358	0.0029356289917509\\
298.364724028334	0.00273315677161489\\
312.571584968824	0.00270233820900338\\
327.454916287773	0.0026804944259002\\
343.046928631492	0.00245343379933199\\
359.381366380463	0.00230805062294239\\
376.493580679247	0.00220211888044925\\
394.420605943766	0.00219939149700611\\
413.201240011533	0.00204110382262477\\
432.876128108306	0.00187399128949102\\
453.487850812858	0.00181207129628984\\
475.081016210279	0.00176748294515106\\
497.702356433211	0.0016981667544931\\
521.400828799968	0.0016280246156725\\
546.227721768434	0.0014782883745478\\
572.236765935022	0.00148680062138543\\
599.484250318941	0.00137169236056505\\
628.029144183425	0.00132450123862287\\
657.933224657568	0.00127503772565695\\
689.261210434969	0.00124273687638327\\
722.080901838546	0.00118877198596931\\
756.463327554629	0.0010788317254713\\
792.482898353917	0.0010850790327606\\
830.217568131974	0.00102831482901849\\
869.749002617783	0.00103505221110187\\
911.162756115489	0.000943837717144151\\
954.548456661835	0.000891447191714654\\
1000	0.000590652181914153\\
};
\addlegendentry{$M=16$, $N_f=20$}

\addplot [color=black, dashed, line width=1.0pt]
  table[row sep=crcr]{%
10	0.0571655490658491\\
10.4761575278966	0.0562449556208259\\
10.9749876549306	0.0534741167755307\\
11.4975699539774	0.0481310348969803\\
12.0450354025878	0.0484124723340398\\
12.6185688306602	0.0456943096677691\\
13.2194114846603	0.044216992829887\\
13.8488637139387	0.0411444679704816\\
14.5082877849594	0.0391683424239998\\
15.1991108295293	0.0385133843735989\\
15.9228279334109	0.0355964068185737\\
16.6810053720006	0.033413067393888\\
17.4752840000768	0.03362279063343\\
18.3073828029537	0.0318996035140772\\
19.1791026167249	0.0307065368154785\\
20.0923300256505	0.0286122593989341\\
21.0490414451202	0.0274532977795184\\
22.0513073990305	0.0252588648613003\\
23.1012970008316	0.0254184286350347\\
24.2012826479438	0.023273195626897\\
25.3536449397011	0.0222185845754912\\
26.5608778294669	0.0225704206037764\\
27.8255940220712	0.0214029899541998\\
29.1505306282518	0.0202640839308226\\
30.5385550883342	0.0187199230406574\\
31.9926713779738	0.0182150317638794\\
33.5160265093884	0.0170356158003929\\
35.1119173421513	0.0160903384634892\\
36.7837977182863	0.0156169386987561\\
38.5352859371053	0.014713829117415\\
40.3701725859655	0.0136384585556278\\
42.292428743895	0.0133073644648005\\
44.3062145758388	0.0128365848820121\\
46.4158883361278	0.0123235620598905\\
48.6260158006535	0.0118891906917435\\
50.9413801481638	0.0111633977306201\\
53.3669923120631	0.0105260339527502\\
55.9081018251222	0.0107927586986648\\
58.5702081805667	0.00944890456535563\\
61.3590727341317	0.00942356911708323\\
64.2807311728432	0.00931149718682933\\
67.3415065775082	0.00845533683490899\\
70.5480231071865	0.00809685691891652\\
73.9072203352578	0.0077991594637092\\
77.4263682681127	0.00731555908121845\\
81.1130830789687	0.00687182027583882\\
84.9753435908644	0.00703969810949096\\
89.0215085445038	0.00633136269340567\\
93.260334688322	0.00601608782112099\\
97.7009957299226	0.00567309174954872\\
102.353102189903	0.00551070513855731\\
107.226722201032	0.00521298566789448\\
112.332403297803	0.00507588221703144\\
117.6811952435	0.00491251943016534\\
123.284673944207	0.00442101711352922\\
129.154966501488	0.00448657349651072\\
135.304777457981	0.00432204563079583\\
141.74741629268	0.00392904217667676\\
148.496826225447	0.00398350865050523\\
155.567614393047	0.00370336226282644\\
162.975083462064	0.00360331258444789\\
170.735264747069	0.00344953399192269\\
178.864952905744	0.0031156261091423\\
187.381742286038	0.0030192591878415\\
196.304065004027	0.00298719802845374\\
205.651230834865	0.00276027967637712\\
215.443469003188	0.00272358523288696\\
225.701971963392	0.00253259835855043\\
236.448941264541	0.00236382615003985\\
247.707635599171	0.00231081538924041\\
259.502421139974	0.00229983561157515\\
271.858824273294	0.00212376738233272\\
284.80358684358	0.0020249370053085\\
298.364724028334	0.00196856091332318\\
312.571584968824	0.00187671623515239\\
327.454916287773	0.00176736829731983\\
343.046928631492	0.0017134554599065\\
359.381366380463	0.00164163344537359\\
376.493580679247	0.0015270544844859\\
394.420605943766	0.00146382106399413\\
413.201240011533	0.00138233949062598\\
432.876128108306	0.00133400922777553\\
453.487850812858	0.00128046985292758\\
475.081016210279	0.00119611619579599\\
497.702356433211	0.0011321329857374\\
521.400828799968	0.00108953942710031\\
546.227721768434	0.00106704051793318\\
572.236765935022	0.00099462142317305\\
599.484250318941	0.000958003116264062\\
628.029144183425	0.000943094237447155\\
657.933224657568	0.000837473733501323\\
689.261210434969	0.000800882571007131\\
722.080901838546	0.000762399373771464\\
756.463327554629	0.000778477247026107\\
792.482898353917	0.000740912621731317\\
830.217568131974	0.000701977750874795\\
869.749002617783	0.000651782419600415\\
911.162756115489	0.000629078798133862\\
954.548456661835	0.00060882658828919\\
1000	0.00043215737874062\\
};
\addlegendentry{$M=32$, $N_f=20$}

\addplot [color=red, line width=1.0pt]
  table[row sep=crcr]{%
10	0.0282577423018896\\
10.4761575278966	0.0288637225168821\\
10.9749876549306	0.0280388083446154\\
11.4975699539774	0.0257361490788526\\
12.0450354025878	0.0246317214398746\\
12.6185688306602	0.023582337421995\\
13.2194114846603	0.021403422428443\\
13.8488637139387	0.0217231358897739\\
14.5082877849594	0.0202181639508668\\
15.1991108295293	0.0196458898529985\\
15.9228279334109	0.0185324911768598\\
16.6810053720006	0.018365900350425\\
17.4752840000768	0.0168169964063068\\
18.3073828029537	0.016116000903916\\
19.1791026167249	0.0158851347659217\\
20.0923300256505	0.0147855211842453\\
21.0490414451202	0.0144640757929935\\
22.0513073990305	0.0134819675996031\\
23.1012970008316	0.0133736587623461\\
24.2012826479438	0.0126448926644954\\
25.3536449397011	0.0119589276667666\\
26.5608778294669	0.0113504050069804\\
27.8255940220712	0.0107019402206254\\
29.1505306282518	0.0099690585914378\\
30.5385550883342	0.00954304270286761\\
31.9926713779738	0.00917232135217631\\
33.5160265093884	0.008846461568585\\
35.1119173421513	0.00879628598291445\\
36.7837977182863	0.00853542402773837\\
38.5352859371053	0.00785313626384729\\
40.3701725859655	0.00761061760813854\\
42.292428743895	0.00717288656948821\\
44.3062145758388	0.00661336441625518\\
46.4158883361278	0.00637824135278757\\
48.6260158006535	0.00613162036184055\\
50.9413801481638	0.00594810710813692\\
53.3669923120631	0.00529974458539318\\
55.9081018251222	0.00536169034681098\\
58.5702081805667	0.00501858470765224\\
61.3590727341317	0.0048460665258457\\
64.2807311728432	0.00458443708828745\\
67.3415065775082	0.00444543348497633\\
70.5480231071865	0.00404621346916902\\
73.9072203352578	0.00407858488352618\\
77.4263682681127	0.00390989395634401\\
81.1130830789687	0.00372291869642167\\
84.9753435908644	0.00358196418705722\\
89.0215085445038	0.00335315421158619\\
93.260334688322	0.00313749427630899\\
97.7009957299226	0.00313683672915782\\
102.353102189903	0.00280179071785356\\
107.226722201032	0.00291649932909333\\
112.332403297803	0.00290913501022432\\
117.6811952435	0.00259365832635386\\
123.284673944207	0.00245556870556066\\
129.154966501488	0.0022648774165187\\
135.304777457981	0.00222020858560636\\
141.74741629268	0.00204772249567245\\
148.496826225447	0.00209890311404682\\
155.567614393047	0.00189802787143304\\
162.975083462064	0.00184305575575719\\
170.735264747069	0.0017454495499345\\
178.864952905744	0.00169006891722115\\
187.381742286038	0.00159369515758429\\
196.304065004027	0.00149712669996149\\
205.651230834865	0.00148094987685141\\
215.443469003188	0.0013681779085348\\
225.701971963392	0.00133134444658585\\
236.448941264541	0.00126818769587503\\
247.707635599171	0.0011829508403756\\
259.502421139974	0.0011293993302135\\
271.858824273294	0.00110623457397236\\
284.80358684358	0.00105670359609239\\
298.364724028334	0.000965695746276196\\
312.571584968824	0.000963941783504902\\
327.454916287773	0.00088658078634361\\
343.046928631492	0.000874109958613242\\
359.381366380463	0.000834574825073572\\
376.493580679247	0.000837681662290402\\
394.420605943766	0.00073399833956213\\
413.201240011533	0.000703121134221021\\
432.876128108306	0.000689370765395508\\
453.487850812858	0.000654228913636769\\
475.081016210279	0.000608258610947333\\
497.702356433211	0.000606930397974662\\
521.400828799968	0.0005677656626579\\
546.227721768434	0.000546056769181719\\
572.236765935022	0.000527566377948238\\
599.484250318941	0.00050058223936904\\
628.029144183425	0.000467673670630082\\
657.933224657568	0.000482804624953164\\
689.261210434969	0.000430930958327116\\
722.080901838546	0.000409305224851607\\
756.463327554629	0.00039018346763083\\
792.482898353917	0.000381895095691139\\
830.217568131974	0.000352419536453684\\
869.749002617783	0.000347216714908588\\
911.162756115489	0.000325695765050731\\
954.548456661835	0.000315055224568639\\
1000	0.000213002347404865\\
};
\addlegendentry{$M=16$, $N_f=40$}

\addplot [color=red, dashed, line width=1.0pt]
  table[row sep=crcr]{%
10	0.020206434618705\\
10.4761575278966	0.0185817238084367\\
10.9749876549306	0.0178364291543114\\
11.4975699539774	0.0185082413834891\\
12.0450354025878	0.0178192943899127\\
12.6185688306602	0.0164283105499331\\
13.2194114846603	0.0151569399133297\\
13.8488637139387	0.0149094519729171\\
14.5082877849594	0.0147538870799178\\
15.1991108295293	0.0132113685666414\\
15.9228279334109	0.0132369069887555\\
16.6810053720006	0.0123167011610514\\
17.4752840000768	0.0108758726384359\\
18.3073828029537	0.0112687641928477\\
19.1791026167249	0.0107221414521152\\
20.0923300256505	0.0102008665979859\\
21.0490414451202	0.0100821849549512\\
22.0513073990305	0.00923579135542609\\
23.1012970008316	0.00854112494689725\\
24.2012826479438	0.00773279112083791\\
25.3536449397011	0.00779747489928868\\
26.5608778294669	0.007281073462747\\
27.8255940220712	0.0071925134714398\\
29.1505306282518	0.006726522115284\\
30.5385550883342	0.00673683944391944\\
31.9926713779738	0.0063405642837474\\
33.5160265093884	0.00601871478066629\\
35.1119173421513	0.00592041466811627\\
36.7837977182863	0.00549833224100079\\
38.5352859371053	0.0056069779861809\\
40.3701725859655	0.00505960007379852\\
42.292428743895	0.00474452581671053\\
44.3062145758388	0.00467380558017666\\
46.4158883361278	0.00438852159866791\\
48.6260158006535	0.00415683593839911\\
50.9413801481638	0.00402781463088289\\
53.3669923120631	0.00383512509325441\\
55.9081018251222	0.00378092244509778\\
58.5702081805667	0.00352093126315001\\
61.3590727341317	0.00341615488164762\\
64.2807311728432	0.00303957195054572\\
67.3415065775082	0.00306096847255914\\
70.5480231071865	0.00289454082264286\\
73.9072203352578	0.00276524817022849\\
77.4263682681127	0.0025700724570193\\
81.1130830789687	0.002476233077605\\
84.9753435908644	0.00231453470149513\\
89.0215085445038	0.00226199685606498\\
93.260334688322	0.00215644444082873\\
97.7009957299226	0.00211999034404085\\
102.353102189903	0.00206662306327718\\
107.226722201032	0.00196300633753606\\
112.332403297803	0.00176731581305506\\
117.6811952435	0.00175905923986072\\
123.284673944207	0.0017469950939291\\
129.154966501488	0.00165575678008918\\
135.304777457981	0.00157047205111474\\
141.74741629268	0.00140205845197009\\
148.496826225447	0.00135131583220284\\
155.567614393047	0.00130349608835531\\
162.975083462064	0.00129909548439728\\
170.735264747069	0.00122996747951001\\
178.864952905744	0.00114295768971013\\
187.381742286038	0.00108305503576547\\
196.304065004027	0.00100799283617323\\
205.651230834865	0.000990871152232121\\
215.443469003188	0.000955847553505395\\
225.701971963392	0.000908934562053195\\
236.448941264541	0.000858108030921345\\
247.707635599171	0.000854577666346258\\
259.502421139974	0.000754981207380853\\
271.858824273294	0.000769307525374177\\
284.80358684358	0.000710258171968334\\
298.364724028334	0.000701121914159279\\
312.571584968824	0.000639700025175743\\
327.454916287773	0.000615694501923617\\
343.046928631492	0.000583972962275078\\
359.381366380463	0.00056415722951561\\
376.493580679247	0.000520988595843608\\
394.420605943766	0.000503836914772349\\
413.201240011533	0.000477002611995723\\
432.876128108306	0.000471410809906968\\
453.487850812858	0.000448865764754466\\
475.081016210279	0.000429534523302411\\
497.702356433211	0.000424611980874333\\
521.400828799968	0.000389708701129208\\
546.227721768434	0.000365468659432653\\
572.236765935022	0.000344458738907584\\
599.484250318941	0.000347132135544404\\
628.029144183425	0.000317731041062944\\
657.933224657568	0.00032089493903577\\
689.261210434969	0.000298349146219444\\
722.080901838546	0.00028087669959386\\
756.463327554629	0.000271914485008718\\
792.482898353917	0.000257821348267684\\
830.217568131974	0.000255218897209761\\
869.749002617783	0.000242476513078923\\
911.162756115489	0.000224803151688964\\
954.548456661835	0.000218227609914102\\
1000	0.000151228304229038\\
};
\addlegendentry{$M=32$, $N_f=40$}

\end{axis}
\end{tikzpicture}%

%% file: BeamSyncPhasePerformanceWithFrequencyOffset.tex
\begin{tikzpicture}

\begin{axis}[%
%width=0.55\textwidth,
%height=0.4\textwidth,
width=0.4\textwidth,
height=0.3\textwidth,
at={(1.083in,0.808in)},
scale only axis,
xmin=-30,
xmax=30,
xlabel style={font=\color{white!15!black}},
xlabel={SNR (dB)},
ymode=log,
ymin=0.0001,
ymax=10,
yminorticks=true,
ylabel style={font=\color{white!15!black}},
ylabel={RMSE (radians)},
axis background/.style={fill=white},
xmajorgrids,
ymajorgrids,
yminorgrids,
legend style={legend cell align=left, align=left, draw=white!15!black}
]
\addplot [color=blue, line width=1.0pt, mark=asterisk, mark options={solid, blue}]
  table[row sep=crcr]{%
-30	1.72106975117401\\
-27	1.62859113207408\\
-24	1.43678283837798\\
-21	1.09221705461125\\
-18	0.612510700397549\\
-15	0.27789356791807\\
-12	0.147298278725691\\
-9	0.0870425527704557\\
-6	0.0532934182910291\\
-3	0.0332435319954959\\
0	0.0208782993999382\\
3	0.0135596881401059\\
6	0.00900672998858603\\
9	0.00613897631622442\\
12	0.00426462838766517\\
15	0.00298158686428141\\
18	0.00210520182454673\\
21	0.00148447896840619\\
24	0.00104720140432771\\
27	0.000741797644802663\\
30	0.000525296006366934\\
};
\addlegendentry{$\Delta=0$ Hz}

\addplot [color=black, dashdotted, line width=1.0pt, mark=diamond, mark options={solid, black}]
  table[row sep=crcr]{%
-30	1.72098419406812\\
-27	1.62842580221252\\
-24	1.4368092026917\\
-21	1.09226741283231\\
-18	0.612988476863861\\
-15	0.277974002355112\\
-12	0.147462829575341\\
-9	0.087347808996191\\
-6	0.0537589315013186\\
-3	0.0340081598936712\\
0	0.0220762771983568\\
3	0.015337349397226\\
6	0.01151535007755\\
9	0.00945610277830806\\
12	0.00834238693100704\\
15	0.0077680399412008\\
18	0.0074773900874499\\
21	0.00733397332933033\\
24	0.00725356850482634\\
27	0.00722039738922142\\
30	0.00719964344820028\\
};
\addlegendentry{$\Delta=0.5$ Hz}

\addplot [color=red, dotted, line width=1.0pt, mark=square, mark options={solid, red}]
  table[row sep=crcr]{%
-30	1.72105450777502\\
-27	1.62837847875126\\
-24	1.43675768549891\\
-21	1.09248839158699\\
-18	0.613066055277671\\
-15	0.278356706046494\\
-12	0.147970094933517\\
-9	0.0882353397420432\\
-6	0.0551809408381495\\
-3	0.0362207804667483\\
0	0.0253118092550179\\
3	0.0197431604479278\\
6	0.016951193152523\\
9	0.0156314028119955\\
12	0.0149705720599213\\
15	0.0146623876728685\\
18	0.0145153177799226\\
21	0.0144406430139113\\
24	0.0144002031694909\\
27	0.0143816621340624\\
30	0.0143712579249014\\
};
\addlegendentry{$\Delta=1$ Hz}

\addplot [color=blue, dashed, line width=1.0pt, mark=o, mark options={solid, blue}]
  table[row sep=crcr]{%
-30	1.72121903698704\\
-27	1.62879099051205\\
-24	1.43664081007625\\
-21	1.09283801562586\\
-18	0.614305598291229\\
-15	0.279753494108391\\
-12	0.150024948627255\\
-9	0.0916954156490611\\
-6	0.0604967476804652\\
-3	0.043946173458337\\
0	0.0354906504080784\\
3	0.0317484878967087\\
6	0.0300979968568329\\
9	0.0293811478685238\\
12	0.0290303542506497\\
15	0.0288721166994087\\
18	0.0288044935308448\\
21	0.0287638161026961\\
24	0.0287384513341365\\
27	0.02873409651965\\
30	0.0287306780584674\\
};
\addlegendentry{$\Delta=2$ Hz}

\end{axis}
\end{tikzpicture}%